\documentclass[fleqn, usenatbib]{mnras}

\usepackage{newtxtext,newtxmath}
\usepackage{pifont}

\usepackage[T1]{fontenc}

\DeclareRobustCommand{\VAN}[3]{#2}
\let\VANthebibliography\thebibliography
\def\thebibliography{\DeclareRobustCommand{\VAN}[3]{##3}\VANthebibliography}

\usepackage{graphicx}	
\usepackage{amsmath}	

\usepackage{listings}
\usepackage{color}
\definecolor{dkgreen}{rgb}{0,0.6,0}
\definecolor{gray}{rgb}{0.5,0.5,0.5}
\definecolor{mauve}{rgb}{0.58,0,0.82}
\definecolor{golden}{rgb}{0.86,0.65,0.01}
\usepackage{tikz}

\newcommand{\shrug}[1][]{%
\begin{tikzpicture}[baseline,x=0.8\ht\strutbox,y=0.8\ht\strutbox,line width=0.125ex,#1]
\def\arm{(-2.5,0.95) to (-2,0.95) (-1.9,1) to (-1.5,0) (-1.35,0) to (-0.8,0)};
\draw \arm;
\draw[xscale=-1] \arm;
\def\headpart{(0.6,0) arc[start angle=-40, end angle=40,x radius=0.6,y radius=0.8]};
\draw \headpart;
\draw[xscale=-1] \headpart;
\def\eye{(-0.075,0.15) .. controls (0.02,0) .. (0.075,-0.15)};
\draw[shift={(-0.3,0.8)}] \eye;
\draw[shift={(0,0.85)}] \eye;
\draw (-0.1,0.2) to [out=15,in=-100] (0.4,0.95); 
\end{tikzpicture}}

\title[Gaia BH1]{A Sun-like star orbiting a black hole}

\author[El-Badry et al.]{Kareem El-Badry$^{1,2,3}$\thanks{E-mail: kareem.el-badry@cfa.harvard.edu}, Hans-Walter Rix$^{3}$, Eliot Quataert$^4$, Andrew W. Howard$^5$, Howard Isaacson$^{6,7}$,  \newauthor Jim Fuller$^5$, Keith Hawkins$^8$, Katelyn Breivik$^9$, Kaze W. K. Wong$^9$, Antonio C. Rodriguez$^5$,  \newauthor Charlie Conroy$^1$, Sahar Shahaf$^{10}$, Tsevi Mazeh$^{11}$, Fr\'{e}d\'{e}ric Arenou$^{12}$,  Kevin B. Burdge$^{13}$, Dolev Bashi$^{11}$,  \newauthor  Simchon Faigler$^{11}$, Daniel R. Weisz$^{6}$, Rhys Seeburger$^{3}$, Silvia Almada Monter$^{3}$, Jennifer Wojno$^{3}$ \\  \\ 
$^{1}$Center for Astrophysics $|$ Harvard \& Smithsonian, 60 Garden Street, Cambridge, MA 02138, USA\\
$^{2}$Harvard Society of Fellows, 78 Mount Auburn Street, Cambridge, MA 02138\\
$^{3}$Max-Planck Institute for Astronomy, K\"onigstuhl 17, D-69117 Heidelberg, Germany\\
$^{4}$Department of Astrophysical Sciences, Princeton University, Princeton, NJ 08544, USA\\
$^{5}$Department of Astronomy, California Institute of Technology, Pasadena, CA 91125, USA\\
$^{6}$Department of Astronomy, University of California, Berkeley, 501 Campbell Hall \#3411, Berkeley, CA 94720, USA \\
$^{7}$Centre for Astrophysics, University of Southern Queensland, Toowoomba, QLD, Australia \\
$^{8}$Department of Astronomy, The University of Texas at Austin, 2515 Speedway Boulevard, Austin, TX 78712, USA \\
$^{9}$Center for Computational Astrophysics, Flatiron Institute, 162 Fifth Ave, New York, NY, 10010, USA \\
$^{10}$Department of Particle Physics and Astrophysics, Weizmann Institute of Science, Rehovot 7610001, Israel \\
$^{11}$School of Physics and Astronomy, Tel Aviv University, Tel Aviv, 6997801, Israel \\
$^{12}$GEPI, Observatoire de Paris, Universit\'{e} PSL, CNRS, 5 Place Jules Janssen, 92190 Meudon, France \\
$^{13}$MIT-Kavli Institute for Astrophysics and Space Research 77 Massachusetts Ave. Cambridge, MA 02139, USA}

\date{\vspace{-1.0cm}}

\pubyear{2022}

\begin{document}
\label{firstpage}
\pagerange{\pageref{firstpage}--\pageref{lastpage}}
\maketitle

\begin{abstract}
We report discovery of a bright, nearby ($G = 13.8;\,\,d = 480\,\rm pc$) Sun-like star orbiting a dark object. We identified the system as a black hole candidate via its astrometric orbital solution from the {\it Gaia} mission. Radial velocities validated and refined the {\it Gaia} solution, and spectroscopy ruled out significant light contributions from another star. Joint modeling of radial velocities and astrometry constrains the companion mass to $M_2 = 9.62\pm 0.18\,M_{\odot}$. The spectroscopic orbit alone sets a minimum companion mass of $M_2>5\,M_{\odot}$; if the companion were a $5\,M_{\odot}$ star, it would be $500$ times more luminous than the entire system. These constraints are insensitive to the mass of the luminous star, which appears as a slowly-rotating G dwarf ($T_{\rm eff}=5850\,\rm K$, $\log g = 4.5$, $M=0.93\,M_{\odot}$), with near-solar metallicity ($\rm [Fe/H] = -0.2$) and an unremarkable abundance pattern. We find no plausible astrophysical scenario that can explain the orbit and does not involve a black hole. The orbital period, $P_{\rm orb}=185.6$ days, is longer than that of any known stellar-mass black hole binary. The system's modest eccentricity ($e=0.45$), high metallicity, and thin-disk Galactic orbit suggest that it was born in the Milky Way disk with at most a weak natal kick. How the system formed is uncertain. Common envelope evolution can only produce the system's wide orbit under extreme and likely unphysical assumptions. Formation models involving triples or dynamical assembly in an open cluster may be more promising. This is the nearest known black hole by a factor of 3, and its discovery suggests the existence of a sizable population of dormant black holes in binaries. Future {\it Gaia} releases will likely facilitate the discovery of dozens more.
\end{abstract}

\begin{keywords}
binaries: spectroscopic -- stars: black holes 
\vspace{-0.5cm}
\end{keywords}



\section{Introduction}
\label{sec:intro}
The Milky Way is expected to contain of order $10^8$ stellar mass black holes (BHs), an unknown fraction of which are in binaries. The inventory of known and suspected BHs consists of about 20 dynamically confirmed BHs in X-ray binaries, an additional $\sim 50$ X-ray sources suspected to contain a BH based on their X-ray properties \citep[e.g.][]{McClintock2006, Remillard2006, Corral-Santana2016}, a few X-ray quiet binaries in which a BH is suspected on dynamical grounds, and an isolated BH candidate discovered via microlensing \citep[][]{Sahu2022, Lam2022, Mroz2022}. In X-ray bright systems, a BH accretes material from a close companion through stable Roche lobe overflow or stellar winds. Based on the distance distribution and outburst properties of known BH X-ray binaries, it has been estimated that of order $10^3$ such systems exist in the Milky Way \citep[][]{Corral-Santana2016} -- only a tiny fraction of the expected total Galactic BH population.

BH X-ray binaries all have relatively short orbital periods. In Roche lobe overflowing systems with low-mass ($\lesssim 1\,M_{\odot}$) main-sequence  donors (``LMXBs''), mass transfer only occurs at periods $P_{\rm orb}\lesssim 1$ day. Systems with massive or evolved donors can overflow their Roche lobes at somewhat longer periods. The longest orbital period for a known BH X-ray binary is 33 days, in GRS 1915+105  \citep[][]{Greiner2001}, where the donor is a red giant. Binary population synthesis models predict that a large fraction of the total BH + normal star binary population may be found in wider binaries \citep[e.g.][]{Breivik2017, Chawla2022}, where there is no significant mass transfer. These longer-period BH binaries are difficult to find because they are not X-ray bright, but they may represent the vast majority of the BH binary population. Searches for dormant BH binaries began even before the identification of the first BHs in X-ray binaries \citep[][]{Guseinov1966, Trimble1969}. The intervening decades have witnessed the proliferation of vast spectroscopic and photometric surveys well-suited for finding dormant BHs, but only a few solid candidates have been identified \citep[e.g.][]{Giesers2018, Giesers2019, Shenar2022, Mahy2022}.
 
Gravitational wave observations have in the last decade also begun to detect large number of binaries containing stellar-mass BHs \citep[e.g.][]{TheLIGOScientificCollaboration2021}. The BHs in these systems likely represent a similarly rare outcome of the binary evolution process to X-ray binaries, but their enormous gravitational wave luminosities during merger allow them to be detected all throughout the Universe. The formation channels of these merging BHs and their evolutionary relation to local BHs in X-ray binaries are still uncertain. 

The {\it Gaia} mission opens a new window on the Galactic binary population -- including, potentially, BHs in binaries -- through large-scale astrometric orbit measurements. {\it Gaia} has been predicted to discover large numbers of BHs in binaries, with the predicted number varying by more than 4 orders of magnitude between different studies \citep[e.g.][]{Mashian2017, Breivik2017, Shao2019, Andrews2019, Wiktorowicz2020, Chawla2022, Shikauchi2022, Janssens2022}. The large dispersion in these predictions reflects both inherent uncertainties in binary evolution modeling and different assumptions about the {\it Gaia} selection function.

The first binary orbital solutions from {\it Gaia} were recently published in the mission's 3rd data release \citep[][]{GaiaCollaboration2022, Arenou2022}, including about 170,000 astrometric solutions and 190,000 spectroscopic solutions. Early assessments of the BH candidate population in these datasets have been carried out for both spectroscopic solutions  \citep[e.g.][]{El-Badry2022_gaia_algols, Jayasinghe2022} and astrometric solutions \citep[e.g.][]{Andrews2022,Shahaf2022}.  The DR3 binary sample represents a factor of $\sim$100 increase in sample size over all previously published samples of binary orbital solutions, and is thus a promising dataset in which to search for rare objects. At the same time, stringent SNR cuts were applied to the sample of orbital solutions that was actually published in {\it Gaia}  DR3. The DR3 binary sample thus represents only a few percent of what is expected to be achievable in the mission's data releases DR4 and DR5.

This paper presents detailed follow-up of one astrometric BH binary candidate, which we found to be the most compelling candidate published in DR3. Our follow-up confirms beyond reasonable doubt the object's nature as binary containing a normal star and at least one dormant BH. The remainder of this paper is organized as follows. Section~\ref{sec:discovery} describes how we identified the source as a promising BH candidate. Section~\ref{sec:rvs} presents the radial velocities (RVs) from archival surveys and our follow-up campaign. In Section~\ref{sec:mass}, we constrain the mass of the unseen companion using the RVs and {\it Gaia} astrometric solution. Section~\ref{sec:Gspec} describes analysis of the luminous star's spectrum, including estimates of the atmospheric parameters and abundance pattern, and the non-detection of a luminous companion. Section~\ref{sec:orbit} describes the system's Galactic orbit, and Section~\ref{sec:xray_radio} discusses X-ray and radio upper limits. We compare the object to other BHs and BH imposters in Section~\ref{sec:discussion}, where we also discuss its evolutionary history. Section~\ref{sec:howmany} discusses constraints on the occurrence rate of wide BH companions to normal stars. Finally, we summarize our results and conclude in Section~\ref{sec:conclusion}. The Appendices provide further details on several aspects of our data and modeling.
 
\section{Discovery}
\label{sec:discovery}
In a search for compact object companions to normal stars with astrometric binary solutions from {\it Gaia}, we considered all 168,065 sources in the \texttt{gaiadr3.nss\_two\_body\_orbit} catalog with purely astrometric (\texttt{nss\_solution\_type = Orbital}) or joint astrometric/spectroscopic  (\texttt{nss\_solution\_type = AstroSpectroSB1}) solutions. These solutions describe the ellipse traced on the sky by each source's $G-$band light centroid due to binary motion. In the \texttt{AstroSpectroSB1} solutions, which are only available for bright ($G\lesssim 13$) sources, RVs and astrometry are fit simultaneously.

Our selection strategy and the candidates we considered are described in Appendix~\ref{sec:other_cands}. In brief, we searched for astrometric solutions with unusually large photocenter ellipses at fixed orbital period, exploiting the facts that (a) massive companions have larger orbits at fixed period due to Kepler's 3rd law, and (b) dark companions produce larger photocenter wobbles than would luminous companions of the same mass \citep[e.g.][]{vandeKamp1975}. This yielded 6 initially promising sources. Individual vetting and spectroscopic follow-up showed that in four of the six sources, the astrometric solutions are spurious, making a BH companion unlikely. In one case, the astrometric solution may be correct, but the luminous star is a giant and the orbital period is longer than the {\it Gaia} DR3 baseline, making the reliability of the solution and the nature of the companion difficult to assess without long-term spectroscopic monitoring. One candidate emerged as very promising, {\it Gaia} DR3 4373465352415301632. We colloquially refer to the source as Gaia BH1.

\subsection{Properties of the luminous source}
\label{sec:lum_star}

\begin{figure*}
    \centering
    \includegraphics[width=\textwidth]{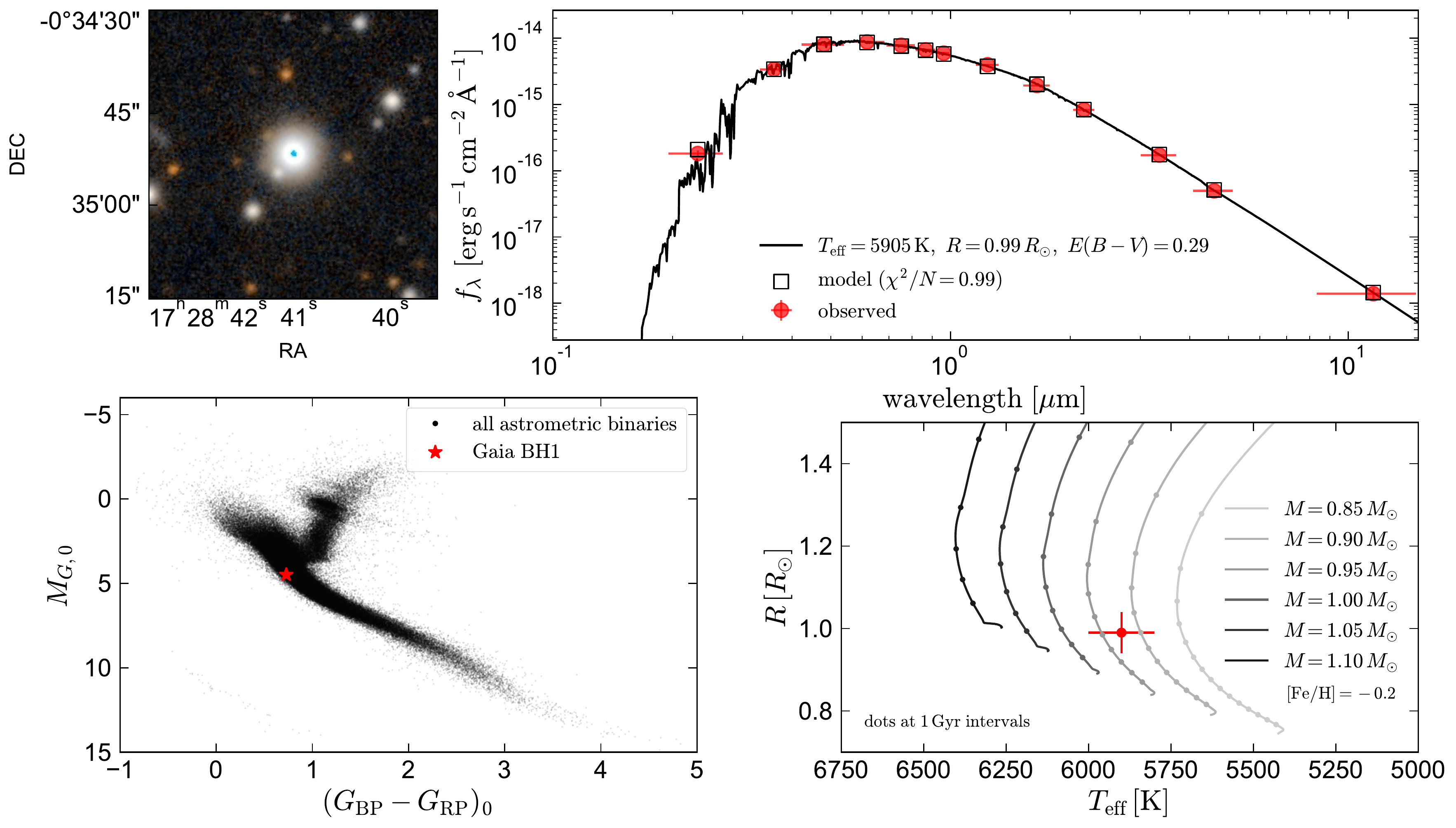}
    \caption{Properties of the luminous star. Upper left shows a 45 arcsecond wide $z$/$g$ PanSTARRS postage stamp centered on the source. Upper right shows the broadband SED and a single-star fit. Red points show observed photometry; open black squares show the integrated fluxes predicted for the model spectrum (black line). Lower left compares the source to the rest of the {\it Gaia} astrometric binary sample on the color-magnitude diagram. Lower right compares the temperature and radius measured from the SED to single-star MIST evolutionary models with [Fe/H] = -0.2 (as measured from spectroscopy), which suggest a mass of $M_\star \approx 0.93\,M_{\odot}$. }
    \label{fig:mosaic}
\end{figure*}

Basic observables of the luminous source are summarized in Figure~\ref{fig:mosaic}. It is a bright ($G=13.77$) solar-type star in Ophiuchus ($\alpha=$\,17:28:41.1; $\delta=-$00:34:52). The field (upper left) is moderately but not highly crowded. The nearest companion detected by {\it Gaia} is 5.1 magnitudes fainter at a distance of 3.9 arcseconds. None of the {\it Gaia}-detected neighbors have parallaxes and proper motions consistent with being bound to the source. In the color-magnitude diagram, the source appears as a solar-type main sequence star. The DR3 orbital astrometric solution puts it at a distance of $477\pm 4$ pc. 

The {\it Gaia} astrometric solution has a photocenter ellipse with semi-major axis $a_0=3.00\pm 0.22$ mas.\footnote{Here and elsewhere in the paper, we quantify uncertainties on parameters derived from the {\it Gaia} solution using Monte Carlo samples from the covariance matrix.} Given the constraints on the source's parallax, this corresponds to a projected photocenter semi-major axis of $a_0/\varpi = 1.44\pm 0.11\,\rm au$. If the photocenter traced the semimajor axis (which it does for a dark companion in the limit of $M_\star/M_2\to 0$), this would imply a total dynamical mass of $M_{\rm tot}= 11.5\pm 2.7\,M_{\odot}$. For larger $M_\star/M_2$ or a luminous secondary, the implied total mass would be larger. If we take the mass of the luminous star to be $M_\star=0.93\pm 0.05 M_{\odot}$ (as implied by its temperature and radius; see below), the astrometric mass ratio function \citep[see][]{Shahaf2019} is $\mathcal{A}=2.32\pm 0.17$. This is  much larger than the maximum value of $\mathcal{A}\approx 0.6$ that can be achieved for systems with main-sequence components, including hierarchical triples. For a dark companion,  this value of $\mathcal{A}$ implies a mass ratio $q=M_2/M_\star = 14.2 \pm 2.8$.

We retrieved photometry of the source in the GALEX NUV band \citep[][]{Martin2005}, SDSS $u$ band \citep[][]{Padmanabhan2008}, PanSTARRS $grizy$ bands \citep[][]{Chambers2016}, 2MASS $JHK$ bands \citep[][]{Skrutskie_2006}, and WISE $W1\,W2\,W3$ bands \citep[][]{Wright_2010}, and fit the resulting SED with a single-star model. We set an uncertainty floor of 0.02 mag to allow for photometric calibration issues and imperfect models.
We predict bandpass-integrated magnitudes using empirically-calibrated model spectral from the BaSeL library \citep[v2.2;][]{Lejeune1997, Lejeune1998}. We assume a \citet{Cardelli_1989} extinction law with total-to-selective extinction ratio $R_V =3.1$, and we adopt a prior on the reddening $E(B-V) = 0.30\pm 0.03$ based on the \citet{Green2019} 3D dust map. We use \texttt{pystellibs}\footnote{\href{https://mfouesneau.github.io/pystellibs/}{https://mfouesneau.github.io/pystellibs/}} to interpolate between model SEDs, and \texttt{pyphot}\footnote{\href{https://mfouesneau.github.io/pyphot/}{https://mfouesneau.github.io/pyphot/}}  to calculate synthetic photometry. We then fit the SED using \texttt{emcee} \citep{emcee2013} to sample from the posterior, with the temperature, radius, metallicity, and reddening sampled as free parameters.

The results are shown in the upper right panel of Figure~\ref{fig:mosaic}. A single-star model yields an excellent fit to the data, with $\chi^2/N_{\rm data} = 0.99$, where $N_{\rm data}$ is the number of photometric points. The inferred temperature and radius correspond to a solar-type star near the main sequence, and evolutionary models then predict a mass of $M_\star \approx 0.93\pm 0.05\,M_{\odot}$. The inferred mass is consistent between MIST \citep{Choi2016} and PARSEC \citep{Marigo2017} models within $0.02\,M_{\odot}$. The fact that the source falls near the main sequence provides independent confirmation that the distance inferred from the {\it Gaia} astrometry is not catastrophically wrong, and suggest that there is no bright companion. 

We inspected the ASAS-SN $V-$ and $g-$band light curves of the source \citep{Kochanek2017}, which contain 3300 photometric epochs over a 10-year period, with a typical uncertainty of 0.03 mag. This did not reveal any significant periodic or long-term photometric variability. The photometry from ZTF \citep[][]{Bellm2019}, which is more precise but has a shorter baseline, also did not reveal any significant variability. 

\section{Radial velocities}
\label{sec:rvs}

\subsection{Archival data}
\label{sec:archival_rvs}
Gaia BH1 was observed twice by the LAMOST survey \citep[][]{Cui2012}, which obtained low-resolution ($R\approx 1800$) spectra in 2017 and 2019. The LAMOST spectra revealed a main-sequence G star with reported $T_{\rm eff} = 5863$, $\log g = 4.36$, and $[\rm Fe/H]=-0.29$.  The two RVs measured by LAMOST were $20.0\pm 4.0$ and $8.9\pm 5.6\,\rm km\,s^{-1}$, both not far from the mean RV of $23.0\pm 2.6$ reported in {\it Gaia} DR3,\footnote{For sources fainter than $G_{\rm RVS}=12$, the {\it Gaia} radial velocity and its uncertainty are calculated from the peak and curvature of the cross-correlation function (CCF), which is constructed as the average of the CCFs from all individual visits. The {\it Gaia} RV thus approximately represents the mode of the RVs at all times the source was observed.} and consistent with no RV variation at all. However, the LAMOST observations both happened to occur at times when the astrometric orbit predicted the luminous star would be near apastron (Figure~\ref{fig:rvfig}), and thus did not rule out the {\it Gaia} astrometric solution. Analysis of the {\it Gaia} scanning law (Appendix~\ref{sec:gost}) showed that most of the {\it Gaia} observations of the source also occurred near apastron. We thus initiated a spectroscopic follow-up campaign.

\subsection{Spectroscopic follow-up}
\label{sec:followup}
We obtained follow-up spectra using several instruments. Details about the observing setup, data reduction, and analysis for each instrument are listed in Appendix~\ref{sec:appendix}. The first follow-up observation yielded an RV of $\approx 64\,\rm km\,s^{-1}$, which was clearly different from the LAMOST and {\it Gaia} RVs and suggested that the {\it Gaia} astrometric solution might be correct. We obtained 39 spectra over the course of 4 months, as described in Appendix~\ref{sec:appendix} and summarized in Figure~\ref{fig:rvfig}. The follow-up RVs span most of the orbit's predicted dynamic range and broadly validate the {\it Gaia} solution.

We measured RVs via cross-correlation with a synthetic template spectrum, which we took from the BOSZ library of Kurucz model spectra \citep{Kurucz_1970, Kurucz_1979, Kurucz_1993} computed by \citet{Bohlin_2017}. The instruments we used have spectral resolutions ranging from $R \approx 4300$ to $R \approx 55000$, and in most cases the data have high SNR ($\gtrsim 20$ per pixel; see Table~\ref{tab:rvs}). Because we used several different instruments, with different line spread functions and wavelength coverage, the RV uncertainty is dominated by uncertainties in wavelength calibration and zeropoint offsets between different spectrographs. When possible, we performed flexure corrections using sky lines to minimize such offsets, and we used telluric and interstellar absorption lines to verify stability and consistency of the wavelength solutions. 

Including calibration uncertainties, the per-epoch RV uncertainties range from 0.1-4  $\rm km\,s^{-1}$. The RV zeropoint is set by telluric wavelength calibration of the Keck/HIRES spectra as described by \citet[][]{Chubak2012}. This brings the RV zeropoint to the  \citet{Nidever2002} scale within $\sim0.1\,\rm km\,s^{-1}$. Together with the two archival LAMOST RVs, our follow-up RVs are sufficient to fully constrain the orbit, even without including information from the astrometric solution (Section~\ref{sec:rvs_only}).

\begin{figure*}
    \centering
    \includegraphics[width=\textwidth]{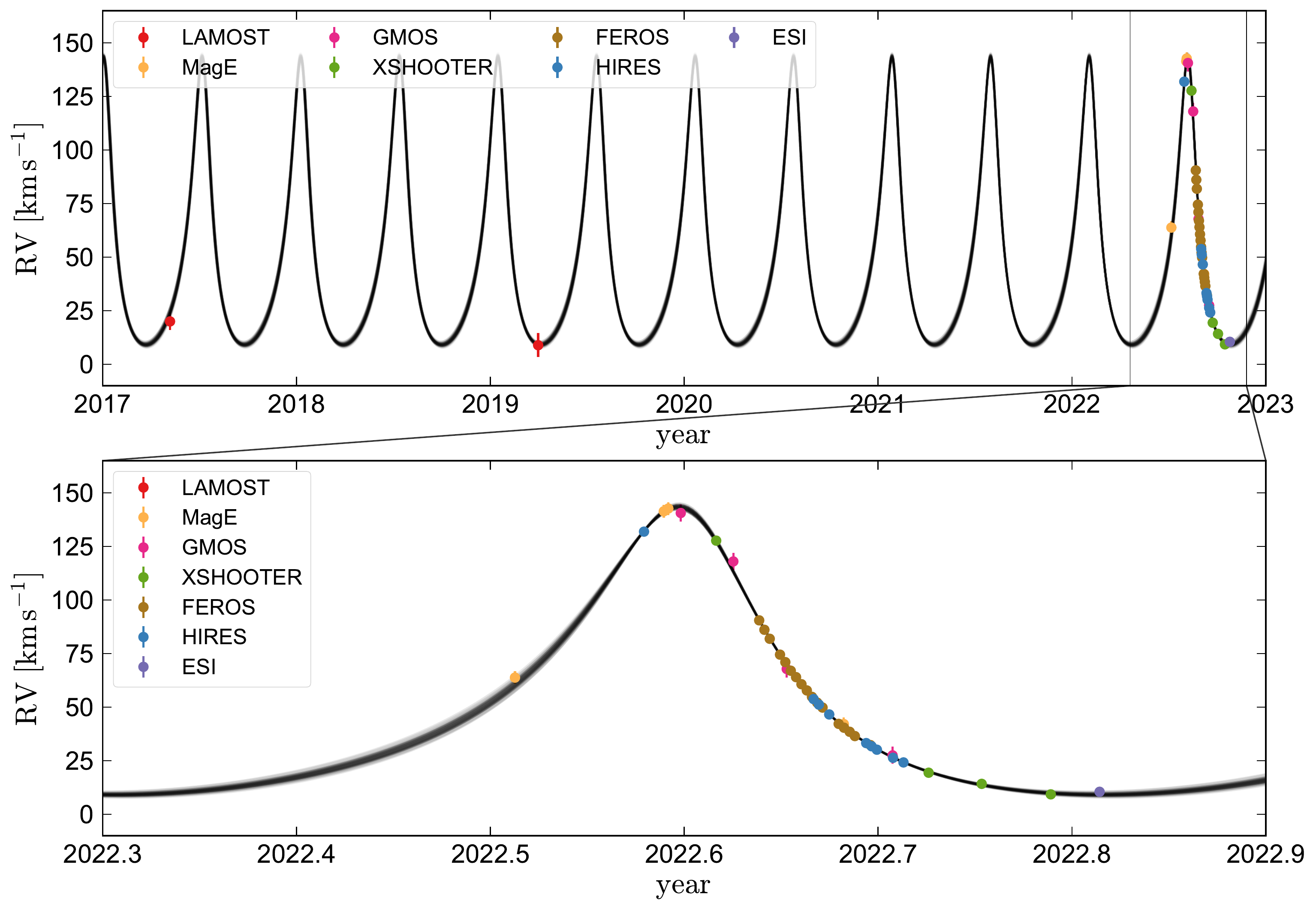}
    \caption{Radial velocities. Points with error bars are measurements; gray lines are draws from the posterior when jointly fitting these RVs and the {\it Gaia} astrometric constraints. Top panel shows all available RVs, including observations by the LAMOST survey in 2017 and 2019; bottom panel highlights our follow-up in 2022. The best-fit solution has a period of 186 days, eccentricity 0.45, and RV semi-amplitude of $67\,\rm km\,s^{-1}$. Together with the inclination constraint from astrometry, this implies a companion mass of $9.62\pm 0.18\,M_{\odot}$. }
    \label{fig:rvfig}
\end{figure*}

\section{Companion mass}
\label{sec:mass}
We explore three different approaches to constraining the orbit and companion mass: (a) simultaneously fitting the RVs and {\it Gaia} astrometric constraints, (b) using only the {\it Gaia} astrometric constraints, and (c) fitting only the RVs.

\subsection{Joint astrometric + RV orbit fitting}
\label{sec:fitting}

The {\it Gaia} orbital solution is parameterized as joint constraints on 12 astrometric parameters: the 5 standard parameters for single-star solutions (\texttt{ra, dec, pmra, pmdec, parallax}) and 7 additional parameters describing the photocenter ellipse (\texttt{period, t\_periastron, eccentricity, a\_thiele\_innes, b\_thiele\_innes, f\_thiele\_innes, g\_thiele\_innes}). The Thiele-Innes coefficients describe the ellipse orientation and are transformations of the standard Campbell orbital elements \citep[e.g.][]{Halbwachs2022}. Individual-epoch astrometric  measurements are not published in DR3. 

We refer to the vector of best-fit astrometric parameters constrained by {\it Gaia} as $\boldsymbol \mu_{\rm ast}$, and to the corresponding covariance matrix as $\boldsymbol \Sigma_{\rm ast}$. The latter is constructed from the \texttt{corr\_vec} column in the \texttt{gaiadr3.nss\_two\_body\_orbit} table. We include in our joint fit the 5 standard astrometric parameters as well as the period, eccentricity, inclination, angle of the ascending node $\Omega$, argument of periastron $\omega$, periastron time, center-of-mass velocity, luminous star mass $M_{\star}$, and companion mass $M_2$. For each call of the likelihood function, we then predict the corresponding vector of astrometric quantities, $\boldsymbol \theta_{\rm ast}$, and corresponding likelihood:

\begin{equation}
    \label{eq:lnL_ast}
    \ln L_{{\rm ast}}=-\frac{1}{2}\left(\boldsymbol  \theta_{{\rm ast}}-\boldsymbol \mu_{{\rm ast}}\right)^{\intercal}\boldsymbol\Sigma_{{\rm ast}}^{-1}\left(\boldsymbol \theta_{{\rm ast}}-\boldsymbol \mu_{{\rm ast}}\right).
\end{equation}
We neglect terms in the likelihood function that are independent of $\boldsymbol \theta_{\rm ast}$. The 5 single-star parameters for our purposes are nuisance parameters (they do not constrain the companion mass), but we include them in our fit as free parameters in order to properly account for covariances in the parameters constrained by {\it Gaia}. When predicting the parameters describing the photocenter ellipse, we assume the companion is dark. We predict the Thiele-Innes coefficients using the standard transformations \citep[e.g.][]{Binnendijk1960}.

We additionally predict the RVs of the luminous star at the array of times at which spectra were obtained, $t_i$. This leads to a radial velocity term in the likelihood, 

\begin{equation}
    \label{eq:lnL_RVs}
    \ln L_{{\rm RVs}}=-\frac{1}{2}\sum_{i}\frac{\left({\rm RV_{{\rm pred}}}\left(t_{i}\right)-{\rm RV}_{i}\right)^{2}}{\sigma_{{\rm RV,}i}^{2}},
\end{equation}
where ${\rm RV}_i$ and ${\rm RV}_{{\rm pred}}\left(t_{i}\right)$ are the measured and predicted RVs.
The full likelihood is then 
\begin{equation}
    \label{eq:lnL_tot}
    \ln L = \ln L_{{\rm ast}} +  \ln L_{{\rm RVs}}.
\end{equation}

A more optimal approach would be to fit the epoch-level astrometric data and the RVs simultaneously, but this will only become possible after epoch-level astrometric data are published in DR4. Since the {\it Gaia} astrometric fits are essentially pure likelihood constraints (i.e., they are not calculated with informative priors), it is possible to mix individual-epoch RV measurements and the {\it Gaia} astrometric constraints, without risk of multiplying priors. 

\begin{figure*}
    \centering
    \includegraphics[width=\textwidth]{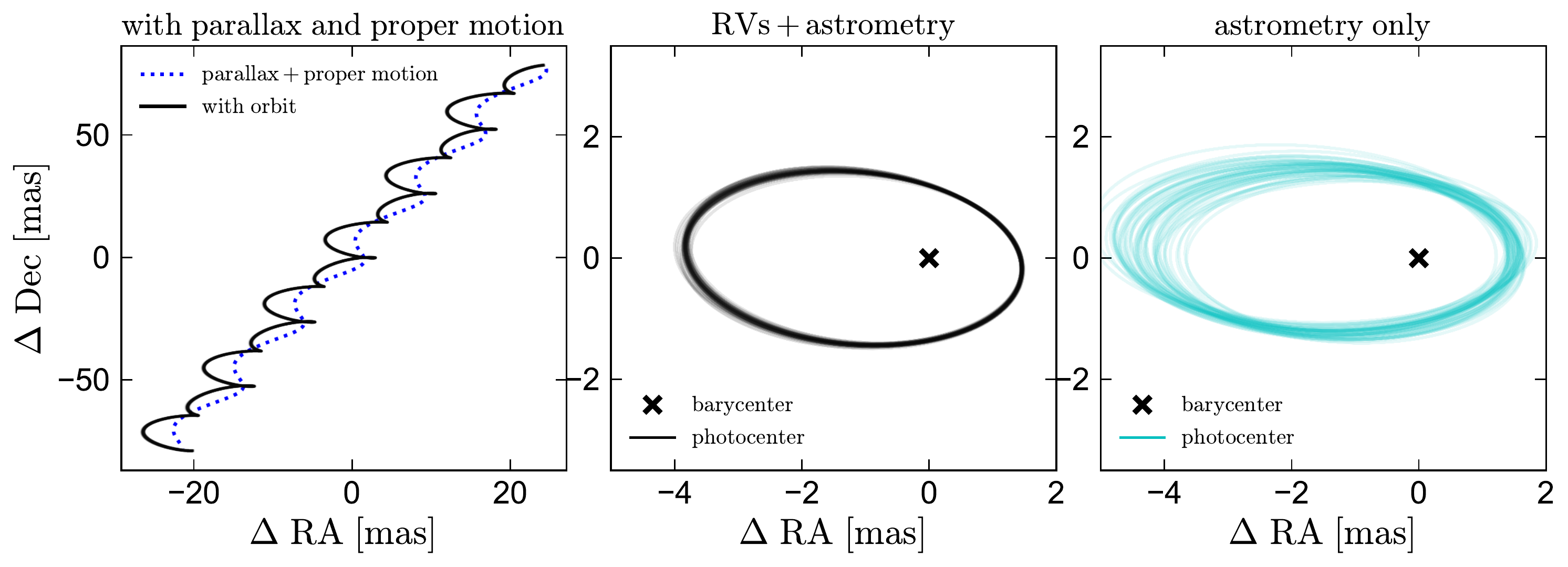}
    \caption{Left panel shows predicted motion of Gaia BH1's photocenter on the sky over a 6-year window. Middle and right panels show the predicted orbit with parallax and proper motion removed. Middle panel shows joint constraints from RVs and astrometry; right panel shows constraints from the {\it Gaia} astrometric solution alone. }
    \label{fig:astrometry}
\end{figure*}

We use flat priors on all parameters except $M_{\star}$, for which we use a normal distribution informed by isochrones, $M_{\star}/M_{\odot}\sim \mathcal{N}(0.93,0.05)$. We sample from the posterior using \texttt{emcee} \citep[][]{emcee2013} with 64 walkers, taking 3000 steps per walker after a burn-in period of 3000 steps. The results of this fitting are visualized in Figures~\ref{fig:rvfig}, \ref{fig:astrometry}, and \ref{fig:corner_plot_comparison}. We infer an unseen companion mass $M_2 = 9.62\pm 0.18\,M_{\odot}$.

Figure~\ref{fig:rvfig} compares the measured RVs to draws from the posterior. The scatter across posterior samples at fixed time is diagnostic of the uncertainty in the RV solution. Unsurprisingly, uncertainties are larger during phases where there are fewer measured RVs. Overall, the fit is good: we obtain a solution that both matches the observed RVs and predicts an astrometric orbit consistent with the {\it Gaia} constraints. This did not have to occur, and it speaks to the quality of the astrometric solution. Our constraints from the joint fit are listed in Table~\ref{tab:system}. Following the {\it Gaia} convention, the periastron time $T_p$ is reported with respect to epoch 2016.0; i.e., JD 2457389.0. 

Figure~\ref{fig:astrometry} shows the predicted astrometric orbit of the photocenter, which we calculate using \texttt{pystrometry} \citep[][]{Sahlmann2019}. The left panel shows the predicted motion of the source on the plane of the sky over a 6-year period. The black line (which is actually many narrow lines representing independent posterior samples) shows the total motion due to proper motion, parallax, and orbital motion. The dotted blue line isolates the contribution due to parallax and proper motion alone. The middle and right panels show the predicted orbital motion, with parallax and proper motion removed, based on the joint astrometry + RV fit (middle) or the pure astrometry fit (right; see below). The orbits predicted in the two cases are similar, but the fit with RVs included is better constrained, and has a slightly smaller photocenter semi-major axis.

\subsection{Purely astrometric constraints}
\label{sec:fitting_astro_only}
We also explored fitting the orbit without the constraints from RVs; i.e., simply removing the $\ln L_{\rm RVs}$ term from Equation~\ref{eq:lnL_tot}. 
The results of this exercise are described in Appendix~\ref{sec:comp_rvs_appendix}, where we also compare the RVs predicted by the astrometry-only solution to the measured RVs. This comparison allows us to assess the reliability of the astrometric solution. 

In brief, we find that the measured RVs are consistent with the purely astrometric solution within its uncertainties. The joint RVs+astrometry fit yields a marginally lower companion mass and more face-on orbit than the best-fit pure astrometry solution, but the astrometric parameter constraints from our joint {\it Gaia}+RVs fit are consistent with the purely astrometric constraints at the 1.6$\sigma$ level (Table~\ref{tab:innes_elememts}).  We also report the purely astrometric constraints in Table~\ref{tab:system}. The joint RVs+astrometry constraints are much tighter than those from astrometry alone.

\subsubsection{Astrometric uncertainties}
\label{sec:bigger_uncertainties}
Although we find no indication that the astrometric solution is unreliable, there is one way in which it is unusual: the uncertainties on the parameters describing the photocenter ellipse are significantly larger than is typical for a source with $G=13.77$. This is reflected in the uncertainty on the photocenter ellipse semi-major axis, $\sigma_{a_0}=0.22\,\rm mas$. The median uncertainty for sources with $13.6 < G< 13.9$ in the \texttt{nss\_two\_body\_orbit} table is only $\sigma_{a_0}=0.026\,\rm mas$, and only 1.4\% of sources in that magnitude range have larger $\sigma_{a_0}$ than Gaia BH1.

The unusually large $\sigma_{a_0}$ appears to be a result of two factors. First, astrometric solutions with larger $a_0$ have larger $\sigma_{a_0}$ (see Appendix~\ref{sec:sigma_a0_appendix}); the median $\sigma_{a_0}$ for sources in the same magnitude range with $2.5 \leq a_0/{\rm mas} \leq 3.5$ is 0.064 mas, and Gaia BH1 is only in the 85th percentile of $\sigma_{a_0}$ for such sources.  Second, as we discuss in Appendix~\ref{sec:gost}, all {\it Gaia} observations of the source that contributed to the DR3 solution covered the same $\approx 50\%$ of the orbit, even though they were spread over more than 5 orbital periods. Despite this, the combination of astrometry and RVs constrains the orbit well, and leads to a tighter constraint on $a_0$ than is achieved for typical {\it Gaia} sources at the same apparent magnitude. In any case, a catastrophic problem with the astrometric solution is firmly ruled out by the RVs, which are consistent with the astrometric solution and require a BH-mass companion even without the astrometric constraints.

\begin{table}
\centering
\caption{Physical parameters and 1$\sigma$ uncertainties for both components of Gaia BH1. The inclination, $i\approx 127\,\rm deg$, is equivalent to $i \approx 53\,\rm deg$ for a spectroscopic orbit. We compare constraints on the orbit based on both astrometry and RVs (3rd block), astrometry alone (4th block), and RVs alone (5th block).  }
\begin{tabular}{lll}
\hline\hline
\multicolumn{3}{l}{\bf{Properties of the unresolved source}}   \\ 
Right ascension & $\alpha$\,[deg] & 262.17120816 \\
Declination & $\delta$\,[deg] & $-0.58109202$ \\
Apparent magnitude & $G$\,[mag] &  13.77 \\
Parallax & $\varpi$\,[mas] & $ 2.09 \pm 0.02 $ \\
Proper motion in RA & $\mu_{\alpha}^{*}$\,[$\rm mas\,yr^{-1}$] & $ -7.70 \pm 0.020 $ \\
Proper motion in Dec & $\mu_{\delta}$\,[$\rm mas\,yr^{-1}$] & $ -25.85 \pm 0.027 $ \\
Tangential velocity & $v_{\perp}\,\left[{\rm km\,s^{-1}}\right]$ & $61.0\pm 0.5$ \\ 
Extinction & $E(B-V)$\,[mag] & $ 0.30 \pm 0.03 $ \\

\hline
\multicolumn{3}{l}{\bf{Parameters of the G star}}  \\ 
Effective temperature & $T_{\rm eff}$\,[K] & $5850 \pm 50 $ \\
Surface gravity   & $\log(g/(\rm cm\,s^{-2}))$  & $4.55 \pm 0.16$  \\
Projected rotation velocity & $v\sin i$\,[km\,s$^{-1}$] &  $< 3.5$ \\
Radius & $R\,[R_{\odot}]$ & $0.99 \pm 0.05$  \\ 
Bolometric luminosity & $L\,[L_{\odot}]$ & $1.06\pm 0.04$ \\ 
Mass &  $M\,[M_{\odot}]$ & $0.93 \pm 0.05$ \\
Metallicity &  $\rm [Fe/H]$ & $-0.2 \pm 0.05$ \\
Abundance pattern &  $\rm [X/Fe]$ & Table~\ref{tab:bacchus} \\

\hline
\multicolumn{3}{l}{\bf{Parameters of the orbit (astrometry + RVs)}}  \\ 
Orbital period & $P_{\rm orb}$\,[days] & $ 185.59 \pm 0.05 $ \\
Photocenter semi-major axis & $a_0$\,[mas] & $ 2.67 \pm 0.02 $ \\
Semi-major axis & $a$\,[au] & $ 1.40 \pm 0.01 $ \\

Eccentricity & $e$ & $0.451 \pm 0.005 $ \\
Inclination  & $i$\,[deg] & $ 126.6 \pm 0.4 $ \\
Periastron time & $T_p$\,[JD-2457389] & $-1.1 \pm 0.7 $ \\
Ascending node angle & $\Omega$\,[deg] & $97.8 \pm 1.0 $ \\
Argument of periastron & $\omega$\,[deg] & $12.8 \pm 1.1 $ \\
Black hole mass & $M_2$\,[$M_{\odot}$] & $9.62 \pm 0.18 $ \\
Center-of-mass RV & $\gamma$\,[$\rm km\,s^{-1}$] & $46.6 \pm 0.6 $ \\

\hline
\multicolumn{3}{l}{\bf{Parameters of the orbit (astrometry only)}}  \\ 
Orbital period & $P_{\rm orb}$\,[days] & $ 185.77 \pm 0.31 $ \\
Photocenter semi-major axis & $a_0$\,[mas] & $ 3.00 \pm 0.22 $ \\
Eccentricity & $e$ & $0.49 \pm 0.07 $ \\
Inclination  & $i$\,[deg] & $ 121.2 \pm 2.8 $ \\
Periastron time & $T_p$\,[JD-2457389] & $-12.0 \pm 6.3 $ \\
Ascending node angle & $\Omega$\,[deg] & $89.6 \pm 3.7 $ \\
Argument of periastron & $\omega$\,[deg] & $-10.3 \pm 11.6 $ \\
Black hole mass & $M_2$\,[$M_{\odot}$] & $12.8 \pm 2.0 $ \\

\hline
\multicolumn{3}{l}{\bf{Parameters of the orbit ( RVs only)}}  \\ 
Orbital period & $P_{\rm orb}$\,[days] & $ 185.6 $ (fixed) \\
 & & $ 184.8 \pm 0.7 $ (if not fixed) \\

RV semi-amplitude & $K_{\star}\,[\rm km\,s^{-1}]$ & $ 66.7 \pm 0.6 $ \\
Periastron time & $T_p$\,[JD-2457389] & $2411.8 \pm 0.2 $ \\
Eccentricity & $e$ & $0.447 \pm 0.005 $ \\
Center-of-mass RV & $\gamma$\,[$\rm km\,s^{-1}$] & $47.1 \pm 0.6 $ \\
Argument of periastron & $\omega$\,[deg] & $13.9 \pm 1.2 $ \\
Mass function & $f_m\,[M_{\odot}]$ & $4.08 \pm 0.08 $ \\

\hline
\end{tabular}
\begin{flushleft}

\label{tab:system}
\end{flushleft}
\end{table}

\subsection{Solution based on radial velocities alone}
\label{sec:rvs_only}
Although the {\it Gaia} astrometric solution provides strong constraints on the binary's orbit, it is useful to consider constraints based only on the measured RVs, which are insensitive to any possible problems with the {\it Gaia} solution. To this end, we fit the RVs with a model with the standard 6 orbital parameters for a single-lined binary. We first search for the maximum-likelihood solution with simulated annealing \citep[e.g.][]{Iglesias-Marzoa2015} and then use MCMC to explore the posterior in the vicinity of the maximum likelihood solution. The dense sampling of RVs in the last 2 months of our observations allows us to robustly constrain the orbit. We first fit the RVs with no period constraint; this yielded a solution with $P_{\rm orb}=184.8\pm 0.7$ days. Since this is consistent with the {\it Gaia} solution, but the {\it Gaia} solution is based on a longer time baseline, we then fixed the period to $P_{\rm orb}=185.6$ days, as inferred in Section~\ref{sec:fitting}. The constraints on the RV solution obtained in this way are reported in Table~\ref{tab:system}, and the best-fit RV curve is shown in Figure~\ref{fig:spec_fm}. The reduced $\chi^2$ is 0.54, suggesting that the fit is good and the RV uncertainties are overestimated on average.

Fitting only the RVs yields an RV semi-amplitude and eccentricity consistent with the predictions of the purely astrometric and astrometric + RV fits. The main difference is that the inclination is unknown in the pure-RV solution, and the uncertainties on other parameters are somewhat larger. The periastron time inferred in this fit is 13 orbital cycles later than the one reported in the {\it Gaia} solution, as it corresponds to the periastron passage in August 2022 that is actually covered by our RVs.

The constraint on the companion mass from RVs can be expressed in terms of the spectroscopic mass function, 

\begin{equation}
    \label{eq:fm_spec}
    f\left(M_{2}\right)_{\rm spec}=M_{2}\left(\frac{M_{2}}{M_{2}+M_{\star}}\right)^{2}\sin^{3}i=\frac{P_{{\rm orb}}K_{\star}^{3}}{2\pi G}\left(1-e^{2}\right)^{3/2}.
\end{equation}
$P_{\rm orb}$, $K_{\star}$, and $e$ are inferred directly from the RVs. Since both the terms $\left(\frac{M_{2}}{M_{2}+M_{\star}}\right)^{2}$ and $\sin^3 i$ must be less than one, $f\left(M_{2}\right)_{\rm spec} = 4.08\pm 0.08$ sets an absolute lower limit on the mass of the unseen companion. This constraint is illustrated in the lower panel of Figure~\ref{fig:spec_fm}, where we plot the best-fit orbit and residuals, and the companion mass required to explain the mass function for different inclinations and $M_\star$. 

The figure shows that the RVs cannot accommodate a companion mass below $4\,M_{\odot}$, independent of the astrometric solution and the mass of the G star. When a plausible mass for the G star is assumed, this limit increases to $5.5\,M_{\odot}$.
We also show the inclination implied by the {\it Gaia} orbital solution. Assuming this inclination and $M_{\star} = 0.93\pm 0.05\,M_{\odot}$ leads to a companion mass of $9.50\pm 0.20\,M_{\odot}$, in good agreement with the value we inferred by simultaneously fitting the RVs and astrometry. The fact that this value is not identical to the $M_2 = 9.62\pm 0.18\,M_{\odot}$ inferred in the joint fit reflects the fact that the joint RV + astrometry fit implies a slightly larger $K_{\star}$. 

\subsubsection{Astrometric mass function}
To appreciate the additional information provided by the astrometric solution, we can consider the astrometric mass function,
\begin{equation}
    \label{eq:fm_ast}
    f\left(M_{2}\right)_{{\rm ast}}=\left(\frac{a_{0}}{\varpi}\right)^{3}\left(\frac{P_{{\rm orb}}}{{\rm yr}}\right)^{-2}=M_{2}\left(\frac{M_{2}}{M_{\star}+M_{2}}\right)^{2}
\end{equation}
where the 2$^{\mathrm{nd}}$ equality holds only for a dark companion. The quantity after the 2$^{\mathrm{nd}}$ equality is equivalent to $f\left(M_{2}\right)_{\rm spec}$ except for a factor of $\sin^3 i$. In the limit of small uncertainties, $f\left(M_{2}\right)_{{\rm ast}}$ sets a dynamical limit on the mass of the companion that is more stringent than the spectroscopic mass function. In particular, $f\left(M_{2}\right)_{{\rm ast}}$ is the mass that the orbit would imply if the G star were a massless test particle.

The constraints from astrometry alone yield $f\left(M_{2}\right)_{{\rm ast}} = 11.2\pm 1.9\,M_{\odot}$. Those from joint fitting of the RVs and astrometry yield $f\left(M_{2}\right)_{{\rm ast}} = 8.00\pm 0.16\,M_{\odot}$.
When we adopt $M_\star = 0.93\pm 0.05\,M_{\odot}$ and solve for $M_2$, the joint constraint yields $M_2 = 9.62 \pm 0.18\,M_{\odot}$. 

\subsubsection{Possibility of underestimated uncertainties}
It is worth considering whether the astrometric uncertainties could be underestimated. Given that the {\it Gaia} single-star astrometric solutions have been shown to have uncertainties underestimated by $\sim 30\%$ near $G=13.8$ \citep[e.g.][]{El-Badry2021_gaia}, and there is modest ($\sim 1.6\sigma$) tension between the joint fit and the pure-astrometry solution (Appendix~\ref{sec:comp_rvs_appendix}) this is not implausible. To explore the possible effects of underestimated astrometric uncertainties on our constraints, we multiplied all the astrometric uncertainties by 2, constructed the covariance matrix assuming the same correlation matrix (this is equivalent to multiplying the covariance matrix by 4), and repeated the joint fit of astrometry and RVs. This yielded a companion mass constraint of $M_2 = 9.54 \pm 0.26\,M_{\odot}$ -- simular to our fiducial result, and still a rather tight constraint.

\begin{figure}
    \centering
    \includegraphics[width=\columnwidth]{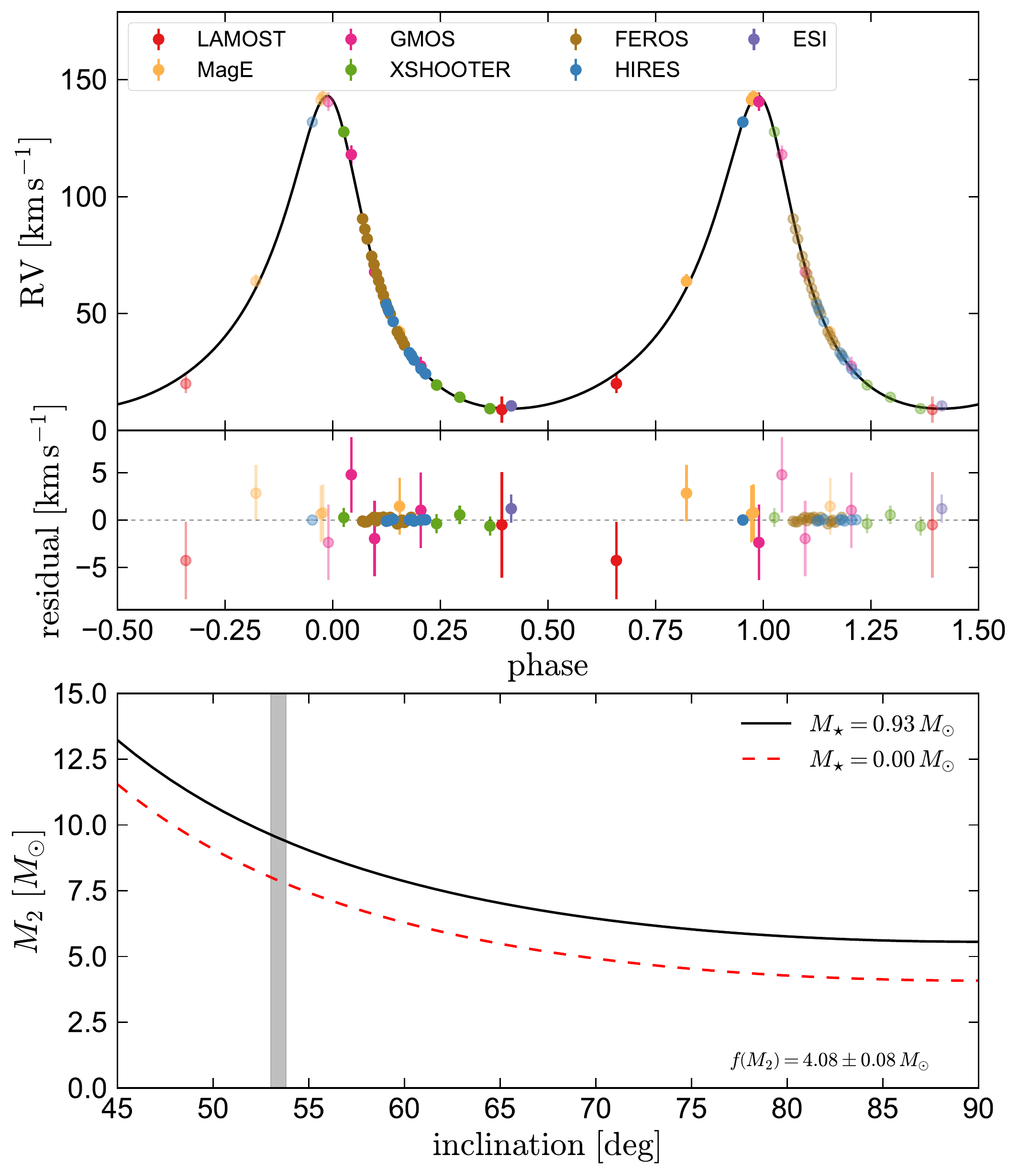}
    \caption{Top panel shows the best-fit orbital solution to the RVs, ignoring the astrometric solution. We show two periods; duplicated datapoints are transparent. Bottom panel shows the resulting constraints on the companion's mass. Black line assumes a mass of $0.93\,M_{\odot}$ for the luminous star as implied by the isochrones; red dashed line assumes (pathologically) that the star is a massless test particle. Shaded region shows the {\it Gaia} astrometric inclination constraint, which corresponds to a companion mass of $9.50\pm 0.20\,M_{\odot}$. Even if the {\it Gaia} astrometric solution were completely wrong and the luminous star had negligible mass, the implied mass of the companion would be $>4\,M_{\odot}$, which is inconsistent with the observed SED and spectra for any luminous companion.  }
    \label{fig:spec_fm}
\end{figure}

\section{Analysis of the G star spectrum}
\label{sec:Gspec}
We analyzed the spectrum of the G star using several different methods. Most of our analysis used the Keck HIRES spectrum obtained on JD 2459791 using the standard  California Planet Survey setup \citep[CPS;][]{Howard2010}. We first fit the spectrum using the empirical \texttt{SpecMatch-Emp} code \citep[][]{Yee2017}, which compares the continuum-normalized rest-frame spectrum to a library of HIRES spectra of FGKM dwarfs and subgiants observed with the standard CPS setup and analyzed with traditional spectroscopic methods. It estimates uncertainties based on the dispersion in parameters of objects in the library with spectra most similar to an observed spectrum. \texttt{SpecMatch-Emp} yielded an effective temperature $T_{\rm eff} = 5801 \pm 110\,\rm K$ and metallicity $\rm [Fe/H] = -0.34 \pm 0.09$.  

\begin{figure*}
    \centering
    \includegraphics[width=\textwidth]{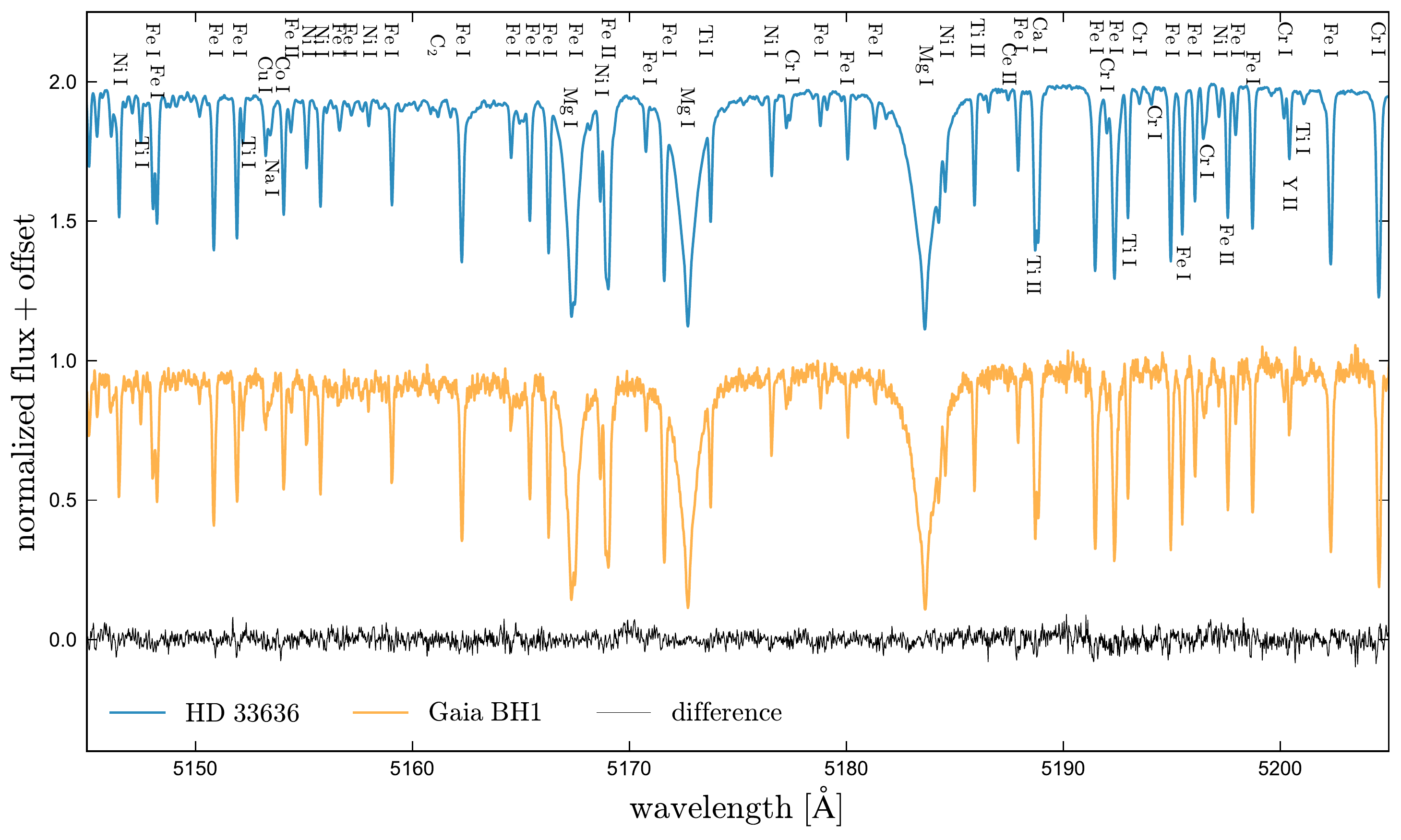}
    \caption{Comparison of the HIRES spectrum of Gaia BH1 (gold) and the standard star HD 33636 (blue). Black line shows the difference between the two spectra. Cutout is centered on the Mg I b triplet; other identifiable lines are labeled. The two spectra are very similar, indicating that the surface properties and abundances of the G star are normal for thin-disk stars in the solar neighborhood. }
    \label{fig:hires_emp}
\end{figure*}

In Figure~\ref{fig:hires_emp}, we compare the HIRES spectrum of Gaia BH1 to that of its closest match in the library, HD 33636. The parameters of that star, as measured by \citet{Valenti2005} from the HIRES spectrum  and synthetic spectral models, are $T_{\rm eff}=5904\,\rm K$, $\log g = 4.43$, $v\sin i = 3.1\,\rm km\,s^{-1}$, and $[\rm Fe/H]=-0.13$. The region of the spectrum shown is centered on the Mg I b triplet and also contains strong lines of the elements Fe, Ni, Ti, Cr, Cu, Ca, and Y, many of which are labeled.  The two spectra are strikingly similar.
This simple comparison suggests that the luminous star in Gaia BH1  is a rather unremarkable solar-type star. Quantitative analysis of the spectrum yielded similar conclusions, as described below.

We also fit the spectrum using the HIRES-adapted implementation of {\it the Cannon} \citep[][]{Ness2015} that was trained by \citet{Rice2020}. {\it The Cannon} uses a library of spectra with known stellar labels (i.e., atmospheric parameters and abundances) to build a spectral model that smoothly interpolates between spectra, predicting the normalized flux at a given wavelength as a polynomial function of labels. This data-driven spectral model is then used to fit for the labels that can best reproduce an observed spectrum through a standard maximum-likelihood method. The {\it Cannon} fit yielded parameters  $T_{\rm eff}=5863\,\rm K$, $\log g = 4.43$, $v\sin i = 1.34\,\rm km\,s^{-1}$, and $[\rm Fe/H]=-0.29$, similar to the values inferred by  \texttt{SpecMatch-Emp}. It also returned constraints on the abundances of 14 metals beside iron.  The full set of labels inferred by {\it the Cannon} can be found in Appendix~\ref{sec:cannon}. 

\begin{table}
\begin{tabular}{lllll}
Parameter & BACCHUS Constraint & $N_{\rm lines}$ \\
\hline
$T_{\rm eff}\,[\rm K]$ & $5883 \pm 27$  &   \\
$\log\left(g/{\rm cm\,s^{-2}}\right)$ & $ 4.55 \pm 0.16$  &   \\
$v_{\rm micro}\,\left[{\rm km\,s^{-1}}\right]$ & $0.79\pm 0.05$ &  \\
$\rm [Fe/H]$& $-0.19 \pm 0.01$ &    \\
$\rm [Na/Fe]$& $-0.07 \pm 0.15$ & 10  \\
$\rm [Mg/Fe]$& $-0.08\pm 0.14$ & 9  \\
$\rm [Al/Fe]$& $0.06 \pm  0.04$ & 2  \\
$\rm [Si/Fe]$& $-0.04 \pm 0.06$& 11  \\
$\rm [Ca/Fe]$& $0.01 \pm 0.05$ & 15  \\
$\rm [Ti/Fe]$& $-0.06 \pm 0.15$ & 39  \\
$\rm [V/Fe]$& $-0.19 \pm 0.21$ & 14  \\
$\rm [Sc/Fe]$& $0.12 \pm 0.05$ & 9  \\
$\rm [Cr/Fe]$& $-0.07 \pm 0.09$ & 15  \\
$\rm [Mn/Fe]$& $-0.20 \pm 0.03$ & 7  \\
$\rm [Co/Fe]$& $-0.22 \pm 0.14$ & 6  \\
$\rm [Ni/Fe]$& $-0.06 \pm 0.11$ & 17  \\
$\rm [Cu/Fe]$& $-0.24 \pm 0.07$ & 5  \\
$\rm [Zn/Fe]$& $-0.24 \pm 0.12$ & 3  \\
$\rm [Y/Fe]$& $-0.23 \pm 0.11$ & 5  \\
$\rm [Zr/Fe]$& $-0.04 \pm 0.07$ & 5  \\
$\rm [Ba/Fe]$& $0.17 \pm 0.04$ & 3  \\
$\rm [La/Fe]$& $0.13 \pm 0.10$ & 4  \\
$\rm [Nd/Fe]$& $-0.01 \pm 0.18$ & 6  \\
$\rm [Eu/Fe]$& $-0.04 \pm 0.10$ & 1  \\
$\rm A(Li)$& $2.30 \pm 0.1$ & 1  \\
\end{tabular}
\caption{Parameters of the G star inferred by BACCHUS. $N_{\rm lines}$ is the number of independent absorption lines used to infer each element. }
\label{tab:bacchus}
\end{table}

To investigate constraints on the G star's projected rotation velocity $v\sin i$, we convolved an $R=300,000$ Kurucz spectrum from the \texttt{BOSZ} library with a range of rotational broadening kernels and with the $R\approx 55,000$ HIRES instrumental broadening kernel, which we model as a Gaussian. We find that we can infer a robust upper limit of $v \sin i < 3.5\,\rm km\,s^{-1}$: a higher value would broaden the lines more than observed. However, the actual value of $v\sin i$ is not well constrained because at lower values of $v \sin i$, instrumental broadening dominates, and the adopted value of $v \sin i$ has little effect on the predicted line profiles. For $R = 1\,R_{\odot}$, this translates to a rotation period $P_{{\rm rot}} \gtrsim14\times\left(\sin i\right)\,{\rm days}$.  Assuming $\sin i \sim 1$ and using a canonical gyrochronological scaling relation for solar-type stars \citep[e.g.][]{Skumanich1972, Mamajek2008}, this implies an age of $\gtrsim 1\,\rm Gyr$. Application of such scaling relations is of course only appropriate if the G star was not previous tidally synchronized and spun down by interaction with the companion. 

\subsection{Detailed abundances}
\label{sec:abundances}
We fit the HIRES spectrum using the Brussels Automatic Code for Characterizing High accUracy Spectra \citep[BACCHUS;][]{Masseron2016} with the same set up as in \cite{Hawkins2020}. BACCHUS enables us to derive the stellar atmospheric parameters, including the effective temperature ($T_{\rm eff}$), surface gravity (log$g$), metallicity ([Fe/H]) and microturblent velocity ($v_{\rm micro}$) by assuming Fe excitation/ionization balance; i.e., the requirement that lines with different excitation potentials all imply the same abundances. For the set up of BACCHUS, we make use of the fifth version of the Gaia-ESO atomic linelist \citep{Heiter2021}.  Hyperfine structure splitting is included for Sc I, V I Mn I, Co I, Cu I, Ba II, Eu II, La II, Pr II, Nd II, Sm II \citep[see more details in][]{Heiter2021}. We also include molecular line lists for the following species: CH \citep{Masseron2014}, and CN, NH, OH, MgH and  C$_{2}$ (T. Masseron, private communication). Finally, we also include the SiH molecular line list from the Kurucz linelists\footnote{http://kurucz.harvard.edu/linelists/linesmol/}. Spectral synthesis for BACCHUS is done using the TURBOSPECTRUM \citep{Alvarez1998, Plez2012} code along with the line lists listed above and the MARCS model atmosphere grid \citep{Gustafsson2008}. The stellar atmospheric parameters derived from BACCHUS can be found in Table~\ref{tab:bacchus}.

Once the stellar parameters were determined, the model atmosphere is fixed and individual abundances were derived using BACCHUS' `abund' module. For each spectral absorption feature, this module creates a set of synthetic spectra, which range between -0.6 $<$ [X/Fe] $<$ +0.6~dex, and performs a $\chi^2$ minimization between the observed and synthetic spectra. The reported atmospheric [X/Fe] abundances are the median of derived [X/Fe] across all lines for a given species. The uncertainty in the atmospheric [X/Fe] is defined as the dispersion of the [X/H] abundance measured across all lines for a species. If only 1 absorption line is used, we conservatively assume a [X/Fe] uncertainty of 0.10~dex. For a more detailed discussion of the BACCHUS code, we refer the reader to Section~3 of both \cite{Hawkins2019, Hawkins2020}. We ran BACCHUS on the full HIRES spectrum, after merging individual de-blazed orders and performing a preliminary continuum normalization using a polynomial spline. Further normalization is performed by BACCHUS. The resulting stellar parameters and abundances are listed in Table~\ref{tab:bacchus}. 

Overall, the abundance pattern of the G star is typical of the thin disk. When comparing the star's position in the [X/Fe] vs. [Fe/H] plane to the population of stars in the solar neighborhood with abundances measured  by \citet{Bensby2014}, it falls within the $\sim 1\,\sigma$ observed scatter for most elements, and within 2$\sigma$ for all. Given  that the G star is expected to have a thin convective envelope (making mixing of accreted material into the interior inefficient), this suggests that it did not suffer much pollution from its companion. 

The measured lithium abundance, $A(\rm Li) = 2.3\pm 0.1$, is also not unusual. For solar-type stars, lithium abundance can be used as an age indicator, because the surface lithium abundance is depleted over time \citep[e.g.][]{Skumanich1972,  Baumann2010}. The age -- $A(\rm Li)$ correlation is well-studied for Sun-like stars \citep[e.g.][]{Ramirez2012, Carlos2016}, and for solar twins, $A(\rm Li) = 2.3$ corresponds to an age of about 1\,Gyr. However, $A(\rm Li)$ varies strongly with both $T_{\rm eff}$ and [Fe/H] (higher $T_{\rm eff}$ and lower [Fe/H] both produce a thinner convective envelope and thus imply an older age at fixed A(Li)), so caution should be taken in applying the relation from solar twins to a star somewhat warmer and more metal-poor than the sun. More robustly, the measured abundance allows  us to rule out youth: the star falls well below the Hyades (age $\approx 600\,\rm Myr$) in the $A(\rm Li)$ vs. $T_{\rm eff}$ plane \citep[][]{Takeda2013}, so we can confidently rule out ages younger than this. Comparison of the G star's effective temperature and radius to isochrones (Figure~\ref{fig:mosaic}) suggests an age of $\gtrsim 4$\,Gyr.  

We find no evidence for pollution of the G star's photosphere by $\alpha$-elements synthesized during the companion's death, as has been reported in some BH companions \citep[e.g.][]{Israelian1999, GonzalezHernandez2011}. The barium abundance is slightly enhanced compared to the solar value, with $[\rm Ba/Fe] = 0.17\pm 0.04$. Strong barium enhancement is often interpreted as a result of accretion of the AGB wind of a binary companion (``barium stars''; \citealt{McClure1990}), usually in stars with white dwarf companions. However, the [Ba/Fe] we measure in Gaia BH1 is not high enough to qualify it as a barium star and is probably unrelated to the companion. We also do not find significant enrichment of other $s-$process elements, or of the $r-$process element europium. 

\subsubsection{Limits on a luminous companion}
\label{sec:reamatch}
As part of the CPS pipeline, the HIRES spectrum of Gaia BH1 was analyzed with the \texttt{Reamatch} tool \citep[][]{Kolbl2015}, which searches for a second peak in the CCF after subtracting the best-fit template. For Gaia BH1, this yielded a null result, with a single narrow peak in the CCF and no evidence of another luminous star. \citet{Kolbl2015} found that for luminous binaries with typical solar-type primaries, \texttt{reamatch} has a $>90\%$ detection rate for companions that contribute $>3\%$ of the light in the optical in a single-epoch HIRES spectrum. These limits of course depend on the spectral type and rotation rate of the secondary, and a rapidly rotating secondary would be harder to detect. However, we find that even a secondary that contributes pure continuum cannot contribute more than 10\% of the light in the HIRES spectrum, or it becomes impossible to achieve a good match to the observed line depths with \texttt{SpecMatch-Emp}. This can be recognized simply from the depth of the observed absorption lines (Figure~\ref{fig:hires_emp}): several lines have depths that reach 10\% of the continuum flux. There cannot be another source contributing more than $\approx 10\%$ of the light at these wavelengths, or the spectrum of the G star would have to reach negative fluxes in the absorption lines to produced the observed total spectrum.

We also find no evidence of a luminous companion at longer or shorter wavelengths. Spectra from X-shooter provide good SNR over the full optical and NIR wavelength range, from 3100 to 24000 \AA, and are consistent with a single G star. This and the GALEX + WISE photometry allow us to rule out pathological companions that are hot and small or cool and large. 

\section{Galactic orbit}
\label{sec:orbit}
\begin{figure*}
    \centering
    \includegraphics[width=\textwidth]{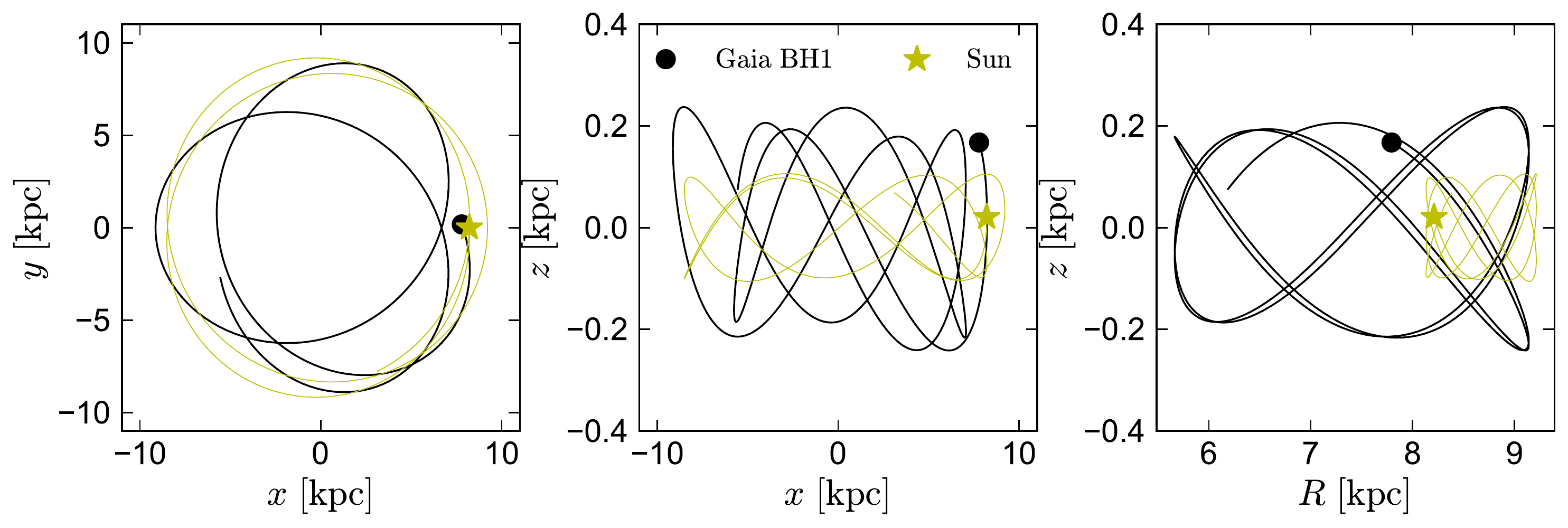}
    \caption{Galactic orbit of Gaia BH1, calculated backward for 500 Myr from the measured proper motion and center-of-mass RV. For comparison, we show the Sun's orbit calculated over the same period. The orbit is typical of a thin-disk star, ruling out a large natal kick or dynamical formation in a globular cluster.}
    \label{fig:galpy}
\end{figure*}

To investigate the past trajectory of Gaia BH1, we used its parallax and proper motion from the {\it Gaia} astrometric binary solution, as well as the center-of-mass radial velocity inferred from the spectroscopic fit, as starting points to compute its orbit backward in time for 500 Myr using \texttt{galpy} \citep[][]{Bovy2015}. We used the Milky Way potential from \citet{McMillan2017}. The result is shown in Figure~\ref{fig:galpy}; for comparison, we also show the orbit of the Sun. 
 The orbit appears typical of a few-Gyr old thin-disk star, with excursions above the disk midplane limited to $\pm 250$ pc. 
The orbit also never comes very close to the Galactic center, and is not aligned with the orbits of any globular clusters (GCs). These conclusion are insensitive to the assumed Galactic potential and uncertainties in the source's proper motion or center-of-mass RV.

\section{X-ray and radio upper limits}
\label{sec:xray_radio}
We checked archival all-sky X-ray and radio surveys to place upper limits on the flux from Gaia BH1. We used the second ROSAT all-sky survey (2RXS) catalog \citep{boller16} and the Very Large Array sky survey (VLASS) Epoch 1 CIRADA catalog \citep{lacy20, gordon20}, respectively. At the position of Gaia BH1, the nearest ROSAT source (unassociated with any optical counterpart) is 2RXS J172552.5-003516, located 26'4".1 away.  The median separation between sources in ROSAT and SDSS sources is about 12" \citep{Agueros09}, which would place Gaia BH1 130$\sigma$ away if a true association. Therefore, assuming no ROSAT detection, we place upper limits on the flux of Gaia BH1 to be the limit of the 2RXS catalog, $\sim 10^{-13}\,\rm erg\,s^{-1}\,cm^{-2}$ \citep{boller16}. This corresponds to a luminosity upper limit of $L_{\rm X} \lesssim 3\times 10^{30}\,\rm erg\,s^{-1}$ in the 0.1--2.4 keV range.

In the radio, the nearest VLASS source (also unassociated with an optical counterpart) is VLASS1QLCIR J172823.79-003035.0, located 6'5".3 away. The beam size of VLASS is 2".5 \citep{lacy20}, which would place Gaia BH1 146$\sigma$ away. Assuming no VLASS detection, we can place upper limits on the flux of Gaia BH1 to be the 3$\sigma$ limit of a single epoch of the VLASS catalog, $\sim 0.36$ mJy, which in the 2-4 GHz range (S-band) corresponds to a flux limit of $\lesssim 7 \times 10^{-18}\,\rm\ erg\,s^{-1}\,cm^{-2}$ \citep{lacy20}. This corresponds to a radio luminosity upper limit of $L_{\rm R} \lesssim 2\times 10^{26}\,\rm erg\,s^{-1}$ in the S-band.

\subsection{Could Gaia BH1 be detected with X-ray or radio observations?}
We briefly consider whether an X-ray or radio detection is expected. A rough estimate of the expected accretion rate onto the BH can be obtained under the assumption that a spherically-symmetric wind from the G star is accreted at the Hoyle-Lyttleton rate:
\begin{equation}
\begin{split}
\dot{M}_{{\rm acc}}	&=\frac{G^{2}M_{{\rm BH}}^{2}\dot{M}_{{\rm wind}}}{v_{{\rm wind}}^{4}a^{2}} \\
	&=6\times10^{-18}\,M_{\odot}\,{\rm yr}^{-1}\left(\frac{M_{{\rm BH}}}{10\,M_{\odot}}\right)^{2}\left(\frac{\dot{M}_{{\rm wind}}}{10^{-14}M_{\odot}\,{\rm yr}^{-1}}\right)\times\\ &\left(\frac{v_{{\rm wind}}}{600\,{\rm km\,s^{-1}}}\right)^{-4}\left(\frac{a}{1\,{\rm au}}\right)^{-2}.
\end{split}
\end{equation}
Here $\dot{M}_{\rm wind}$ is the wind mass-loss rate of the G star and $v_{\rm wind}$ is the wind's velocity. The predicted $\dot{M}_{\rm acc}$ is very low: less than $10^{-10} \dot{M}_{\rm edd}$, where $\dot{M}_{{\rm edd}}\sim 2\times 10^{-7}\,M_{\odot}\,{\rm yr^{-1}}\left(M_{{\rm BH}}/\left[10\,M_{\odot}\right]\right)$ is the Eddington rate with 10\% efficiency. Assuming accretion onto the BH results in an X-ray luminosity $L_{{\rm X}}=\eta\dot{M}_{\rm acc}c^{2}$, the expected flux at Earth is 
\begin{equation}
\begin{split}
F_{{\rm X}}	&=\frac{\eta\dot{M}c^{2}}{4\pi d^{2}}\\
	&=1.1\times10^{-15}\,{\rm erg\,s^{-1}\,cm^{-2}}\left(\frac{\eta}{0.1}\right)\left(\frac{M_{{\rm BH}}}{10\,M_{\odot}}\right)^{2}\left(\frac{\dot{M}_{{\rm wind}}}{10^{-14}M_{\odot}\,{\rm yr}^{-1}}\right) \\ &\times \left(\frac{v_{{\rm wind}}}{600\,{\rm km\,s^{-1}}}\right)^{-4}\left(\frac{a}{1\,{\rm au}}\right)^{-2}\left(\frac{d}{500\,{\rm pc}}\right)^{-2}.
\end{split}
\end{equation}
For radiatively efficient accretion with $\eta\sim 0.1$, this is within the flux limits of a deep {\it Chandra} ACIS observation, which are of order $10^{-16}\rm \,erg\,s^{-1}\,cm^{-2}$.
Unfortunately, models for advection-dominated accretion flows predict very low radiative efficiencies at the relevant accretion rates \citep[e.g.][]{Narayan1995, Quataert1999}. For example, extrapolating the models of \citet{Sharma2007} yields an expected $\eta \sim 10^{-6}$, corresponding to $F_{\rm X} \sim 10^{-20}\rm \,erg\,s^{-1}\,cm^{-2}$. We conclude that an X-ray detection would be surprising, though it cannot hurt to look.

The expected radio luminosity can be estimated by extrapolating the radio--X-ray correlation for quiescent BH X-ray binaries \citep[e.g.][]{Merloni2003, Gallo2006}. This predicts a radio flux of order $10\,\mu$Jy at 8\,GHz for $\eta = 0.1$ (below the VLASS limit, but detectable with the VLA), or $\sim 1\,\rm nJ$ for $\eta = 10^{-6}$ (not detectable in the near future). Uncertainty in the G star's mass loss rate (which could plausibly fall between $10^{-14}$ and $10^{-13}\,M_{\odot}\,\rm yr^{-1}$) and wind velocity near the BH (plausibly between 300 and 600 $\rm km\,s^{-1}$) leads to more than an order of magnitude uncertainty in these estimates, in addition to the uncertainty in $\eta$.


\section{Discussion}
\label{sec:discussion}

\subsection{Comparison to recent BH imposters}
\label{sec:imposters}
Many recent dormant BH candidates in binaries have turned out not to be BHs, but rather (in most cases) mass-transfer binaries containing undermassive, overluminous stars in short-lived evolutionary phases. Gaia BH1 is different from these systems in several ways.

First, the evidence for a BH companion does not depend on the mass of the luminous star, as it did in LB-1, HR 6819, and NGC 1850 BH1, which all contained low-mass stripped stars  masquerading as more massive main-sequence stars \citep[][]{Shenar2020, Bodensteiner2020, El-Badry2021, El-Badry2022}. In Gaia BH1, the mass of the luminous star could be zero, and the mass of the companion would still vastly exceed the limit allowed by the SED for normal stellar companion, the Chandrasekhar mass, and the maximum neutron star mass.

Second, the BH nature of the companion does not depend on a low assumed inclination, as was for example the case in NGC 2004 \#115 \citep[][]{Lennon2021, El-Badry2022_ngc2004}, LB-1, and NGC 1850 BH1. The inclination of Gaia BH1 could be edge-on, and RVs alone would still require the companion to be a BH. In addition, reliable constraints on the inclination are available from the {\it Gaia} astrometric solution; such constraints have not been available for any previous candidates.

Third, there is no evidence of ongoing or recent mass transfer (i.e., ellipsoidal variability or emission lines from a disk), as there was in all the candidates discussed above except NGC 2004 \#115, as well as other recent candidates with red giant primaries \citep[e.g.][]{Jayasinghe2021, El-Badry2022_unicorns}. The orbital period is long enough, and the luminous star small enough, that there is no plausible evolutionary scenario in which it was recently stripped and either component is in an unusual evolutionary state. A stripped-star scenario is also disfavored by the unremarkable optical spectrum and surface abundances of the G star, and by its measured surface gravity ($\log [g/(\rm cm\,s^{-2})]=4.55\pm 0.16$), which implies a mass $M_\star =10^{\log g}R^{2}/G=1.27\pm0.54\,M_{\odot}$.

Finally, the luminous star in Gaia BH1 has a luminosity of only $1\,L_{\odot}$, which is hundreds to thousands of times fainter than the luminous stars in all the  BH imposters discussed above. This makes it vastly harder to hide any luminous companion in the SED: any normal star with a mass exceeding $\approx 0.4\,M_{\odot}$ would easily be detected in the spectrum. 

\subsection{Nature of the unseen companion}
\label{sec:bh_or_cand}
Considering just the RVs and the properties of the G star, it seems incontrovertible that the companion is a $> 5\,M_{\odot}$ dark object.  Given the good agreement between the observed RVs and those predicted by the astrometric solution, we are also inclined to trust the inclination constraint from the astrometric solution, and thus the most likely companion mass is $\sim 10\,M_{\odot}$.

The companion has not been detected in X-rays or at radio wavelengths, and it lacks other observables associated with accreting BHs (e.g. outbursts or rapid flickering). An X-ray or radio detection would be unexpected given the weak stellar wind expected from the G star and wide separation. Constraints on the nature of the companion thus come down to probabilistic arguments conditioned on its inferred mass and low luminosity.

The mass of the companion is consistent, for example, with a single $10\,M_{\odot}$ BH, 2 lower-mass BHs, 5 neutron stars, 10 massive white dwarfs, or 200 brown dwarfs. Scenarios involving more than two objects inside the G star's orbit are difficult to assemble and unlikely to be dynamically stable. As we discuss in Section~\ref{sec:evol_history}, a scenario in which the dark object is a close binary with total mass $\sim 10\,M_{\odot}$ (containing 2 BHs or a BH and a neutron star) may be plausible. We thus proceed under the assumption that the unseen companion is either a single BH or a binary containing  2 compact objects, at least one of which is a BH.

\subsection{Comparison to other BHs}
\label{sec:population_comparison}

\begin{figure*}
    \centering
    \includegraphics[width=\textwidth]{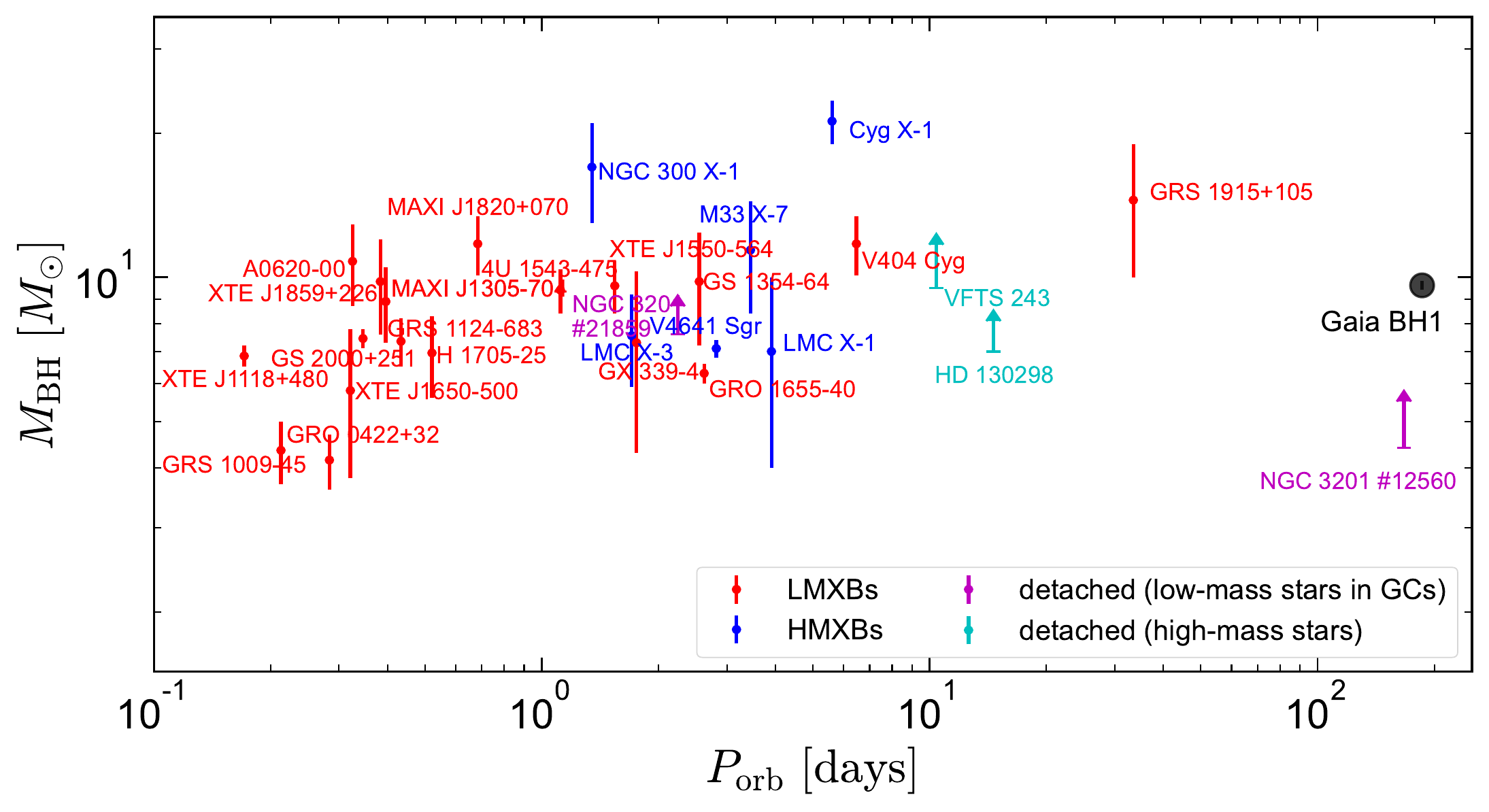}
    \caption{Comparison of Gaia BH1 (black) to known BH X-ray binaries. Red and blue symbols correspond to accreting BHs with low- and high-mass companions. Magenta symbols show detached binaries in the globular cluster NGC 3201, and cyan points show detached binaries in which the luminous star is a high-mass ($\gtrsim 20\,M_{\odot}$) star. The properties of Gaia BH1 are most similar to the detached spectroscopic binary NGC 3201 \#12560 \citep[][]{Giesers2018}. However, that system is in a globular cluster and thus probably formed through a different channel. The  closest system from an evolutionary standpoint is perhaps the LMXB GRS 1915+105, which has a red giant donor and was likely detached when it was a main-sequence star. }
    \label{fig:xrbs}
\end{figure*}

Figure~\ref{fig:xrbs} compares Gaia BH1 to other stellar-mass BHs with known masses and orbits in the literature. Red and blue points show low- and high-mass X-ray binaries, whose parameters we take from \citet{Remillard2006} and the \texttt{BlackCAT} catalog of X-ray transients introduced by \citet{Corral-Santana2016}. We take the most recent mass estimate for Cyg X-1 from \citet[][]{Miller-Jones2021}.
We also show the binaries VFTS 243 (in the LMC; \citealt{Shenar2022}) and HD 130298 (in the Milky Way; \citealt{Mahy2022}), which are both single-lined binaries containing $\sim 25\,M_{\odot}$ O stars and unseen companions suspected to be BHs. Finally, we show in magenta two binaries in the GC NGC 3201 discovered with MUSE \citep[][]{Giesers2018, Giesers2019} with suspected BH companions. For the detached systems besides Gaia BH1, the inclination is unknown, so only a lower limit on the BH mass can be inferred. Other detached BH candidates exist in the literature \citep[e.g.][]{Qian2008, Casares2014, Khokhlov2018, Thompson2019, Gomez2021}, but we do not include them because we consider their status as BHs more uncertain. Similarly, there are other X-ray bright binaries proposed to contain BHs that we do not include (e.g., Cyg X-3, SS 433) because a neutron star is not fully ruled out. We exclude IC 10 X-1 because the BH's mass is very uncertain \citep{Laycock2015}. We note that for many of the X-ray binaries included in the figure, mass estimates across different studies differ by significantly more than the reported uncertainties in individual studies. 

Unsurprisingly, the X-ray bright systems are concentrated at short periods, as they involve accretion from the companion. All the systems with low-mass donors (red symbols) and $P_{\rm orb} \gtrsim 0.5\,\rm days$ have donors that are somewhat evolved and overflow their Roche lobes. The X-ray bright systems with OB-type companions are at slightly longer periods and are fed by wind accretion from donors that nearly fill their Roche lobes. The detached systems with OB primaries, VFTS 243 and HD 130298, are at only slight longer periods than the HMXBs, and thus might evolve to form systems similar to Cyg X-1. 

The orbital period and companion star in Gaia BH1  are most similar to in NGC 3201 \#12560, the long-period BH candidate discovered in the GC NGC 3201 with MUSE  \citep[][]{Giesers2018}. That system has an orbital period of 167 days, eccentricity 0.61, and spectroscopic mass function of $3.25\,M_{\odot}$, with an $\approx 0.81\,M_{\odot}$ luminous star. Since that system's inclination is unknown, the companion mass could be any value above $4.5\,M_{\odot}$. The important difference, however, is that NGC 3201 \#12560 is in a GC, where dynamical capture, exchange, and disruption of binaries are all efficient. Two apparent BHs were discovered in NGC 3201 among only 3553 stars with multi-epoch RVs. This implies a BH companion rate of order 1000 times higher than that in the local Galactic field (Section~\ref{sec:howmany}). Such a dramatic enhancement in the rate of BH companions in GCs can be most readily understood if the binaries are formed dynamically through capture or exchange \citep[e.g.][]{Fabian1975, Rasio2000, Kremer2018}, and in any case, the long-term survival of a primordial binary with a 167 day orbit in such a dense stellar system is unlikely. 

Given the orbit of Gaia BH1 in the Galactic disk, formation from an isolated binary seems somewhat more likely than a dynamical formation channel. In this sense, the system is qualitatively different from any other binary shown in Figure~\ref{fig:xrbs}. The most closely related system may be the BH + giant star ``microquasar'' GRS 1915+105 \citep[$P_{\rm orb}=33$ days;][]{Castro-Tirado1992, Greiner2001}, which may have formed through a similar channel to Gaia BH1 but with a significantly closer separation, or through a completely different channel.

\subsection{Evolutionary history}
\label{sec:evol_history}

The progenitor of the BH very likely had a mass of at least $20\,M_{\odot}$. For example, the models of \citet{Sukhbold2016} predict that at solar metallicity, an initial mass of $30-50\,M_{\odot}$ is required to produce a helium core mass of $\gtrsim 9\,M_{\odot}$ (their Figure 19). Such a star would reach a radius of order 10\,au as a supergiant if it were allowed to evolve in isolation, which is significantly larger than the present-day separation of the G star and BH. This suggests that the two stars likely interacted prior to the formation of the BH. Given the extreme mass ratio, this interaction is expected to lead to a common envelope (CE) episode, in which the G star was engulfed by the envelope of the massive star, ultimately spiraling in and ejecting some or all of the envelope. Given the short lifetimes of massive stars, the G star would still be contracting toward the main sequence as the BH progenitor completed its evolution.

It is not clear that a $\lesssim 1\,M_{\odot}$ star can survive a CE episode with a $\gtrsim 20\, M_{\odot}$ companion. Indeed, many calculations in the literature find that ejection of the massive star's envelope by a low-mass star is impossible because there is not enough available orbital energy \citep[e.g.][]{PortegiesZwart1997, Podsiadlowski2003, Justham2006}. The same challenge applies to modeling the formation of BH LMXBs, whose evolutionary channels are still poorly understood. We first review the ``standard'' formation channel through a common envelope event, and then discuss possible alternatives.

\subsubsection{Common envelope channel}
In the CE scenario, the initial separation of the BH progenitor and the G star would have been in the range of 5-15\,au \citep[][]{PortegiesZwart1997}. The binary would have emerged from the CE episode as a $\leq 1\,M_{\odot}$ star in a close orbit with a $\gtrsim 10\,M_{\odot}$ helium core, which might also have retained some of its envelope. The standard $\alpha \lambda$ CE prescription predicts that the ratio of the separations before and after the common envelope episode is \citep[e.g.][]{Webbink1984}
\begin{equation}
    \frac{a_{f}}{a_{i}}=\frac{M_{{\rm c}}}{M_{1}}\left(1+\frac{2M_{{\rm e}}}{\alpha\lambda r_{L}M_{2}}\right)^{-1},
\end{equation}
where $M_{\rm c}$ and $M_{\rm e}$ are the mass of the BH progenitor's core and envelope, $M_1=M_{\rm c} + M_{\rm e}$ is the total mass of the progenitor, $r_L$ is its Roche lobe radius in units of the semimajor axis, and $M_2$ is the mass of the G star. $\alpha$ and $\lambda$ are dimensionless parameters, respectively describing the fraction of the G star's orbital energy that goes into ejecting the BH progenitor's envelope, and the binding energy of the BH progenitor's envelope.  If we take $M_{\rm c} = 10\,M_{\odot}$, $M_{\rm e} = 20\,M_{\odot}$, $r_{L}=0.65$, $M_2= 1\,M_{\odot}$, $\alpha=1$, and $\lambda= 0.5$ (a rather optimistic choice of $\lambda$, e.g. \citealt{Klencki2021}), we obtain $a_f/a_i = 0.0025$, such that for an initial separation of $a_i = 2000\,R_{\odot}$, the final separation is $a_f = 5\,R_{\odot}$.
This is just wide enough that the final configuration {\it might} be stable and avoid a merger, but only for a narrow range of $a_i$. For lower values of $\alpha \lambda$, there are no initial separations that both lead to Roche lobe overflow of the BH progenitor and avoid a merger \citep[e.g.][]{PortegiesZwart1997}. In any case, the predicted final separation is much closer than the observed $a \approx 300\,R_{\odot}$.

Stripped of its envelope, the helium core is expected to drive a prodigious Wolf-Rayet wind \citep[e.g.][]{Woosley1995}, which will both diminish the final mass of the compact remnant and widen the orbit. Exactly how much mass is lost in this stage is uncertain \citep[e.g.][]{Woosley1995, Fryer2001}, and attempts to measure the relevant wind mass loss rates observationally are complicated by factors such as clumping \citep[e.g.][]{Smith2014, Shenar2022_wr}. In Gaia BH1, a strong wind cannot have persisted for very long, or the helium core mass would have fallen below the required $10\,M_{\odot}$.
Eventually, the helium core will collapse to a BH, perhaps accompanied by some additional mass loss and/or an asymmetric natal kick. 

The wide orbit of Gaia BH1 and the unremarkable surface abundances of the G star strongly suggest that there has not been significant mass transfer from the G star to the BH. We can thus -- in contrast with the situation for typical BH LMXBs \citep[e.g.][]{Podsiadlowski2002, Pfahl2003, Justham2006, Fragos2015} -- rule out a scenario in which the luminous star was initially significantly more massive than today and has been stripped down to sub-solar mass by Roche lobe overflow. We can similarly rule out a scenario in which the initial mass of the BH was lower than today, and it grew significantly by accretion. 

The orbital eccentricity, $e \approx 0.45$, places a constraint on mass loss and natal kicks during the BH's formation. If the natal kick was purely due to mass loss (and not an asymmetric stellar death) and the orbit was circular before the BH formed \citep[e.g.][]{Blaauw1961}, then the predicted final eccentricity would be

\begin{equation}
    \label{eq:blaauw}
    e=\frac{\Delta M_{{\rm BH}}}{M_{{\rm BH}}+M_{\star}},
\end{equation}
such that mass loss of $\Delta M_{\rm BH}\approx 4.8\,M_{\odot}$  would be required to fully explain the observed eccentricity. Here $\Delta M_{\rm BH}$ represents any mass lost during the BH progenitor's death that does not fall into the BH. The net velocity imparted to the center-of-mass in this case is $v_{{\rm sys}}=e\times v_{{\rm BH\,prog}}$, where $v_{{\rm BH\,prog}}$ is the orbital velocity of the BH progenitor at the time of its death. Irrespective of the orbital period at the time, $v_{{\rm BH\,prog}}$ was likely rather low given the extreme mass ratio, and so (symmetric) mass loss alone is unlikely to have resulted in a system velocity larger than $\sim 10\,\rm km\,s^{-1}$. Loss of $\gtrsim 4\,M_{\odot}$ from the system at the time of the BH's formation seems somewhat unlikely in the CE channel because the core should collapse rapidly and most of the envelope should have been removed by the CE episode, but the details of how the CE episode would proceed are uncertain. 

On the other hand, if a kick occurred mainly due to asymmetries during the BH progenitor's death (e.g., asymmetric ejecta or neutrino emission), the final eccentricity depends on the kick velocity and direction, and on the period at the end of the CE episode \citep[e.g.][]{Brandt1995, Hurley2002}. In the limit of no mass loss, the systemic velocity induced by a kick is simply $v_{{\rm kick}}\times q/\left(1+q\right)$, which in the limit of $q\gg 1$ is close to $v_{\rm kick}$. Here $q=M_2/M_{\star}$. 

Forming Gaia BH1 via a CE is quite challenging for two reasons: (1) the G star may not have had enough orbital energy to eject the envelope of its much more massive companion (this is generically a problem for BH LMXB formation models, but is particularly problematic here because the G star cannot have been born with a higher mass than observed today), and (2) if CE ejection {\it was} successful, the post-CE separation is expected to be significantly closer ($\sim 5\,R_{\odot}$) than the wide orbit observed today ($\sim 300\,R_{\odot}$) under standard assumptions for CE ejection efficiencies. A fine-tuned natal kick to the BH could in principle widen the orbit, but this would also make the final orbit highly eccentric and impart a large systemic velocity to the binary's center of mass, both of which are not observed. For example, assuming an optimistic post-CE separation of $20\,R_{\odot}$ and random kick orientations, we find \citep[e.g.][]{Brandt1995} that kicks of order $200\,\rm km\,s^{-1}$ are required to widen the orbit to $a\gtrsim 200\,R_{\odot}$. Such kicks result in a minimum eccentricity of 0.90 and a median of 0.95, both much larger than observed.  

Given the seemingly impossible nature of forming Gaia BH1 through the CE channel under standard assumptions, we performed a study of more extreme assumptions for CE and BH natal kicks to determine the required combination of assumptions to produce a binary with the observed properties of Gaia BH1. We follow the methods introduced in \citet{Wong2022} which uses COSMIC, a rapid binary population synthesis code, to explore all possible combinations of Zero Age Main Sequence (ZAMS) properties, CE assumptions, and natal kick assumptions; for an in-depth discussion of how COSMIC evolves binary-star populations, see \citet{Breivik2020}. 

Instead of selecting a single set of assumptions for CE and natal kicks, we use \texttt{emcee} to sample the posterior of the masses, orbital period, and eccentricity of the binary hosting Gaia BH1 with combinations of the ZAMS mass, orbital period, and eccentricity, as well as the CE ejection efficiency ($\alpha$) and the BH natal kick strength ($v_{\rm{kick}}$) and direction ($\theta$, $\phi$) and the mean anomaly of the orbit at the time of core collapse ($\mathcal{M}$) as model parameters. We place uniform priors on all ZAMS binary parameters as well as the CE and natal kick assumptions. We assign wide limits for the ZAMS primary mass prior between $1\,M_{\odot}$ and $150\,M_{\odot}$ and narrow limits for the ZAMS secondary mass prior between $0.5\,M_{\odot}$ and $2\,M_{\odot}$ since the G star is not expected to have significantly gained or lost mass during the binary evolution. The orbital period and eccentricity priors are limited to $500$ days and $6000$ days and $0$ and $1$ respectively. We limit our prior on $\alpha$ to be between $0.1$ and $20$ and our prior on $v_{\rm{kick}}$ to be between $0\,\rm{km\,s^{-1}}$ and $300\,\rm{km\,s^{-1}}$ based on the relatively wide and moderately eccentric orbital configuration of Gaia BH1's binary host. We place uniform priors on the unit sphere for the natal kick angles, $\theta$, $\phi$, and all allowed values of the mean anomaly, $\mathcal{M}$, at core collapse ($0$--$360^{\deg}$). Finally, we fix the metallicity of all simulated binaries to $Z=0.63Z_{\odot}$, where $Z_{\odot}=0.02$, consistent with $\rm [Fe/H] = -0.2$ under the assumption that the G star follows the solar abundance pattern.

Since the age of the observed G star is not known to great precision, we require the properties of our simulated BH binaries to match the observed properties of the binary hosting Gaia BH1 just after the formation of the BH. This choice does not affect our results because the G star's presence on the main sequence implies that tides are not expected to alter the binary's orbit between the formation of the BH and the present day. 

We initialize $1024$ walkers uniformly over the prior space in ZAMS mass, orbital period, eccentricity, natal kick strength and direction, and CE ejection efficiency. We evolve each walker for $100000$ steps, thin the chains by a factor of $10$, and retain the final $2000$ steps of each chain to ensure a proper burn-in of our sampling.

\begin{figure*}
    \centering
    \includegraphics[width=\textwidth]{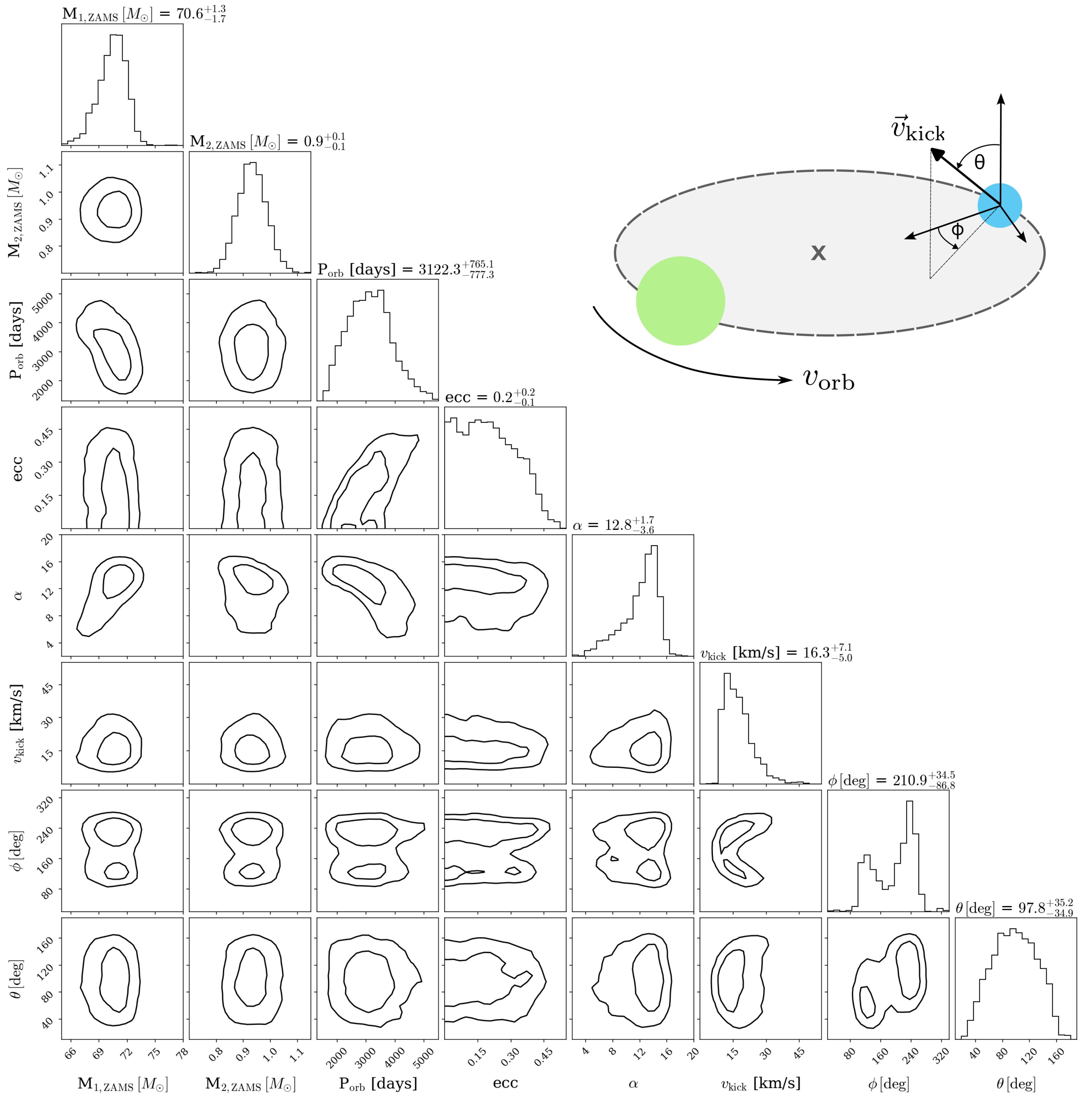}
    \caption{The combination of sampled ZAMS component masses, initial orbital period, and initial eccentricity with uncertain binary evolution model parameters for CE ($\alpha$) and BH natal kicks ($v_{\rm{kick}}$, $\phi$, $\theta$) which evolve to produce BHs in binaries with properties similar to Gaia BH1 and its binary host. We do not show distributions for the mean anomaly ($\mathcal{M}$) because there is not a significant preferred angle nor any appreciable correlations with any other parameters. The contours show the $50^{\rm{th}}$ and $90^{\rm{th}}$ percentiles of each distribution. The diagram in the upper right corner illustrates the orbital configuration and natal kick direction for a Wolf-Rayet BH progenitor (blue) at core collapse in a circular orbit with a G star (green). Our models strongly prefer small natal kicks at BH formation and extremely efficient CE ejection efficiency.}
    \label{fig:CE_params_corner}
\end{figure*}

We find that the preferred evolutionary pathway is for binary to begin with an orbital period between $1600$--$4800$ days depending on the initial eccentricity, which ranges between $0$--$0.5$. The BH progenitor's ZAMS mass is preferred to be between $65$--$75\,M_{\odot}$, while the G star's ZAMS mass is preferred to be between $0.8$--$1.1\,M_{\odot}$. From this initial configuration, the BH progenitor loses roughly $20\,M_{\odot}$ due to strong stellar winds which widens the binary out to periods of $\sim7000$ days. As the BH progenitor evolves off the main sequence and begins core helium burning, its envelope expands and tides work to circularize and shrink the binary back to orbital periods between near $\sim3000$ days. Near the end of core helium burning, the BH progenitor fills it's Roche lobe and the mass transfer becomes dynamically unstable and forms a CE due to both the convective envelope of the BH progenitor and the orbital evolution which drives the binary components close together due to their highly asymmetric mass ratios. The BH progenitor's envelope is fully stripped in the CE, leaving behind a $\sim20\,M_{\odot}$ Wolf-Rayet star in a $\sim20\,\rm{day}$ orbit with the G star. The Wolf-Rayet star's strong wind mass loss widens the orbit to roughly $80\,\rm days$. Finally the BH progenitor reaches core collapse and explodes which both reduces the BH progenitor mass from $\sim12\,M_{\odot}$ to the mass of Gaia BH1 and imparts a natal kick that increases the eccentricity and orbital period to reach the present day observed values. 

Figure~\ref{fig:CE_params_corner} shows the combination of ZAMS parameters and binary evolution assumptions which produce Gaia BH1-like binaries whose masses, orbital period, and eccentricity match Gaia BH1's present day properties for the $50^{\rm{th}}$ and $90^{\rm{th}}$ percentiles of each distribution. Due to the moderately wide orbit of Gaia BH1, our models prefer low natal kick strengths with $v_{\rm{kick}}<45\,\rm{km\,s^{-1}}$, which is consistent with Gaia BH1's orbit lying near the plane of the Galactic disk. There is a slight  preference for kicks centered in the plane of the orbit ($\theta\sim90^{\circ}$), though there is support over nearly the entire allowed range. A correlation between $\phi$ and $v_{\rm{kick}}$ exists such that larger kicks prefer azimuthal angles near $90^{\circ}$ and $270^{\circ}$. This is due to a higher likelihood of the binary to remain bound for larger natal kicks when the orbital and kick velocities are pointed in opposite directions. We do not show samples for the mean anomaly since there is a near equal preference for all values of $\mathcal{M}$. This is not unexpected since the binary's orbit is circularized during the CE, thus fixing a constant orbital speed.

While the ZAMS masses are not correlated with the ZAMS orbital period, ZAMS eccentricity, or natal kick parameters, there is a correlation between the ZAMS primary mass and the CE ejection efficiency. This is expected since an increased primary mass will have a larger envelope which requires a larger, more efficient, $\alpha$ to produce a successful CE ejection. The ZAMS orbital period is also correlated with $\alpha$; longer initial periods require lower, less efficient $\alpha$'s to shrink the orbit to the proper separation after the CE. 

Our models prefer extremely efficient CE ejections with $\alpha$ constrained to always be larger than $5$ with a preference for $\alpha=14$. We remind the reader that $\alpha$ represents the fraction of the G star's orbital energy that goes into unbinding the BH progenitor's envelope. Values of $\alpha$ significantly larger than 1 signify that the liberated orbital energy alone is insufficient to eject the envelope, and that some other source of energy is required. While there are large uncertainties in the binding energy of the BH progenitors in our models for which $\lambda\sim0.5$, it is highly unlikely that such uncertainties will change $\lambda$ by more than a factor of $10$. Furthermore, while it is possible for $\alpha$ values to exceed unity due to additional energy sources like recombination, it is unlikely that such sources will dominate the energy budget \citep{Ivanova2015, Ivanova2018}. It is also possible that CE proceeds differently than the standard $\alpha\lambda$ prescription \citep[e.g.][]{Hirai2022}, though we leave this for a future study. We thus conclude that Gaia BH1 can be formed through standard binary evolution with a common envelope only under extreme (and likely unphysical) assumptions about how the common envelope evolution proceeds. 

In this sense,  Gaia BH1 is reminiscent of the several known pulsars in wide, moderately eccentric binaries ($P_{\rm orb}=80-400$ days) with companions that appear to be low-mass main-sequence stars \citep[e.g.][]{Phinney1991, Champion2008, Parent2022}. These systems are also too wide to form via a common envelope, but too close to have avoided one. It is possible that the companions are white dwarfs, but their nonzero eccentricities are difficult to explain in this scenario. Perhaps they formed through a similar channel to Gaia BH1. We consider possible alternative formation scenarios below.

\subsubsection{Formation from a progenitor that never became a giant}
\label{sec:no_common_env}
A CE event can be avoided if the BH progenitor never became very large.
Models predict that at sufficiently high masses ($M\gtrsim 50\,M_{\odot}$, depending on wind prescriptions and rotation rates), stars do not expand to become red supergiants, but instead lose their hydrogen envelopes to winds and outbursts during and shortly after their main-sequence evolution, reaching maximum radii of order $50\,R_{\odot}$ \citep[e.g.][]{Humphreys1994, Higgins2019}. A scenario in which the BH progenitor in Gaia BH1 never became a red supergiant has the attractive feature that it could seem to avoid a CE episode. It requires, however, that the G star formed uncomfortably close to the BH progenitor. The orbit would have expanded by a factor of $\sim 5$ due to the BH progenitor's mass loss, as the quantity $a\times(M_1 + M_2)$ is conserved under adiabatic orbit evolution \citep{Jeans1924}. This in turn requires an initial separation of $\lesssim 60\,R_{\odot}$ and an initial period of less than 10 days. It seems somewhat unlikely that a pre-main sequence solar-mass star could form and survive so close to a massive star, but observational constraints on the existence of such companions to O stars -- which would have luminosity ratios $\gtrsim 10^5$ and angular separations of order 0.1 mas at typical Galactic distances -- are extraordinarily difficult to obtain \citep[e.g.][]{Rizzuto2013, Sana2014}.

A related set of models allow rapidly-rotating massive stars to avoid becoming red supergiants due to mixing, such that a chemical gradient never forms and a majority of the star is converted to helium \citep[``quasi-homogeneous evolution''; e.g.][]{Maeder1987, Woosley2006}. Such a scenario could also avoid the problems with the CE channel, because the G star could have been born in an orbit close to the current separation, never interacting with the BH companion. The BH progenitor could have also had a somewhat lower initial mass, perhaps down to $15\,M_{\odot}$. However, most models predict that such evolution only occurs at low metallicity, because winds remove too much angular momentum at high metallicity \citep[e.g.][]{Yoon2006, Brott2011}. The $\rm [Fe/H] = -0.2$ we measure for the G star thus presents a challenge for this scenario.

\subsubsection{Formation through three-body dynamics}
Mechanisms for forming BH LMXBs have been proposed involving tidal capture of inner binaries in hierarchical triples \citep[e.g.][]{Naoz2016} or very wide BH + normal star binaries perturbed by encounters with field stars \citep[e.g.][]{Michaely2016}. In both cases, tidal effects can bring the binary into a close orbit only if a periastron passage occurs within which the star comes within a few stellar radii of the BH. If this occurs, tides can efficiently dissipate orbital energy, locking the star into a close orbit and forming a LMXB. Such a scenario seems unlikely to have operated in Gaia BH1, because the observed periastron distance is too wide for tides to be important. 

\subsubsection{Formation in a dense cluster}
In dense environments like the cores of GCs, binaries with separations and eccentricities similar to Gaia BH1 can be formed via exchange encounters \citep[e.g.][]{Kremer2018}.
Formation in a GC is disfavored for Gaia BH1 due to its thin-disk like orbit and high metallicity. 

Another possibility is that the Gaia BH1 system was assembled dynamically in an open cluster. Recent work \citep{Fujii2011,Sana2017,Ramirez-Tannus2021} suggests that many massive star binaries are initially formed in long-period orbits, but are subsequently hardened due to dynamical encounters over millions of years. In the case of Gaia BH1, the binary may have formed at longer periods, allowing the BH progenitor to evolve into a red supergiant and then collapse into a BH within $\sim \! 3 \, {\rm Myr}$. Dynamical interactions after the BH formation could then shrink the orbital separation to the value observed today. Alternatively, the BH progenitor may have evolved separately from the G star in Gaia BH1. After BH formation, a dynamical exchange within the birth cluster could have allowed it to capture the G star into the current orbit \citep{Banerjee2018,Banerjee2018b}. The binary system could have simultaneously or subsequently been ejected from the cluster at low velocity, or simply remained bound after the cluster dissolved \citep{Schoettler2019,Dinnbier2022}. 

\citet{Shikauchi2020} recently investigated the dynamical formation of BH + main sequence binaries in open clusters and predicted that only a small fraction of such binaries observable by {\it Gaia} would have formed dynamically. However, their simulations did not include any primordial binaries, and thus did not account for binaries formed through exchange interactions. Further work investigating this formation channel is warranted.

\subsubsection{Formation in a hierarchical triple}
\label{sec:hierarchical}

Given the problems with binary formation models noted above, triple or multi-body formation channels may have operated in Gaia BH1. One possibility is that the BH progenitor was originally in a compact binary ($P_{\rm orb} \! \lesssim \! 20 \, {\rm d}$) composed of two massive stars. As the stars evolve, they transfer mass to each other, preventing either star from evolving into a red supergiant. A simple possibility is that the inner binary merged during unstable Case B mass transfer, and the merger product never swelled into a red supergiant \citep{Justham2014}, allowing the G star (originally an outer tertiary) to survive. 

Alternatively, the inner binary may have undergone stable mass transfer. Consider a binary with $M_1 \! \sim \! 35 \, M_\odot$ and $M_2 \! \sim \! 20 \, M_\odot$ in an orbit with $P_{\rm orb} \! \sim \! 4\, {\rm days}$. As the primary evolves, its envelope is stripped during stable Case A mass transfer, producing a helium star which then collapses into a BH with $M_{\rm BH} \! \sim \! 10 \, M_\odot$. During the mass transfer, the orbital period and secondary mass only change slightly. A very similar evolution is thought to produce observed HMXBs such as Cyg X-1 \citep{qin2019}. The secondary then evolves and has its envelope stripped during a second phase of stable Case A mass transfer, forming a helium star.
Depending on the strength of its stellar wind and the core-collapse process, this helium star could collapse into a second BH, or explode and produce a neutron star, which may or may not remain bound. A similar model was proposed by \citet{Shenar2022} for the binary VFTS 243. At the end of the process we are left with a BH and possibly a second compact object in a short-period orbit, with a still-bound outer tertiary. 

To test this idea, we constructed binary models with MESA \citep{Paxton_2011} using the problem setup described in \citet{fuller2019}. Assuming non-conservative mass transfer (as expected from the high mass transfer rates achieved during Case A mass transfer), we find very similar binary evolution histories to those discussed above. For example, if the second phase of mass transfer begins with a $20 \, M_\odot$ donor and a $10 \, M_\odot$ BH in an orbital period of 4 days, it ends with a $5 \, M_\odot$ helium star in a $9.6 \, {\rm day}$ orbit such that the orbit of the G star in Gaia BH1  could plausibly remain stable. As in the channels with single BH progenitors that never became giants, this scenario requires the G star to have formed quite close ($\lesssim 100\,R_{\odot}$) from the inner binary. Given this and the expected orbital evolution of the inner binary due to mass transfer and winds, the parameter space of triples that can form a system like Gaia BH1, while not empty, is likely rather small. In the triple scenario it would also be somewhat surprising that the only good candidate we identified from {\it Gaia} DR3 has a (relatively) short period, because the effective search volume is larger for long periods (up to 1000 days; Section~\ref{sec:howmany}), and hierarchical triples with long outer periods would also be stable for a wider range of inner binary parameters. 

While the triple model may seem unlikely given the rarity of BH HMXBs, it is important to note that the HMXB phase lasts at most $\lesssim 10^6 \, {\rm yr}$, while the subsequent quiescent phase can easily last more than $10^{\rm 10} \, {\rm yr}$. If a significant fraction of HMXBs host outer low-mass tertiary stars, this could account for the high space-density of objects like Gaia BH1 relative to HMXBs.

A prediction of this channel is that Gaia BH1 might still harbor a binary compact object, whose orbital motion would induce potentially observable perturbations to the RVs of the outer G star on half the orbital period of the inner binary. \cite{Hayashi2020} show that the amplitude of these perturbations would scale as $ \left(P_{{\rm orb,\,inner}}/P_{{\rm orb,\, outer}}\right)^{7/3}$; in Gaia BH1, the expected perturbation semi-amplitude ranges from $\sim 2\,\rm m\,s^{-1}$ for a 4 day inner period, to $\sim 20\,\rm m\,s^{-1}$  for a 10 day inner period, to $\sim 250\,\rm m\,s^{-1}$ for a 30-day inner period. Since the G star is bright and slowly-rotating, these amplitudes are within the realm of detectability.  In addition, an inner binary would cause precession of the G star's orbit, which could be measurable via time-evolution of the orbit's orientation and RVs. The predicted minimum precession period ranges from about 35 years for a 30-day inner period, to 350 years for a 4-day inner period (and still longer for shorter inner periods). The RV semi-amplitude of this precession would be large (of order the observed semi-amplitude;  $70\,\rm km\,s^{-1}$), but a long observing time baseline would be required to detect it.

\section{How many similar objects exist?}
\label{sec:howmany}
Here we consider what the discovery of one dormant BH in a binary implies about the broader population. The precision of the conclusions we can draw is of course limited  by the Poisson statistics of $N=1$. We will show, however, that {\it Gaia} has observed tens of millions of sources around which a BH companion could have been detected, if one existed. This makes possible robust inference about the occurrence rate of BH companions, as long as the selection function of the astrometric catalog can be understood. 

As described by \citet{Halbwachs2022}, the cuts placed on astrometric solutions published in {\it Gaia} DR3 were quite conservative, designed to reduce false positives at the expense of completeness. Attempts to fit astrometric orbits were made only for sources satisfying \texttt{ruwe > 1.4} (indicating a poor fit with a single-star solution) that were observed in at least 12 visibility periods. Sources with evidence of marginally resolved companions or contaminated photometry were likewise discarded. Sources for which a satisfactory fit could be achieved with an accelerating solution were not fit with an orbital solution. After orbital solutions were fit, those which did not satisfy all of the following three conditions were discarded:

\begin{align}
    \label{eq:cut}
    \frac{\varpi}{\sigma_{\varpi}}	&>\frac{20000\,{\rm days}}{P_{{\rm orb}}} \\
    \label{eq:acut}
    \frac{a_{0}}{\sigma_{a_{0}}}	&>\frac{158}{\sqrt{P_{{\rm orb}}/{\rm day}}} \\
    \label{eq:ecut}
    \sigma_{e}	&<0.079\ln\left(P_{{\rm orb}}/{\rm day}\right)-0.244
\end{align}
Here $\sigma_e$ is the uncertainty on the orbital eccentricity. As shown by \citet{Halbwachs2022}, Equation~\ref{eq:cut} removes a large fraction of otherwise acceptable solutions. After Equation~\ref{eq:cut} is applied, Equations~\ref{eq:acut} and \ref{eq:ecut} remove a relatively small number of additional solutions. 

All these cuts are more aggressive for binaries with shorter periods than those with longer periods. 
The astrometric solution for Gaia BH1 has $\varpi/\sigma_{\varpi} = 120$ and $20000/P_{\rm orb} = 107$. That is, it just narrowly passes the cut in Equation~\ref{eq:cut}, and would have been excluded if it were 10\% more distant. Similarly, it has $a_0/\sigma_{a_0} = 13.6$; if this quantity were smaller than $158/\sqrt{P_{\rm orb}}= 11.6$, it would have been excluded by Equation~\ref{eq:acut}. The fact that Gaia BH1 narrowly escaped the quality cuts applied to the whole catalog has two implications. 

First, there are likely other similar systems that could be discovered with only slightly looser cuts. We estimate the number of detectable systems explicitly in Section~\ref{sec:surveyvol}. Second, BH companions may be more common at shorter periods than at longer periods.  {\it Gaia} DR3 was much more sensitive to binaries with long orbital periods (up to the observing baseline, $\approx 1000$ days) than to those with short periods: Equation~\ref{eq:cut} will exclude systems with $\varpi/\sigma_{\varpi} < 100$ at $P_{\rm orb} = 200$ days, but only those with $\varpi/\sigma_{\varpi} < 20$ at  $P_{\rm orb} = 1000$ days. At fixed absolute magnitude, this translates to a factor of $\approx 3$ larger distance limit, and a factor of $\approx 10$ larger survey volume (Section~\ref{sec:surveyvol}). 

Gaia BH1 has a shorter orbital period than 95\% of all objects in DR3 with astrometric solutions. The fact that it appears to be the only credible BH with $P_{\rm orb} \lesssim 1000$ days, while at longer-periods ($P_{\rm orb} = 400-1000$ days), BH companions could have been detected around a much larger number of sources, thus points to a relative paucity of longer-period binaries consisting of a BH and a normal (low-mass) star.

\subsection{Effective survey volume of the astrometric binary sample}
\label{sec:surveyvol}

We now explore around how many sources {\it Gaia} could have detected a BH companion if one existed. Since ``detectability'' depends on the astrometric uncertainties, this requires a noise model. We describe such a model in Appendix~\ref{sec:sigma_a0_appendix}.

The actual uncertainties for a given source will depend on factors such as the orbital eccentricity, orientation, and {\it Gaia} scanning pattern, but we neglect these details. 
We assume that at $P_{\rm orb} < 1000$ days, the astrometric uncertainties $\sigma_{\varpi}$ and $\sigma_{a_0}$ depend only on apparent magnitude and $a_0$. We constrain the dependence on both parameters empirically using the 170,000 astrometric binary solutions published in DR3. The median astrometric uncertainty for sources with orbital solutions in DR3 ranges from $\sigma_{\varpi}=0.020$ mas at $G< 14$, to 0.045 mas at $G = 16$, to 0.08 mas at $G = 17$. The median uncertainty in $a_0$ is larger by a factor of $\approx 1.5$ on average. The astrometric uncertainties also depend on $a_0$, increasing by about a factor of 2 from $a_0=0.3$ mas to $a_0 = 3$ mas. This reduces the number of BH companions that could be detected, since BHs produce larger $a_0$ at fixed $P_{\rm orb}$ than luminous companions.

We now assign a hypothetical $10\,M_{\odot}$ dark companion to each source in the {\it Gaia} catalog and ask whether it could have been detected at a given period. When calculating the expected $a_0$, we assign all sources on the main sequence a mass based on their absolute magnitude, and all giants a mass of 1 $M_{\odot}$. We apply the same cuts on apparent magnitude, photometric contamination, and image parameter metrics that were applied to sources in the actual {\it Gaia} binary catalog \citep[][]{Halbwachs2022}. We calculated the expected uncertainties, $\sigma_{\varpi}$ and $\sigma_{a_0}$, based on the mean and observed scatter of these quantities among all astrometric solutions with the same $G$ and $a_0$, as described in Appendix~\ref{sec:sigma_a0_appendix}. We assume that no binaries with periods within $\pm 15\%$ of one year can be detected. 

\begin{figure}
    \centering
    \includegraphics[width=\columnwidth]{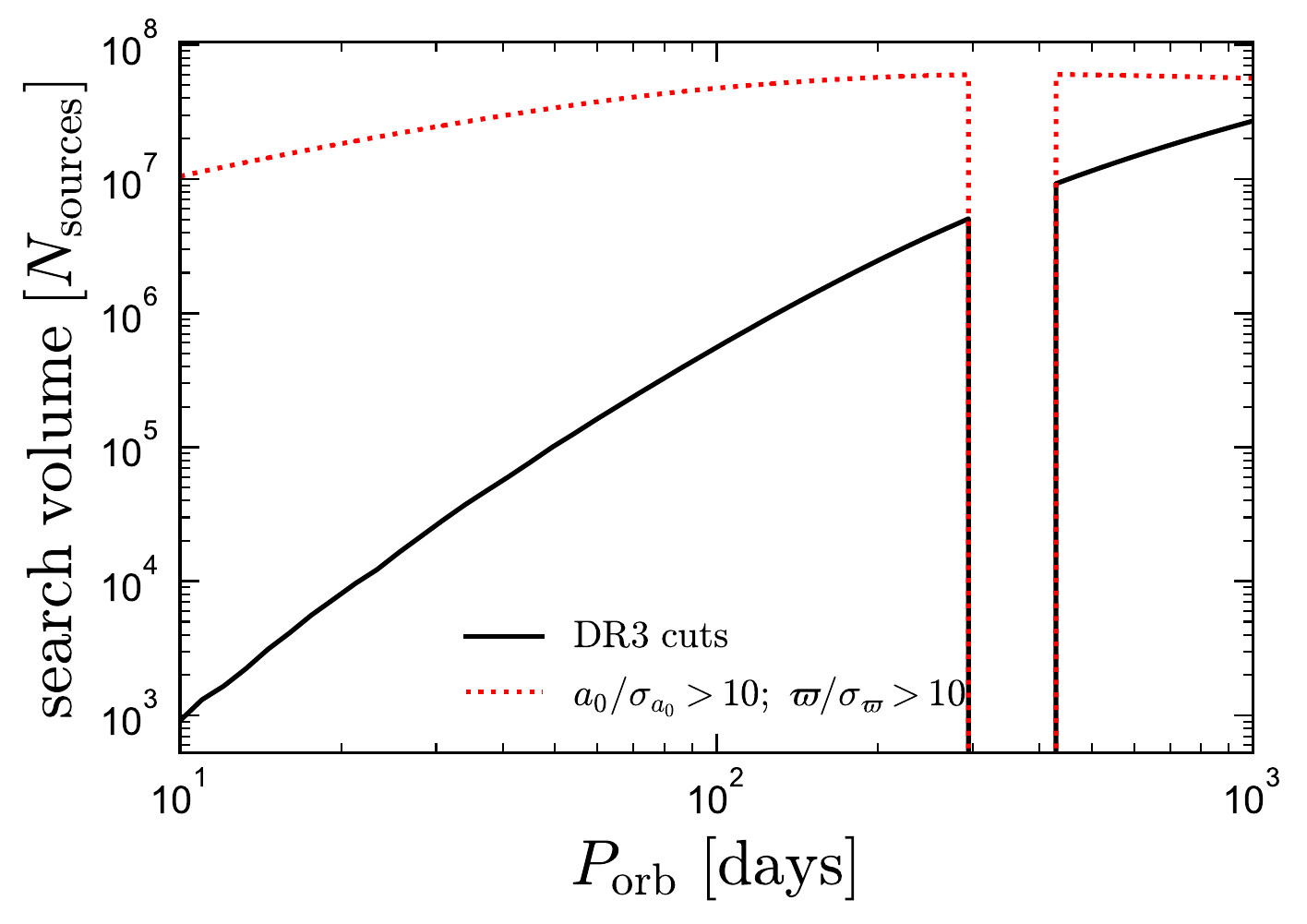}
    \caption{Number of sources in {\it Gaia} DR3 in which a $10\,M_{\odot}$ dark companion could be detected. Black line shows the cuts employed in DR3 (Equations~\ref{eq:cut} and~\ref{eq:acut}), which remove most short-period binaries. Dashed red line corresponds to simply requiring $a_0/\sigma_{a_0} > 10$ and $\varpi/\sigma_{\varpi} > 10$; this results in a much larger searchable volume at short periods. Binaries with periods near 1 year are excluded, being difficult to detect with astrometry.  }
    \label{fig:volume_different_cuts}
\end{figure}

Figure~\ref{fig:volume_different_cuts} shows the number of sources in the \texttt{gaia\_source} catalog for which a $10\,M_{\odot}$ dark companion would likely pass different quality cuts. The black line corresponds to the detection thresholds actually employed in DR3. As discussed above, there is a strong selection bias against shorter orbital periods, which were only accepted at extremely high $\varpi/\sigma_{\varpi}$.

In DR4, {\it Gaia} is planning to publish epoch-level astrometry for all sources, irrespective of whether they satisfy cuts like Equation~\ref{eq:cut}. This will make it possible to identify sources similar to Gaia BH1 at larger distances. The dotted red line in Figure~\ref{fig:volume_different_cuts} shows the effective search volume for the looser -- but still relatively conservative -- threshold of a minimum astrometric SNR of 10. 
Using this threshold instead of  Equations~\ref{eq:cut} and ~\ref{eq:acut} would increase the search volume for $10\,M_{\odot}$ companions at a period of 186 days by a factor of  20. Gains at shorter periods would be even more dramatic. Of course, removing these cuts is likely to result in a large number of spurious solutions -- this is why they were applied in the first place. Inspection of the epoch-level astrometric measurements, coupled with spectroscopic follow-up of the most promising candidates, will be necessary and sufficient to eliminate spurious astrometric BH candidates.

In addition, the factor of two improvement in observing baseline between DR3 and DR4 will make DR4 have more precise astrometry and be sensitive to orbits with periods as large as 2000 days, for which the search volume is also larger. While extrapolating future detection rates from the current search criteria and one detection is risky, we conclude that there is good reason to expect dozens more astrometric BHs from {\it Gaia} DR4. The number could also be higher if the occurrence rate of (wide) BH companions is significantly higher among massive stars than among solar-type stars, as is expected theoretically \citep[e.g.][]{Langer2020, Janssens2022}. Few massive stars received orbital solutions in DR3, because most are too distant to pass the stringent parallax uncertainty cut of Equation~\ref{eq:cut}.

One can infer a rough lower limit on the number of objects similar to Gaia BH1 that exist in the Milky Way. 
Figure~\ref{fig:volume_different_cuts} shows that the effective search volume at 186 days is about $2.5\times 10^6$ sources for which a $10\,M_{\odot}$ companion would be detectable with DR3 cuts. Translating this into a constraint on the occurrence rate of BH companions in a given period range requires an assumed BH companion period distribution, which is obviously very uncertain. If we assume that a fraction $f$ of all low-mass stars have a $10\,M_{\odot}$ BH companion with $100 < P_{\rm orb}/{\rm days} < 300$ and the period distribution is log-uniform in this range (${\rm d}N/{\rm d}\log P_{{\rm orb}}={\rm const}$), then we find that 1 detection implies $f\approx 4\times 10^{-7}$; i.e., an occurrence rate of order 1 in 2.5 million for BH companions to low-mass stars with $100 < P_{\rm orb}/{\rm days} < 300$. This is a constraint on the average incidence across a range of stellar masses, sampled in a way that is different from the mass function. A large majority of the sources in this search volume are main-sequence stars with masses between 0.5 and 2\,$M_{\odot}$. Multiplying by $\sim 10^{11}$ low-mass stars, this would translate to $\sim 40000$ similar systems in  the Milky Way -- a factor of $\sim 40$ larger than the expected number of BH X-ray binaries \citep[][]{Corral-Santana2016}, but still less than 0.1\%  of the expected total number of BHs.  

We can compare the occurrence rate of Gaia BH1-like systems to the number of stars formed that are massive enough to become BHs. For a \citet{Kroupa2001} IMF, roughly 2 out of every 1000 stars formed has $M>20\,M_{\odot}$. Thus, if 1 out of every 5000 stars with $M>20\,M_{\odot}$ has a solar-type companion in a Gaia BH1-like orbit when it becomes a BH, this could produce the population implied by Gaia BH1, with 1 in 2.5 million low-mass stars having a BH companion with $100 < P_{\rm orb}/{\rm days} < 300$. The ``1 in 5000'' factor could reflect a dearth of extreme mass ratio companions to massive stars, and/or the fact that a majority of such companions do not evolve to become intact BH + normal star binaries at these separations. 

The lack of a solid detection at $P_{\rm orb} = 400-1000$ days implies a lower BH companion incidence rate there. Assuming a log-uniform period distribution as above, the probability of detecting at least one BH companion in this period range would be $\gtrsim 50\%$ for  $f > 4\times 10^{-8}$; i.e., if the BH companion occurrence rate were more than 1 in 25 million in this period range. As above, this limit depends sensitively on the assumed period distribution.

Limits at shorter periods are weaker, given the implicit period bias of the selection cuts. For example, the number of sources for which a $10\,M_{\odot}$ companion in a 60-day orbit would be detected astrometrically in DR3 is only 0.3 million. At such short periods, the (lack of) detections in {\it Gaia} SB1 (spectroscopic) solutions becomes a more stringent limit. DR3 contains 5.9 million sources satisfying \texttt{rv\_method\_used} = 1 (meaning that RVs were measured from individual epochs) and \texttt{rv\_nb\_transits} > 10 (enough epochs that an orbital solution could plausibly be inferred). About 2.5 million of these are near the main sequence. Orbital solutions with $K_\star > 250\,\rm km\,s^{-1}$ were removed from DR3 \citep{Arenou2022}, but this will only remove BH companions to solar-type stars at $P_{\rm orb}\lesssim 3$ days. This suggests that BHs could have been detected as companions to about 2.5 million sources at periods ranging from 3 to 1000 days, greatly exceeding the astrometric search volume at short periods. A non-detection of BH companions in the DR3 SB1 sample would seem to rule out a large increase in the BH companion incidence rate toward short periods. However, it is still unclear from the published information how SB1 solutions were processed and what cuts were ultimately applied to them. It is possible, for example, that close, tidally synchronized companions to BHs would have had too large $v\sin i$ for RVs to be reliably measured by the {\it Gaia} RVS pipeline.

\subsection{Prospects for detecting similar systems with spectroscopy and photometry}
Time-domain spectroscopic surveys provide another promising avenue for detecting more BH + normal star binaries like Gaia BH1. Indeed, Gaia BH1 was observed twice by the LAMOST survey; the source would have been easily recognizable as a promising BH candidate if it had been observed $\sim$ 10 times with suitable cadence. {\it Gaia} will eventually provide $>20$ epochs of spectroscopy for most bright, ``normal'' stars with $G \lesssim 12.5$. The calculations above suggest that at most a few BHs similar to Gaia BH1 will satisfy  $G \lesssim 12.5$. Many-epoch surveys from the ground are needed to reach targets that are fainter or harder to measure RVs for (e.g. hot and rapidly-rotating stars). Most current wide-field surveys do not obtain enough epochs per target to constrain the orbital period, but more targeted surveys focused on binarity do \citep[e.g.][]{Giesers2019, Wang2021, Shenar2022_survey}. We note that, except in cases where orbital periods can be constrained with photometry or astrometry, spectroscopic surveys that obtain $\gtrsim 10$ epochs for a modest number of targets are more useful for characterizing binaries than those obtaining 1-3 spectra for a larger number of targets. 

Besides astrometry and spectroscopy, another avenue for detecting Gaia BH1-like binaries is through photometry. Ellipsoidal variation is not currently detectable (the expected variability amplitude is $\approx 5\times 10^{-7}$; \citealt{Morris1993}) but will become detectable when the G star becomes a red giant. In the meantime, Gaia BH1-like systems could be detected through self-lensing \citep[e.g.][]{Masuda2019}. If a binary like Gaia BH1 were viewed edge-on, the G star would brighten by $\approx 5\%$ each time it moved behind the BH. The typical duration of this brightening would be 3-4 hours; i.e., a duty cycle of about $10^{-3}$. Self-lensing would only be detectable for the $\approx 0.3\%$ of binaries with sufficiently edge-on inclinations to be eclipsing, but given the large expected magnification amplitude, a self-lensing search could extend to significantly fainter targets than are accessible with spectroscopic surveys. Given that several self-lensing binaries containing white dwarfs were detected by {\it Kepler} with longer periods and much smaller amplitudes than expected for a system like Gaia BH1 \citep[][]{Kruse2014, Kawahara2018}, future missions like {\it Earth 2.0} \citep[][]{Ge2022} may detect rarer systems containing BHs.  

\subsection{Other Gaia astrometric candidates}
\label{sec:others}
A few other recent works have searched {\it Gaia} DR3 for dormant BH + normal star binaries. \citet{Andrews2022} analyzed the astrometric binary sample and identified 24 candidates for having BH or neutron star companions. They excluded Gaia BH1 from their sample as probably spurious because its orbital period is close to $3\times$ the {\it Gaia} precession period. Our follow-up has shown that the {\it Gaia} solution was nevertheless robust.

The candidates \citet[][]{Andrews2022} do consider all have rather low masses for BHs. Only one, {\it Gaia} DR3 6328149636482597888, has an inferred companion mass above $2\,M_{\odot}$ at the 2$\sigma$ level. We have begun spectroscopic monitoring of this source, for which \citet{Andrews2022} reported $M_2 = 2.71_{-0.36}^{+1.50}\,M_{\odot}$ (2$\sigma$ uncertainties), and we find it to be a metal-poor turn-off star ($\rm [Fe/H] \approx -1.5$) on a halo-like orbit. Andrews et al. assumed a mass $M_\star = 1.21\pm 0.2$ (2$\sigma$ uncertainties) for the luminous star, but we find this to be an overestimate. Fitting the SED, we find $T_{\rm eff} = 6150 \pm 100\,\rm K$ and $R = 1.58 \pm 0.05\,R_{\odot}$, which for $\rm [Fe/H] = -1.5$ MIST models, we find to correspond to $M_\star = 0.78\pm 0.05\,M_{\odot}$. Adopting this mass for the luminous star and fitting the {\it Gaia} astrometric solution, we find $M_2 = 2.25_{-0.26}^{+0.61}\,M_{\odot}$ (2$\sigma$ uncertainties), consistent with a neutron star. The source's RVs thus far show a linear trend consistent with the astrometric solution. This is promising, but the {\it Gaia} orbital period is 736 days, so RV monitoring over an extended period is required to confirm the reliability of the {\it Gaia} solution, and more analysis is required to determine the nature of the companion. A BH seems unlikely. 

Among the lower-mass candidates identified by Andrews et al., which they propose the be neutron stars, a majority are significantly blueward of the main-sequence. This could indicate that (a) they have unresolved white dwarf companions, (b) they have low metallicities, or (c) the astrometric solutions are spurious.  Our spectroscopic follow-up of these objects has shown that a majority do not have unusually low metallicities. Spurious astrometric solutions or white dwarf companions both remain plausible explanations. The candidates in their sample that are not blueward of the main-sequence are all sufficiently luminous ($L\gtrsim 1\,L_{\odot}$) that white dwarf companions would likely escape detection. Among the neutron-star candidates, we find that when reliable mass estimates for the luminous stars are adopted from SED fitting, we in several cases infer companion masses significantly below the Chandrasekhar limit.

\citet{Shahaf2022} also recently searched for compact object companions among the {\it Gaia} DR3 astrometric solutions. They identified 8 sources with inferred companion masses above $2\,M_{\odot}$, including Gaia BH1. Three of the remaining seven sources, including all the sources with inferred $M_2$ above $3\,M_{\odot}$, are ruled out by our spectroscopic follow-up, and likely have spurious astrometric solutions. The remaining candidates in their sample have sufficiently long periods that the astrometric solutions could not yet be tested robustly. Their sample also contains a few dozen neutron star candidates, most with periods ranging from 400 to 1000 days. 

One neutron star candidate, {\it Gaia} DR3 5136025521527939072, was presented by \citet{Arenou2022}: it has an \texttt{AstroSpectroSB1} solution which was validated by their RV follow-up, and also by our follow-up. However, we find that this source is {\it also} metal-poor and on a halo-like orbit. Fitting the SED and spectra, we find ${[\rm Fe/H]} = -1.3, T_{\rm eff}= 6400\,{\rm K}, R = 1.17\,R_{\odot}$, and $M_\star = 0.77\pm 0.05\,M_{\odot}$, lower than the $M_\star = 1.2\,M_{\odot}$ assumed by \citet[][]{Arenou2022}. Adopting this $M_\star$ and fitting the {\it Gaia} solution, we obtain $M_2= 1.23\pm 0.06\,M_{\odot}$, which is lower than the $M_2 = 1.5\,M_{\odot}$ inferred by \citet[][]{Arenou2022}, and consistent with a white dwarf. This object and {\it Gaia} DR3 6328149636482597888 highlight another potential contaminant for astrometric compact object searches: moderately evolved, low-metallicity stars can be easily mistaken for higher-mass stars near the zero-age main sequence.

Further analysis of the \citet{Andrews2022} and \citet{Shahaf2022} candidates is warranted, as some could indeed host neutron star or very low-mass BH companions. RV follow-up is essential for determining the reliability of the astrometric solutions. As our follow-up has already shown (Appendix~\ref{sec:other_cands}), the probability of encountering a spurious solution (or simply one that is in the tails of the noise distribution) is high when one selects objects in sparsely-populated regions of parameter space. For companions with masses consistent with neutron stars, other models (e.g., two white dwarfs) are also harder to rule out. 

We conclude that {\it Gaia} DR3 astrometric solutions contain {\it one} unambigous dormant BH, although we cannot exclude the possibility that up to a few low-mass ($\lesssim 3\,M_{\odot}$) BHs also exist in the dataset. For BHs above $5\,M_{\odot}$, Gaia BH1 is the only solid candidate in the astrometric catalog among sources near the main-sequence (see Appendix~\ref{sec:other_cands}; there is one source with a red giant primary that {\it may} host a BH companion, but RV followup over a few-year period is required to vet it). Among {\it Gaia} DR3 spectroscopic binaries, there are almost certainly no solid BH companions to main-sequence stars \citep[][]{El-Badry2022_gaia_algols,  Jayasinghe2022}, but some neutron star candidates may turn out to be robust. BH companions to giants might still exist in the catalog, but are more difficult to vet. 

We note that if any of the neutron star or BH candidates discussed above turn out to be robust, their evolutionary histories will be a puzzle. Like Gaia BH1, they have orbits that are too wide to be understood as a result of a CE event, but too close to accommodate a red supergiant. In addition, their wide orbits could not survive the typical $\sim 250\,\rm km\,s^{-1}$ natal kicks expected for neutron stars. The existence of Gaia BH1 and these other candidates might thus point to a different dominant channel for forming moderately wide compact object + normal star binaries, such as the hierarchical triple scenario discussed in Section~\ref{sec:hierarchical}.

While this manuscript was under review, a preprint analyzing the same system was posted by \citet{Chakrabarti2022}. Their conclusions are broadly consistent with ours. The best-fit BH mass they report, $M_{2}= 11.9_{-1.6}^{+2.0}\,M_{\odot}$, is marginally higher than the value we found, $M_2 = 9.62\pm 0.18\,M_{\odot}$. We suspect that the difference in our results comes mainly from the fact that their RVs do not include any measurements near periastron, and thus do not constrain the RV amplitude well. This also leads to larger uncertainties in the BH mass.  Their best-fit solution predicts an RV $\approx 20\,\rm km\,s^{-1}$ larger than our measurements near periastron, so we consider their constraints less reliable than those reported here. Nevertheless, it is encouraging that an independent analysis, based on independent RV data, reached similar conclusions about the nature of the Gaia BH1 system. 

\section{Summary and conclusions}
\label{sec:conclusion}
We have presented the discovery of Gaia BH1, a new binary system consisting of a $\sim 10\,M_{\odot}$ dark object orbited by an otherwise unremarkable Sun-like star (Figure~\ref{fig:mosaic}). The system was first identified as a promising BH candidate based on its {\it Gaia} astrometric solution. Our follow-up spectroscopic observations validated the astrometric solution and allowed us to place tight dynamical constraints on the mass of the dark object (Figure~\ref{fig:rvfig}), which are summarized in Table~\ref{tab:system}. Even without consideration of the astrometric solution, the RVs imply a companion mass that can only be attributed to a BH (Figure~\ref{fig:spec_fm}; the $5\sigma$ lower limit on the spectroscopic mass function is $f\left(M_{2}\right)_{{\rm spec}}>3.68\,M_{\odot}$). Scenarios not involving a BH are firmly ruled out by the object's large mass and stringent limits on the light contributions of a luminous companion. 

Gaia BH1 is different in several ways from other known BHs in binaries. It is nearby ($d = 480\,\rm pc$) and bright ($G=13.8$). Together with the lack of contamination from an accretion disk and the fact that the luminous star is cool and slowly rotating, this makes it possible to study the system in greater detail in the optical than is possible for BH X-ray binaries. For example, we measure chemical abundances for 22 elements in the G star's photosphere (Table~\ref{tab:bacchus}) -- and find them entirely unremarkable. 
The fact that the orbit is measured with both astrometry and radial velocities (RVs) allows for tight constraints on the BH mass, with no inclination degeneracy. Several high-precision RV instruments are capable of measuring RVs for the G star with $\sim 1\,\rm m\,s^{-1}$ precision.

The system's properties are difficult to explain with evolutionary models for isolated binaries. The maximum radius of the BH progenitor is expected to have been much larger than the G star's current orbit, leading to an episode of common envelope evolution in which the G star spiraled in through the BH progenitor's envelope and ultimately ejected it. However, a 1 $M_{\odot}$ star is unlikely to have had enough orbital energy to eject the $\sim 20\,M_{\odot}$ envelope of a red supergiant, and would be in danger of merging with the core under standard CE assumptions. This is a generic problem for formation models of BH low-mass X-ray binaries (LMXBs). For Gaia BH1, there is an addition challenge: the orbit today is much wider than expected after a common envelope episode, which would be predicted to leave the G star just a few solar radii from the BH progenitor. One cannot invoke strong natal kicks to the BH to widen the orbit again in this scenario, as this would produce a highly eccentric orbit and high space velocity.

The system's evolution may be better-explained in models in which the G star  was initially a wide tertiary companion to a close binary containing two massive stars. In this case, interactions between the two stars could have prevented either one from expanding to become a red supergiant, such that the G star could have formed in an orbit similar to its current orbit (but somewhat tighter) and remained there ever since. In this case, the dark object could be a binary composed of two lower-mass BHs.  High-precision RV follow-up  offers the tantalizing possibility of testing this scenario. 
Alternatively, the binary may have formed through an exchange interaction. 
Its thin-disk-like Galactic orbit, modest eccentricity, and near-solar metallicity suggest that it did not form dynamically in a globular cluster, but dynamical formation in an open cluster that has since dissolved is more plausible.  

Gaia BH1 is about 3 times nearer to Earth than the next-closest BH known, the LMXB A0620-00 \citep[][]{McClintock1986}, whose {\it Gaia} parallax implies a distance of $1.44\pm 0.25$ kpc. This suggests that wide BH binaries like Gaia BH1, while being harder to detect, are significantly more common than BHs in close binaries with ongoing accretion. Given the conservative cuts applied to astrometric binary solutions published in {\it Gaia} DR3, it is likely that DR4 and DR5 will identify dozens of similar systems.  Gaia BH1 will likely remain one of the nearest and brightest BHs at periods $P_{\rm orb}\lesssim 1000$ days, but even nearer BH binaries may be discovered in longer-period orbits, which could not yet be characterized with DR3 data.

It is both interesting and irksome that the first unambiguous BH discovered by {\it Gaia} seems to be somewhat of an oddball, and not easily accommodated by standard binary evolution models. One interpretation is that -- among solar-type stars, which make up the bulk of the population to which {\it Gaia} is sensitive -- common envelope evolution leads to the merger of most would-be BH + normal star binaries. In this case, only systems formed through alternative channels will survive. It is clear that BH X-ray binaries -- which today are more compact versions of systems like Gaia BH1 -- only narrowly escape a merger if they form through a common envelope, and represent a rare outcome of binary evolution. BH + normal star binaries with $P_{\rm orb} \gtrsim 10$ years can avoid a common envelope, but (a) {\it Gaia} is not yet sensitive to them, and (b) they may be disrupted by natal kicks. BH + massive star binaries are more likely to both survive common envelope and undergo stable mass transfer, but the number of massive stars accessible with {\it Gaia} astrometric orbits is thus far modest. In any case, population demographics of the detached BH + normal star binary population will become clearer in the next decade, as more systems are discovered.

Because of its long orbital period and close distance, the angular size of Gaia BH1's orbit on the sky is more than 10 times larger than any other known BH binary. This makes the system an excellent target for interferometric follow-up. Observations at X-ray and radio wavelengths are also warranted. Although the accretion rate of the G star's wind by the BH is expected to be low and any detectable emission would be faint (Section~\ref{sec:xray_radio}), Gaia BH1 provides a unique opportunity to test imperfectly-understood models of accretion flows at very low accretion rates. 

In the future, the G star will become a red giant. After an X-ray bright period during which it transfers mass to the BH -- perhaps appearing as a wider version of the red giant + BH binary GRS 1915+105 -- the system will terminate its evolution as a wide BH + white dwarf binary. If the system turns out to host an inner BH binary, this evolution may be interrupted by its merger, with the resulting recoil potentially disrupting the outer binary.

\section*{Acknowledgements}
We thank Morgan Macleod, E. Sterl Phinney,  Rohan Naidu, Lieke van Son, Mathieu Renzo, Tom Maccarone, Josh Simon, and Jeff Andrews for helpful discussions, and Vedant Chandra for help with data reduction. We are grateful to Yuri Beletsky, Sam Kim, Angela Hempel, and Régis Lachaume for observing help. 
HWR acknowledges the European Research Council for the ERC Advanced Grant [101054731].
This research made use of pystrometry, an open source Python package for astrometry timeseries analysis \citep[][]{Sahlmann2019}. This work made use of Astropy,\footnote{http://www.astropy.org} a community-developed core Python package and an ecosystem of tools and resources for astronomy \citep{AstropyCollaboration2022}.

This project was developed in part at the Gaia Fête, held at the Flatiron Institute's Center for Computational Astrophysics in June 2022, and in part at the Gaia Hike, held at the University of British Columbia in June 2022. 
This work has made use of data from the European Space Agency (ESA) mission
{\it Gaia} (\url{https://www.cosmos.esa.int/gaia}), processed by the {\it Gaia}
Data Processing and Analysis Consortium (DPAC,
\url{https://www.cosmos.esa.int/web/gaia/dpac/consortium}). Funding for the DPAC
has been provided by national institutions, in particular the institutions
participating in the {\it Gaia} Multilateral Agreement.

Based on observations obtained at the international Gemini Observatory, a program of NSF’s NOIRLab, which is managed by the Association of Universities for Research in Astronomy (AURA) under a cooperative agreement with the National Science Foundation on behalf of the Gemini Observatory partnership: the National Science Foundation (United States), National Research Council (Canada), Agencia Nacional de Investigación y Desarrollo (Chile), Ministerio de Ciencia, Tecnología e Innovación (Argentina), Ministério da Ciência, Tecnologia, Inovações e Comunicações (Brazil), and Korea Astronomy and Space Science Institute (Republic of Korea).

\section*{Data Availability}
Data used in this study are available upon request from the corresponding author. 



\bibliographystyle{mnras}

\appendix

\section{Observations and data reduction}
\label{sec:appendix}

\subsection{MagE}
\label{sec:MagE}
We observed Gaia BH1 4 times using the Magellan Echellette spectrograph \citep[MagE;][]{Marshall2008}  on the 6.5m Magellan Clay telescope at Las Campanas Observatory. All observations were carried out with the 0.7 arcsec slit and exposure times ranging from 300 to 600 seconds. This yielded a typical SNR of 100 per pixel at 5,500\,\AA, spectral resolution $R\approx 5,400$, and wavelength coverage of $3,500-11,000$\,\AA. To minimize drift in the wavelength solution due to flexure, we obtained a ThAr arc on-sky immediately after each science exposure. 

We reduced the spectra using \texttt{PypeIt} \citep[][]{Prochaska_2020}, which performs bias and flat field correction, cosmic ray removal, wavelength calibration, sky subtraction, extraction of 1d spectra, and heliocentric RV corrections. To assess the stability of the wavelength solution, we observed several standard stars, measured their RVs using the same procedure we used for the science targets, and compared the results to their measured {\it Gaia} DR3 RVs. We found that the MagE-measured RVs were $1.5\pm 2\,\rm km\,s^{-1}$ larger than those measured by {\it Gaia}, so we adopted a conservative uncertainty of $3\,\rm km\,s^{-1}$ for all the MagE RVs. 
 
\subsection{GMOS}
\label{sec:GMOS}
We observed Gaia BH1 4 times using the Gemini Multi-Object Spectrograph \citep[GMOS;][]{Hook2004} on the 8.1m Gemini-North telescope on Mauna Kea (program GN-2022B-DD-202). We used the R831\_G5302 grating with a 0.5 arcsecond slit and central wavelength of 5200 \AA, leading to wavelength coverage from 4000 to 6400\,\AA\,\,and resolution $R\approx 4,400$. We obtained a CuAr arc on-sky immediately after each science exposure. We used 342 second exposures, yielding a typical SNR of 90 per pixel at 6000\,\AA. 

We reduced the data using \texttt{PypeIt}, which required construction of a new template for the R831 grating at blue wavelengths. Cross-correlation with a template yielded formal RV uncertainties of order $1\,\rm km\,s^{-1}$, but we found variations of up to $10\,\rm km\,s^{-1}$ between RVs measured at the blue and red ends of the spectrum. We adopted a RV uncertainty of $4\,\rm km\,s^{-1}$ for the GMOS RVs, which was also validated by observations of Gaia BH1 with higher-resolution instruments that occurred nearly concurrently with some of the GMOS observations. 

\subsection{X-shooter}
\label{sec:XSHOOTER}

We observed Gaia BH1 4 times using the X-shooter spectrograph \citep[][]{Vernet2011} on the 8.2m UT3 telescope at the VLT on Cerro Paranal (program 2109.D-5047). We used the 0.5, 0.4, and 0.4 arcsec slits on the UVB, VIS, and NIR arms. This setup yielded a resolution of $R\approx 9,700$ in the UVB arm, $R\approx 18,400$ in the VIS arm, and $R\approx 11,600$ in the NIR arm, with near-continuous wavelength coverage from 3100-24000\,\AA. We reduced the spectra using \texttt{pypeit}. 

We generally measured RVs using the NIR spectra, whose wavelength solution is calculated from telluric OH lines in the science spectra, yielding a maximally stable wavelength solution. For the first epoch, we inadvertently used a 0.9 arcsec slit in the NIR arm, yielding a lower spectral resolution. We measured the RV from the VIS arm for this epoch. We adopt an uncertainty of $1\,\rm km\,s^{-1}$ on all the X-shooter RVs.

\subsection{HIRES}
\label{sec:HIRES}
We observed Gaia BH1 9 times using the High Resolution Echelle Spectrometer \citep[HIRES;][]{Vogt1994} on the 10m Keck I telescope on Maunakea. The data were obtained and reduced using the standard California Planet Survey setup \citep[CPS;][]{Howard2010}, including use of the C2 decker (0.86 arcseconds $\times$ 14 arcseconds), which yields spectra with $R\approx 55,000$ and wavelength coverage over most of 3700--8000\,\AA, with 2 gaps 100--200\,\AA\,wide. We used 600 second exposures, yielding SNR of order 40 per pixel at 6000\,\AA. The CPS reduction includes sky-subtraction using the long C2 decker.

As part of the CPS pipeline, absolute RVs are measured using the method described by \citet{Chubak2012}: to correct for instrumental shifts in the wavelength zero-point, the telluric A and B molecular oxygen absorption bands of each star are cross-correlated with those of RV standard stars, whose adopted RVs are matched to the scale of \citet{Nidever2002} and other RV standard catalogs. The resulting RVs are robust to $0.1\,\rm km\,s^{-1}$. We use these RVs throughout our analysis. Most of our observations were performed through an iodine cell, making it possible to measure higher-precision RVs from these data in the future, after a higher-SNR template spectrum has been obtained and the method of \cite{Butler1996} can be applied.

\subsection{FEROS}
\label{sec:FEROS}
We observed Gaia BH1 17 times with the Fiberfed Extended Range Optical Spectrograph \citep[FEROS;][]{Kaufer1999} on the 2.2m ESO/MPG telescope at La Silla Observatory. We used $2\times 2$ binning in order to reduce readout noise (allowing us to measure RVs down to $G\approx 16$); this resulted in an effective resolution $R\approx 37000$ measured from sky lines. We took 2400 second exposures and achieved a typical SNR of 20 per pixel. We reduced the data with the standard ESO MIDAS pipeline, which performs bias-subtraction, flat fielding, wavelength calibration, and sky subtraction, and merges the spectra from different orders. 

On some nights, the wavelength solution drifted from the one established by the afternoon arcs by up to a few $\rm km\,s^{-1}$. We corrected this by cross-correlating an arc spectrum (which is taken during each exposure with a second fiber) with a reference arc spectrum taken on the first night of our observations. Based on observations of an RV standard with the same setup, we adopted and uncertainty of $0.3\,\rm km\,s^{-1}$ for all the FEROS observations, except in one case where poor conditions yielded a formal uncertainty of  $0.5\,\rm km\,s^{-1}$. Near-concurrent FEROS and HIRES observations of the source allowed us to verify that the two RV zeropoints are consistent within this tolerance.

\subsection{ESI}
\label{sec:ESI}
We observed Gaia BH1 one time with the Echellette Spectrograph and Imager \citep[ESI;][]{Sheinis2002} on the 10m Keck II telescope on Maunakea. We used a 300 second exposure with the 0.3 arcsec slit, yielding a resolution $R\approx 12000$ and SNR = 60, with useful wavelength coverage of 3900-10000\,\AA. We reduced the data using the MAuna Kea Echelle Extraction (MAKEE) pipeline, which performs bias-subtraction, flat fielding, wavelength calibration, and sky subtraction, and we refined the wavelength solution using sky lines. Based on RV measurements of RV standards on the same night, we adopted an uncertainty of $1.5\,\rm km\,s^{-1}$.

\section{Gaia observations}
\label{sec:gost}

As discussed in Section~\ref{sec:bigger_uncertainties}, the astrometric uncertainties for the parameters describing the orbit of Gaia BH1 are unusually large for the star's apparent magnitude. Here we show that this is likely a consequence of the {\it Gaia} scanning law. 

Epoch-level astrometric data is not published in DR3, but one can determine when a source was observed using the {\it Gaia} observation scheduling tool (GOST).\footnote{https://gaia.esac.esa.int/gost/} Given a source position, GOST returns a list of observation times when the scanning law predicts that a source will transit across the {\it Gaia} focal plane. It is not guaranteed that {\it Gaia} actually obtains data on each source at the predicted times, as the tool does not account for gaps between CCDs and issues that cause temporary gaps in the datastream, such as micrometeor impacts. However, for Gaia BH1, GOST predicts 21 visibility periods (i.e., groups of observations separated from other groups by at least 4 days), and the \texttt{gaia\_source} table reports that 21 visibility periods were indeed used in the astrometric solution. This suggests that the scan times predict by GOST are likely a good approximation of reality. 

\begin{figure*}
    \centering
    \includegraphics[width=\textwidth]{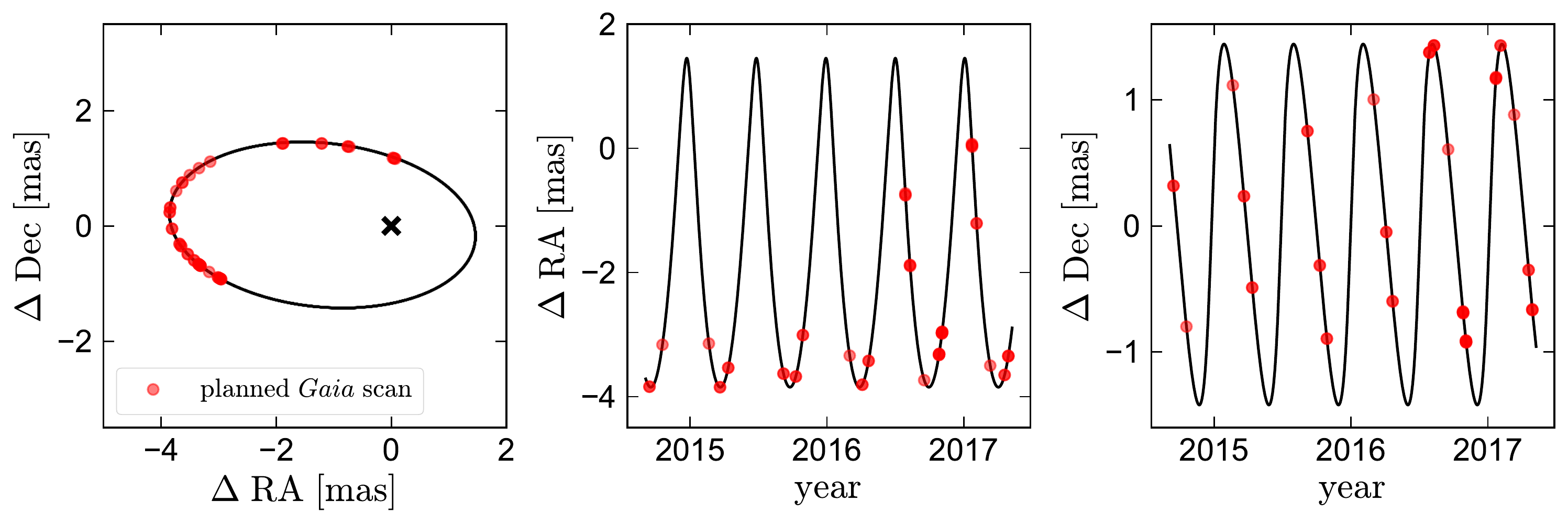}
    \caption{Predicted observation times of Gaia BH1 from the {\it Gaia} observation scheduling tool (GOST). Black line shows the best-fit photocenter orbit. Red points show the predicted photocenter positions at the times when GOST predicts {\it Gaia} would have observed the source. Note that we do not have access to the actual measured $\Delta \rm RA$ and $\Delta \rm Dec$ values; only to the predicted scan times. GOST predicts that {\it Gaia} observed the source in 21 visibility periods which all fall on one side of the orbit. This may be a result of the fact that the orbital period is a near-integer multiple of {\it Gaia's} 63-day precession period, and almost exactly half a year.}
    \label{fig:scantimes}
\end{figure*}

We visualize these times in Figure~\ref{fig:scantimes}. It is clear that although {\it Gaia} DR3 covered more than 5 orbital period of Gaia BH1, the phase coverage is far from uniform, and only about half of the orbit in phase was covered by astrometric data. This may be a consequence of the fact that the orbital period is a near-integer multiple of the {\it Gaia} precession period, and close to half a year. 
Our RV follow-up shows that despite this limitation, {\it Gaia} was able to measure an astrometric orbit that was basically correct. However, the incomplete phase coverage make it unsurprising that the uncertainties in the astrometric parameters are larger than usual for a source of this brightness. 

GOST predicts that the north side of the orbit will remain poorly sampled even in {\it Gaia} DR4. We note that a few strategically timed observations with {\it HST}, {\it JWST}, GRAVITY+, or MICADO could fill in this part of the orbit and yield improved astrometric precision if analyzed together with the epoch-level {\it Gaia} astrometric data that will become available in DR4.

\section{Comparison of the astrometric and RV solutions}
\label{sec:comp_rvs_appendix}
\begin{figure*}
    \centering
    \includegraphics[width=\textwidth]{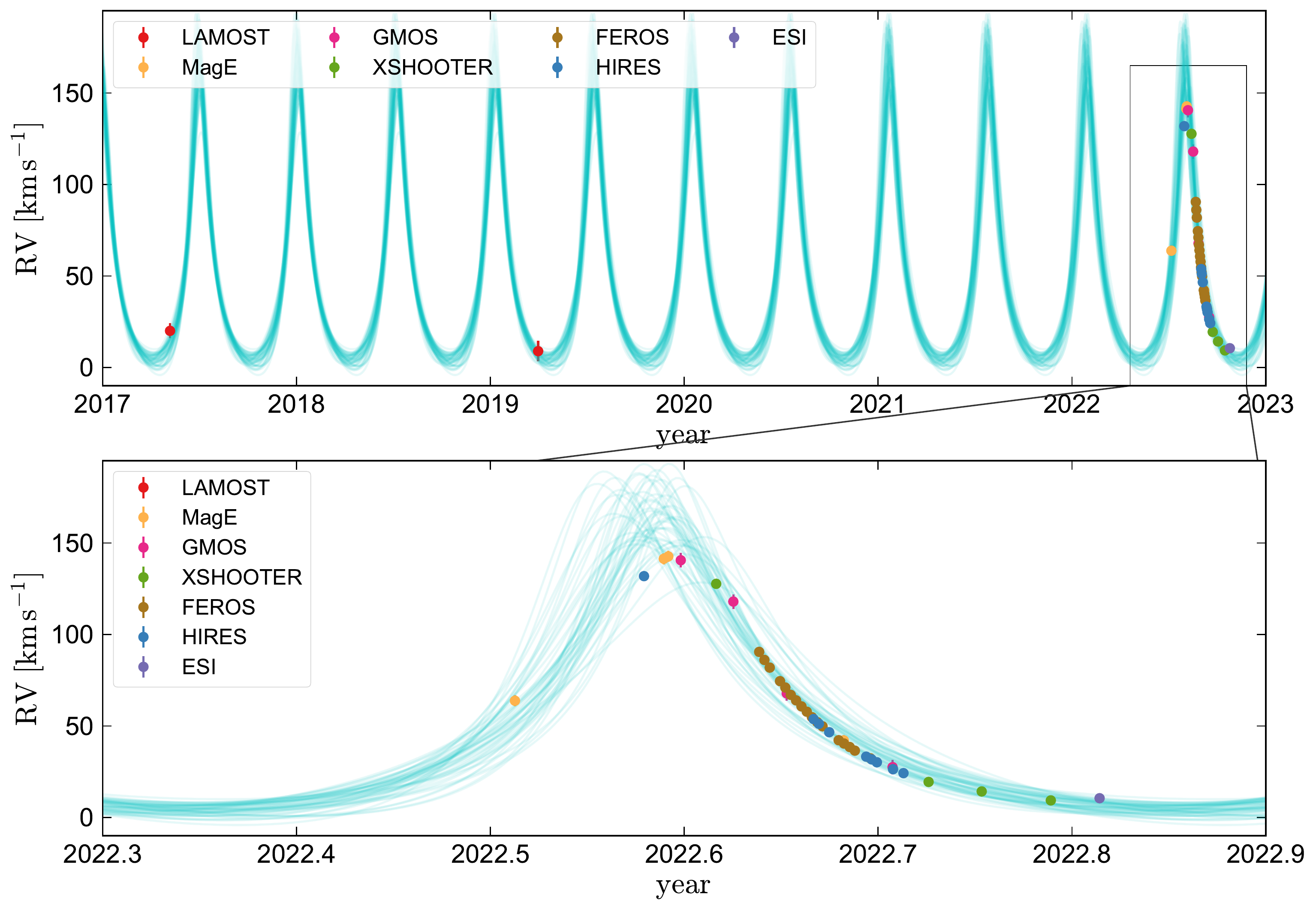}
    \caption{Comparison of measured RVs (same as Figure~\ref{fig:rvfig}) to predictions from the {\it Gaia} astrometric solution alone. The center-of-mass RV (which is not constrained by astrometry) is set to the value inferred from RVs, but the cyan lines otherwise do not know about the measured RVs. The uncertainty in the predictions is larger than when RVs are included in the fit, but the measured RVs are generally consistent with the predictions, validating the {\it Gaia} solution. }
    \label{fig:rvs_gaia_only}
\end{figure*}

As discussed in Section~\ref{sec:mass}, the tightest constraints on the Gaia BH1 orbit can be obtained through joint fitting of the RVs and astrometric constraints. However, it is also useful to compare the RVs predicted by the {\it Gaia} solution alone to the measured RVs, in order to assess whether or not the astrometric solution is consistent with the RVs. Figure~\ref{fig:rvs_gaia_only} compares the measured RVs to predictions based on the {\it Gaia} astrometric solution, which are generated from posterior draws of the fit without RVs. Because the astrometric solution does not constrain the center-of-mass velocity, we set it to $\rm 45\,km\,s^{-1}$ (consistent with the measured value) in all cases. 

\begin{figure*}
    \centering
    \includegraphics[width=0.9\textwidth]{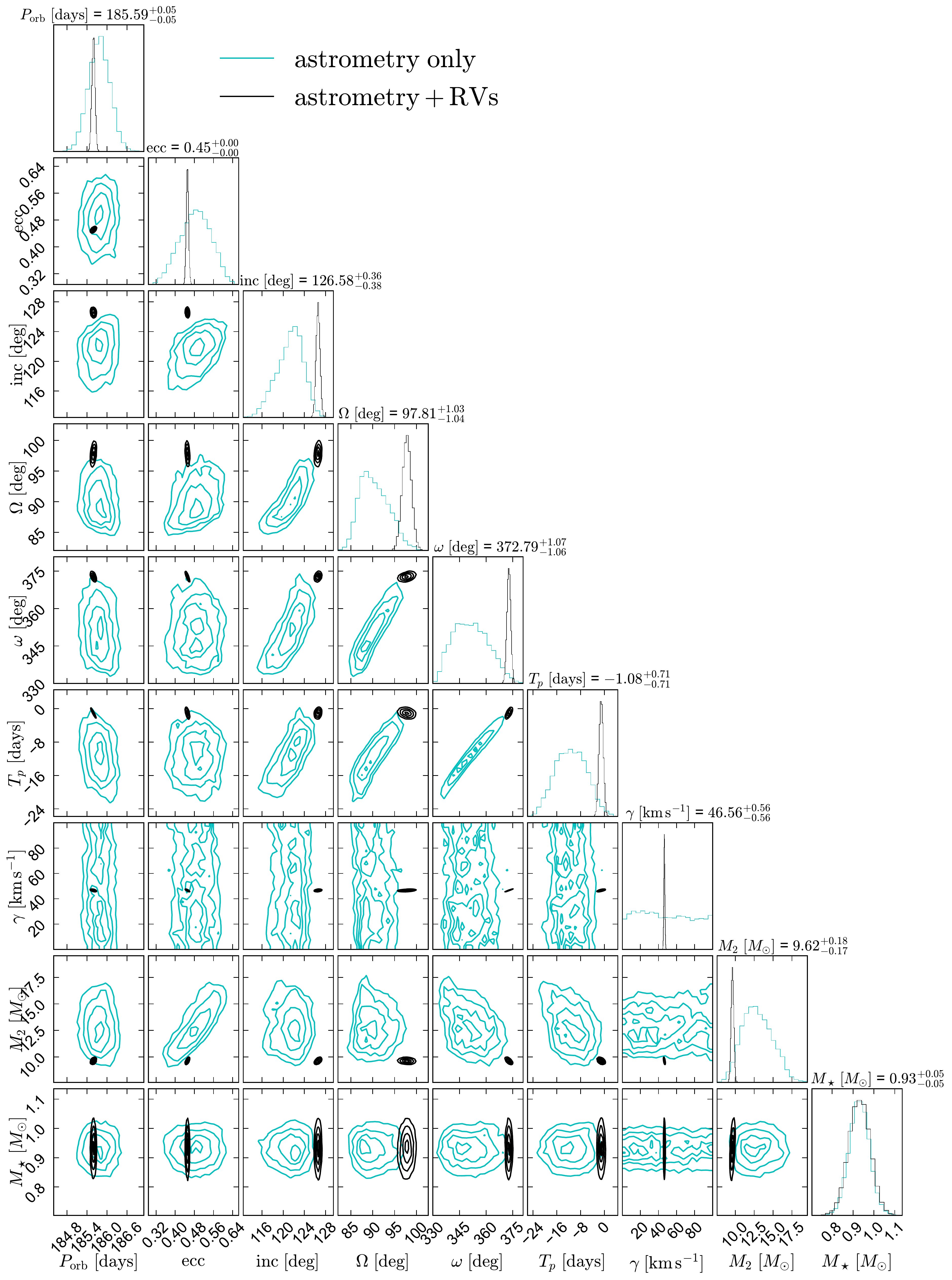}
    \caption{Comparison between constraints obtained from joint fitting of RVs and astrometry (black) and constraints from astrometry alone (cyan). The two sets of constraints are generally consistent, but those with RVs are tighter. Numbers on the diagonals reflect the joint astrometry + RVs constraints. Both sets of constraints are listed in Table~\ref{tab:system}. }
    \label{fig:corner_plot_comparison}
\end{figure*}

The measured RVs are basically consistent with the {\it Gaia}-only solution, in the sense that some posterior draws go through all the RVs. Figure~\ref{fig:corner_plot_comparison} and Table~\ref{tab:innes_elememts} also show that all inferred parameters are consistent within $\lesssim 2\sigma$. This suggests that the {\it Gaia} solution and its uncertainties are reliable, despite the incomplete orbit coverage discussed in Appendix~\ref{sec:gost}. 

 \begin{table*}
\begin{tabular}{lllll}
Parameter & units &  Astrometry-only constraint  & Joint RVs+astrometry constraint & How many sigma discrepant? \\
\hline
\texttt{a\_thiele\_innes} & mas & $-0.26\pm 0.17$ & $-0.01 \pm 0.05$ & 1.1 \\
\texttt{b\_thiele\_innes} & mas & $2.93\pm 0.18$  & $2.62\pm 0.02$    & 1.6 \\
\texttt{f\_thiele\_innes} & mas & $1.52\pm 0.16$  & $1.61\pm 0.02$   & 0.5 \\
\texttt{g\_thiele\_innes} & mas & $0.53\pm 0.55$  & $-0.36\pm 0.05$  & 1.5 \\ 
\texttt{period}         & days& $185.77 \pm 0.31$ &$185.59\pm 0.05$ & 0.5 \\ 
\texttt{eccentricity}   &     & $0.49\pm 0.07$  & $0.45\pm 0.01$ & 0.5 \\
\texttt{t\_periastron}   & days& $-12.0\pm 6.3$  & $-1.1\pm 0.7$  & 1.6 \\

\end{tabular}
\caption{Comparison of the {\it Gaia} astrometric constraints on Gaia BH1's photocenter ellipse to constraints from the joint RVs+astrometry fit. All parameters are consistent at the 1.6$\sigma$ level. }
\label{tab:innes_elememts}
\end{table*}

\section{Radial velocities}
\label{sec:RVs_appendix}
Radial velocities are listed in Table~\ref{tab:rvs}.
All times are reported as mid-exposure times in the heliocentric frame (HJD UTC). RVs are also in the heliocentric frame and their uncertainties include uncertainties in wavelength calibration. The reported SNR is at the wavelength range we used for RV measurement, which in most cases was near $5400$ \AA. 

\begin{table*}
\begin{tabular}{lllll}
HJD UTC & Radial velocity ($\rm km\,s^{-1}$) & Instrument & Resolution $R$ & SNR \\
\hline
2457881.2899 & $20.0\pm 4.0$  & LAMOST & 1800 & 6 \\
2458574.3663 & $8.9 \pm 5.6$  & LAMOST & 1800 & 55 \\
2459767.6226 & $63.8 \pm 3$   & MagE  & 5400 & 110 \\
2459791.9186 & $131.90 \pm 0.1$  & HIRES  & 55000 & 40 \\
2459795.6461 & $141.4 \pm 3$  & MagE  & 5400 & 100 \\
2459796.4995 & $142.7 \pm 3$  & MagE  & 5400 & 100 \\
2459798.8399 & $140.6 \pm 4$  & GMOS  & 4300 & 80 \\
2459805.5101 & $127.7 \pm 1.0$  & X-shooter & 18400 & 55 \\
2459808.7388 & $118.0 \pm 4$  & GMOS  & 4300 & 75 \\
2459813.6045 & $90.5 \pm 0.3$   & FEROS & 37000 & 20 \\
2459814.5874 & $86.1 \pm 0.3$   & FEROS & 37000 & 15 \\
2459815.5927 & $81.9 \pm 0.3$   & FEROS & 37000 & 19 \\
2459817.5278 & $74.5 \pm 0.3$   & FEROS & 37000 & 20 \\
2459818.5266 & $71.0 \pm 0.3$   & FEROS & 37000 & 20 \\
2459818.7870 & $67.8 \pm 4$  & GMOS  & 4300 & 85 \\
2459819.5543 & $67.0 \pm 0.3$   & FEROS & 37000 & 20 \\
2459820.5465 & $64.0 \pm 0.3$   & FEROS & 37000 & 21 \\
2459821.5669 & $60.7 \pm 0.3$   & FEROS & 37000 & 22 \\
2459822.5745 & $57.8 \pm 0.3$   & FEROS & 37000 & 22 \\
2459823.5422 & $54.8 \pm 0.3$   & FEROS & 37000 & 21 \\
2459823.8525 & $53.76 \pm 0.1$ & HIRES & 55000 & 40 \\
2459824.5305 & $52.1 \pm 0.3$   & FEROS & 37000 & 18 \\
2459824.8516 & $51.18 \pm 0.1$ & HIRES & 55000 & 40 \\
2459825.5361 & $49.8 \pm 0.3$   & FEROS & 37000 & 20 \\
2459826.7920 & $46.59 \pm 0.1$ & HIRES & 55000 & 40 \\
2459828.5677 & $42.2 \pm 0.3$   & FEROS & 37000 & 20 \\
2459829.5373 & $42.1 \pm 3$    & MagE & 5400 & 80 \\
2459829.5768 & $40.4 \pm 0.3$   & FEROS & 37000 & 17 \\
2459830.6452 & $38.5 \pm 0.3$   & FEROS & 37000 & 10 \\
2459831.6223 & $36.5 \pm 0.5$   & FEROS & 37000 & 5 \\
2459833.7523 & $33.23 \pm 0.1$   & HIRES & 55000 & 40 \\
2459834.5509 & $32.3 \pm 0.3$   & FEROS & 37000 & 11 \\
2459834.7691 & $31.74 \pm 0.1$   & HIRES & 55000 & 40 \\
2459835.7678 & $30.14 \pm 0.1$   & HIRES & 55000 & 40 \\
2459838.8082 & $26.35 \pm 0.1$   & HIRES & 55000 & 40 \\
2459838.7208 & $27.5 \pm 4$  & GMOS & 4300 & 78 \\
2459840.7729 & $24.20 \pm 0.1$   & HIRES & 55000 & 40 \\
2459845.5069 & $19.4 \pm 1.0$  & X-shooter & 11600 & 45 \\
2459855.5012 & $14.2 \pm 1.0$  & X-shooter & 11600 & 45 \\
2459868.5128 & $9.3  \pm 1.0$  & X-shooter & 11600 & 40 \\
2459877.6978 & $10.5  \pm 1.5$  & ESI & 12000 & 60 \\

\end{tabular}
\caption{Radial velocities.}
\label{tab:rvs}
\end{table*}

\section{Other candidates}
\label{sec:other_cands}

\begin{figure*}
    \centering
    \includegraphics[width=\textwidth]{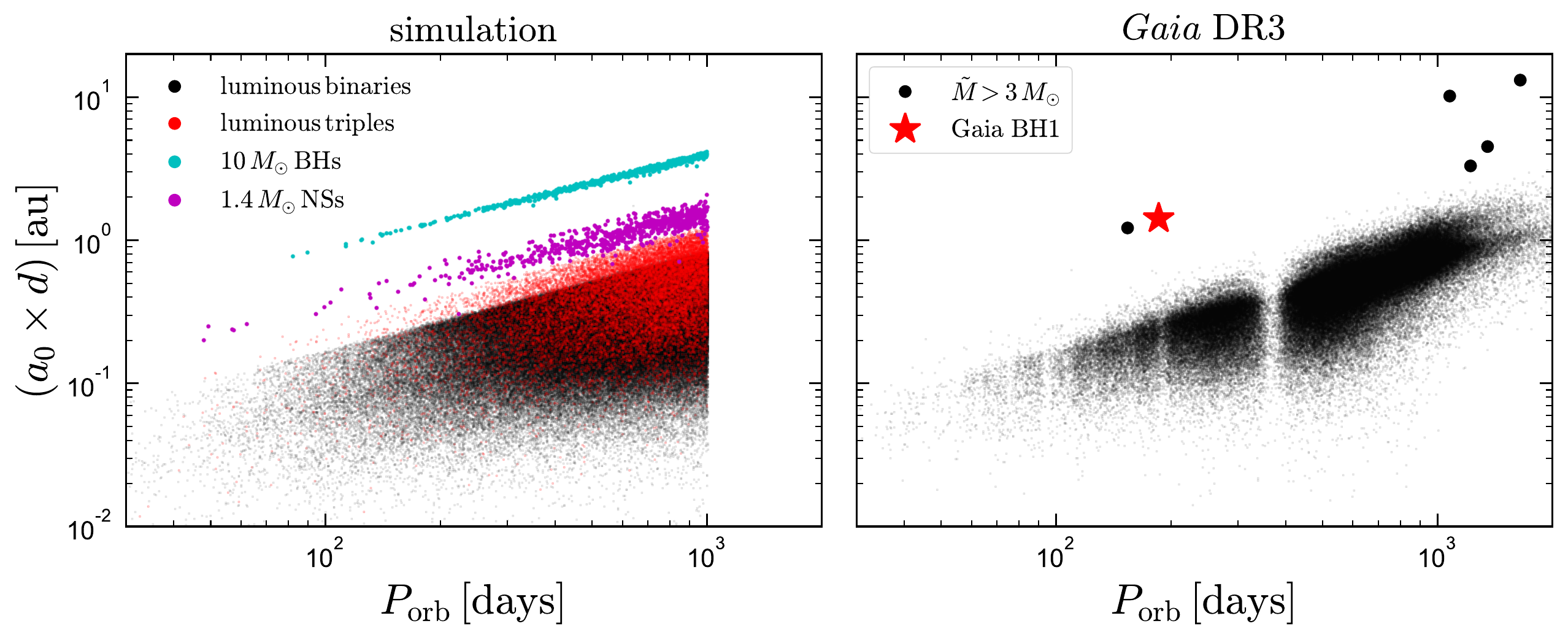}
    \caption{Parameter space within which we selected BH candidates. Left panel shows a simulation, in which we model a population of luminous (main-sequence) binaries, triples, and compact object + normal star binaries subject to the cuts imposed on binary solutions published in DR3. The relative number of objects in the three populations is chosen arbitrarily. Binaries containing dark companions fall at larger separations at fixed period than binaries containing two (or three) luminous stars. Right panel shows the same quantities for binaries published in DR3. Red star shows Gaia BH1; black points show other initially promising candidates, which we discuss in the text. }
    \label{fig:keplerplot}
\end{figure*}

Figure~\ref{fig:keplerplot} illustrates our strategy for selecting BH candidates among the {\it Gaia} astrometric solutions. We show that binaries containing a BH and a low-mass main-sequence star are expected to have much larger photocenter wobble than binaries containing two luminous stars. 

The left panel shows a simulation. Here we begin with all sources in the \texttt{gaia\_source} catalog with \texttt{parallax\_over\_error} > 50. We then attached a simulated binary population to these sources, drawing periods from the canonical lognormal distribution for solar-type stars \citep[][]{Raghavan2010} and mass ratios from a uniform distribution. For each simulated binary, we then calculate the expected photocenter semi-major axis and its uncertainty, and we retain only the binaries that satisfy the cuts imposed on the binary solutions published in DR3. We also generate a population of hierarchical triples, drawing both the inner and outer mass ratios from uniform distributions. We calculate the flux ratios using the $G-$band mass-magnitude relation tabulated by \citet{Janssens2022}. Finally, we generate a population of BH + normal star and neutron star + normal binaries, by assigning $10\,M_{\odot}$ and $1.4\,M_{\odot}$ dark companions to a random subset of the luminous sources, again drawing periods from the canonical lognormal. This is {\it not} meant to represent a realistic distribution of orbital periods, masses, or mass ratios, but only to show the expected photocenter orbit size at a given period.

In the right panel, we show the same observables for all objects in DR3 with astrometric solutions (\texttt{Orbital} and \texttt{AstroSpectroSB1} solutions). As expected, most solutions fall in the region of parameter space that we expect to be populated by luminous binaries and triples. For each binary, we calculate $\tilde{M}$, a quantity which is related to, but not equal to, the total dynamical mass: 

\begin{equation}
    \label{eq:Mtilde}
    \tilde{M}=\frac{4\pi^2}{P_{{\rm orb}}^{2}G}\left(\frac{a_{0}}{\varpi}\right)^{3}.
\end{equation}

For a binary with a luminous star and a dark companion, $\frac{a_{0}}{\varpi}=a\frac{q}{1+q}$, where $q = M_2/M_\star$ and $a$ is the semi-major axis. Thus, $\tilde{M}=M_{{\rm tot}}\left(\frac{q}{1+q}\right)^{3}$. For BH companions to low-mass stars, $q/(1+q)\approx 1$, and thus $\tilde{M}$ is only slightly less than $M_{\rm tot}$. On the other hand, for a binary with two luminous stars, $\frac{a_{0}}{\varpi}=a\delta_{q \ell}$, where
\begin{align}
    \label{eq:delta_ql}
    \delta_{q\ell}=\left|\frac{q/\left(1+\ell\right)-\ell/\left(1+\ell\right)}{\left(1+q\right)}\right|,
\end{align}
and $\ell = F_2/F_1$ is the ratio of the two stars' $G-$band fluxes \citep[e.g.][]{Penoyre2022}. For binaries with two main-sequence stars, $\delta_{q\ell }\lesssim 0.35$ \citep[e.g.][]{Shahaf2019}, and thus in general $\tilde{M}\lesssim 0.05 M_{\rm tot}$.  $\tilde{M}$ is thus a good quantity with which to select massive dark companions to low-mass stars, though more sophisticated methods exist that are sensitive to lower-mass BHs and BH companions to massive stars \citep[e.g.][]{Shahaf2019, Shahaf2022}. 

With this in mind, we selected objects from {\it Gaia} DR3 with $\tilde{M}>3\,M_{\odot}$. This yielded 6 objects, which are highlighted in the right panel of Figure~\ref{fig:keplerplot} and enumerated in Table~\ref{tab:cands}. Sources brighter than $G_{\rm RVs} = 12$ have multi-epoch RVs measured (but not yet published) by {\it Gaia}. The min-to-max range of these RVs is published as \texttt{rv\_amplitude\_robust}. Comparing this to the expected full RV amplitude implied by the {\it Gaia} solution provides a useful diagnostic of whether the solution is likely to be correct.

Each solution is accompanied by a \texttt{goodness\_of\_fit} value. As discussed by \citet{Halbwachs2022}, this quantity is formally expected to follow a normal distribution, $\mathcal{N}(0,1)$, with values significantly larger than 1 indicative of a problematic solution. However, this may not hold in practice, since the uncertainties in epoch-level measurements can be underestimated. We find that the distribution of  \texttt{goodness\_of\_fit} values displays a sharp discontinuity at $G=13$. Fainter values approximately follow a $\mathcal{N}(0,1.5)$ distribution, such that a value much larger than 1 is indeed unusual. However, for brighter solutions, the distribution peaks at $\approx 3.5$, and 13\% of solutions of have \texttt{goodness\_of\_fit} > 10. We thus expect significantly larger  \texttt{goodness\_of\_fit} values to be (potentially) consistent with a good solution for bright sources.

\newcommand{\xmark}{\ding{55}}%
\newcommand{\cmark}{\ding{51}}%

 \begin{table*}
 \begin{tabular}{lllllllll}
 Gaia DR3 source ID & $G$ &  $\tilde{M}$ & $P_{\rm orb}$  & $(a_{0}\times d)$  & GoF & \texttt{rv\_amplitude\_robust} & expected\,RV\,amplitude & Verdict \\
    & [mag] &  $[M_{\odot}]$ & $[\rm days]$  & $[\rm au]$  &   &[$\rm km\,s^{-1}$] & [$\rm km\,s^{-1}$] &   \\
 \hline
 3640889032890567040 & 9.2 &  $122\pm 47$   &  $1076\pm 12$    &$10.2 \pm 1.3$ & 10.3    & 12.9 & $674\pm 154$ & \textcolor{red}{\xmark} \\
 4467000291193143808 & 15.5 &  $119\pm 71$   &  $1647\pm 520$   & $13.4\pm 2.2$ & 4.9     & & $358\pm 182$ & \textcolor{red}{\xmark} \\
 4373465352415301632 & 13.8 &  $11.5 \pm 2.5$  &  $185.8\pm 0.3$  & $1.44\pm 0.10$ & 0.3  & & $165 \pm 20$ &\textcolor{green}{\cmark} \\
 6281177228434199296 & 11.3 &  $10.8 \pm 1.6$  &  $153.9 \pm 0.4$ & $1.24\pm 0.06$ & 8.0  & 20.1 & $171\pm 10$ & \textcolor{red}{\xmark} \\
 5870569352746779008 & 12.3 & $6.7 \pm 0.5$   &  $1352 \pm 45$ & $4.52\pm 0.13$ & 3.1     & 37.0 & $50\pm 2$ & \textcolor{orange}{\shrug} \\
 3664684869697065984 & 11.6 & $3.5 \pm 2.6$  &  $1220 \pm 233$  & $3.35 \pm 1.13$ & 3.6   & 18.3 & $54\pm 14$ & \textcolor{red}{\xmark} \\
 \end{tabular}
 \label{tab:cands}
 \caption{BH candidates highlighted in Figure~\ref{fig:keplerplot}. GoF is the \texttt{goodness\_of\_fit} parameter, for which large values indicate a potentially problematic solution (see text for details).  \texttt{rv\_amplitude\_robust} is the min-to-max range in the individual-epoch {\it Gaia} RV measurements (only available for bright sources); ``expected RV amplitude'' is the min-to-max RV amplitude predicted by the astrometric solution. Sources are discussed individually in the text. }
 \end{table*}
 
We comment briefly on each source:

{\it 3640889032890567040}: Poor GoF and implausibly high companion mass. Ruled out by low \texttt{rv\_amplitude\_robust}.

{\it 4467000291193143808}: Period of 1647 days is significantly longer than {\it Gaia} DR3 baseline. Implausibly high companion mass; poor GoF for the source magnitude. 

{\it 4373465352415301632}: The subject of this paper. 

{\it 6281177228434199296}: Ruled out by our RV follow-up. Low \texttt{rv\_amplitude\_robust} also makes the astrometric solution implausible. 

{\it 5870569352746779008}: Most plausible candidate beside Gaia BH1, but still uncertain. $P_{\rm orb}$ is longer than the {\it Gaia} DR3 observing baseline. The GoF is significantly larger than 1, but is not unusual for sources in the relevant range of $G$ magnitude. The CMD position is consistent with a lower giant, or possibly a red clump star.  The inclination is relatively low ($35\pm 1$ deg), so the nature of the companion hinges on the correctness of the astrometric solution. That is, if the {\it Gaia} solution is correct, RVs alone will only yield a spectroscopic mass function near $1\,M_{\odot}$, which is not high enough to unambiguously rule out a luminous companion to a red giant. The reliability of the astrometric solution will be easier to assess in DR4, when epoch-level astrometric data are published and a full orbit has been covered. The measured \texttt{rv\_amplitude\_robust} is not too far from the expected peak-to-peak amplitude. We have initiated a spectroscopic follow-up campaign, on which we will report in future work. Thus far, the RVs are consistent with the \texttt{AstroSpectroSB1} solution. Further RV monitoring is warranted and necessary. This source was recently investigated in detail by \citet{Tanikawa2022}, who found it to be the only credible BH candidate in {\it Gaia} DR3 with both astrometric and spectroscopic {\it Gaia} data, but did not carry out follow-up. If the solution turns out to be credible, the source will present a similar puzzle to Gaia BH1: the BH progenitor would likely not have fit inside the current orbit, but the orbit is much too wide to be a result of common envelope evolution. The current orbital separation is such that if the luminous star is a core helium burning red clump star, mass transfer via winds and/or stable Roche lobe overflow would have occurred prior to the helium flash. 

{\it 3664684869697065984}: Ruled out by APOGEE RVs.

For a $10\,M_{\odot}$ BH companion, the cut of $\tilde{M}>3\,M_{\odot}$ would be satisfied for luminous stars with $M_\star \lesssim 8\,M_{\odot}$. For higher-mass luminous stars, the AMRF, $\mathcal{A}$ statistic designed by \citet{Shahaf2019} is more sensitive. We verified that no objects in the \texttt{gaiadr3.nss\_two\_body\_orbit} catalog with  photometric masses above $5\,M_{\odot}$ have $\mathcal{A} > 0.6$, as would be required to identify a dark companion with high confidence.

\section{HIRES Cannon labels}
\label{sec:cannon}
The labels inferred by the {\it Cannon} fit to the HIRES spectrum of Gaia BH1 are listed in  Table~\ref{tab:cannon}. The method is not expected to yield reliable formal uncertainties on labels, because at high SNR the errors are dominated by systematics in the spectral model rather than photon noise. We thus report the cross-validation error achieved by \citet{Rice2020} for each label when testing the model, which provides a rough lower limit on the uncertainties in each label. 

The atmospheric parameters and abundances reported by {\it the Cannon} are quite similar to those we find with BACCHUS (Section~\ref{sec:abundances}), which we adopt as our fiducial values. Note that {\it the Cannon} reports [X/H], while for the BACCHUS fit we reported [X/Fe] = [X/H] - [Fe/H]. 

\begin{table}
\begin{tabular}{lll}
Parameter & Best-fit & Cross-validation error \\
\hline
$T_{\rm eff}\,[\rm K]$ & 5863 &  56  \\
$\log\left(g/{\rm cm\,s^{-2}}\right)$ &  4.43  &  0.09  \\
$v\sin i\,\left[{\rm km\,s^{-1}}\right]$ & 1.3  &  0.87 \\
$\rm [C/H] $& -0.24 &  0.05  \\
$\rm [N/H] $& -0.37 &  0.08  \\
$\rm [O/H] $& -0.10 &  0.07  \\
$\rm [Na/H]$& -0.33 &  0.05  \\
$\rm [Al/H]$& -0.28 &  0.04  \\
$\rm [Mg/H]$& -0.23 &  0.04  \\
$\rm [Si/H]$& -0.26 &  0.03  \\
$\rm [Ca/H]$& -0.26 &  0.03  \\
$\rm [Ti/H]$& -0.21 &  0.04  \\
$\rm [V/H] $&  -0.22 & 0.06   \\
$\rm [Cr/H]$& -0.30 &  0.04  \\
$\rm [Mn/H]$& -0.50 &  0.05  \\
$\rm [Fe/H]$& -0.29 &  0.03  \\
$\rm [Ni/H]$& -0.34 &  0.04  \\
$\rm [Y/H] $&  -0.31 & 0.08  \\
\end{tabular}
\caption{Parameters of the G star, from the HIRES cannon. The cross-validation error is taken from \citet{Rice2020} and represents an estimate of the uncertainty in each label.}
\label{tab:cannon}
\end{table}

\section{Dependence of Gaia astrometric uncertainties on observables}
\label{sec:sigma_a0_appendix}

\begin{figure*}
    \centering
    \includegraphics[width=\textwidth]{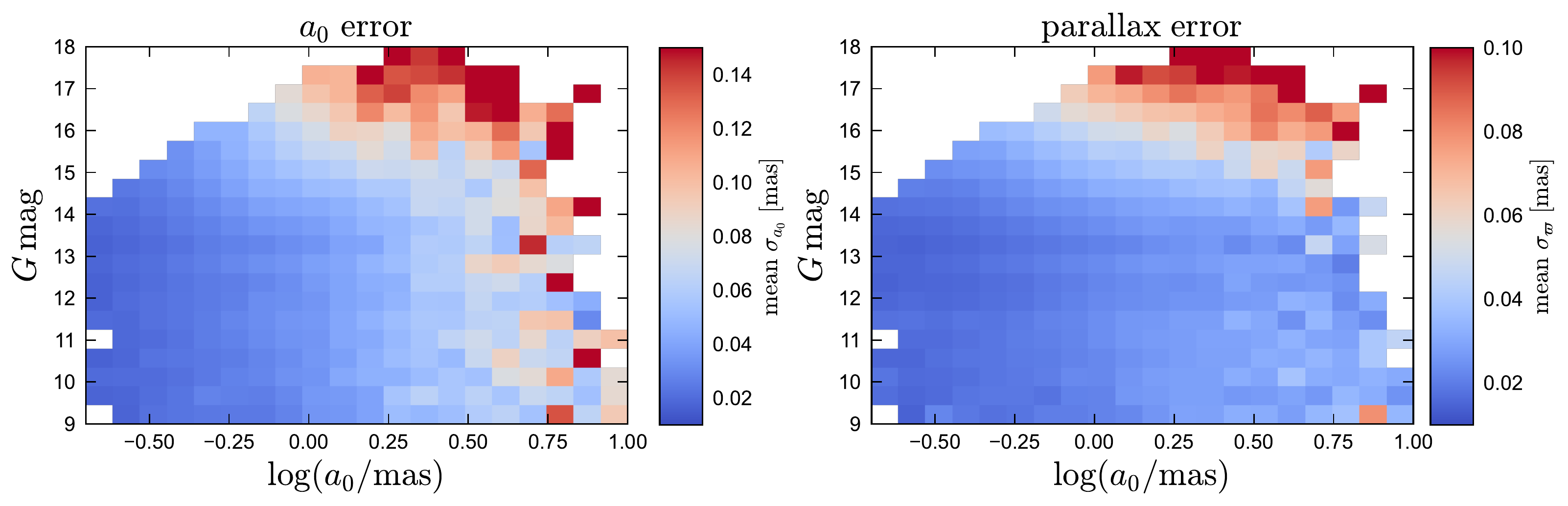}
    \caption{Mean uncertainty in photocenter semi-major axis  ($a_0$; left) and parallax (right) for all {\it Gaia} DR3 astrometric solutions with $P_{\rm orb} < 1000\,\rm days$. Fainter sources and sources with large $a_0$ have larger astrometric uncertainties. The correlation between $a_0$ and the astrometric uncertainties appears to be only partially a result of period-dependent quality cuts imposed on sources in DR3 (see text). We incorporate an explicit $a_0$-dependence when predicting the astrometric uncertainties of a hypothetical binary observed by {\it Gaia}. }
    \label{fig:a0_err}
\end{figure*}

In Figure~\ref{fig:a0_err}, we show the mean uncertainty in $a_0$ and $\varpi$ for DR3 astrometric binary solutions as a function of apparent magnitude and $a_0$. We only include binaries with best-fit $P_{\rm orb} < 1000$ days, which is roughly the time baseline of observations used to calculate DR3 solutions. The figure shows that typical astrometric uncertainties -- especially the uncertainty in $a_0$ -- depend both on apparent magnitude and on $a_0$. At fixed apparent magnitude, binaries with $a_0 \approx 3\,\rm mas$ have uncertainties in $a_0$ and $\varpi$ that are respectively about $2.5\,\times$ and $1.7\,\times$ larger than sources with  $a_0 \approx 0.3\,\rm mas$.

\begin{figure*}
    \centering
    \includegraphics[width=\textwidth]{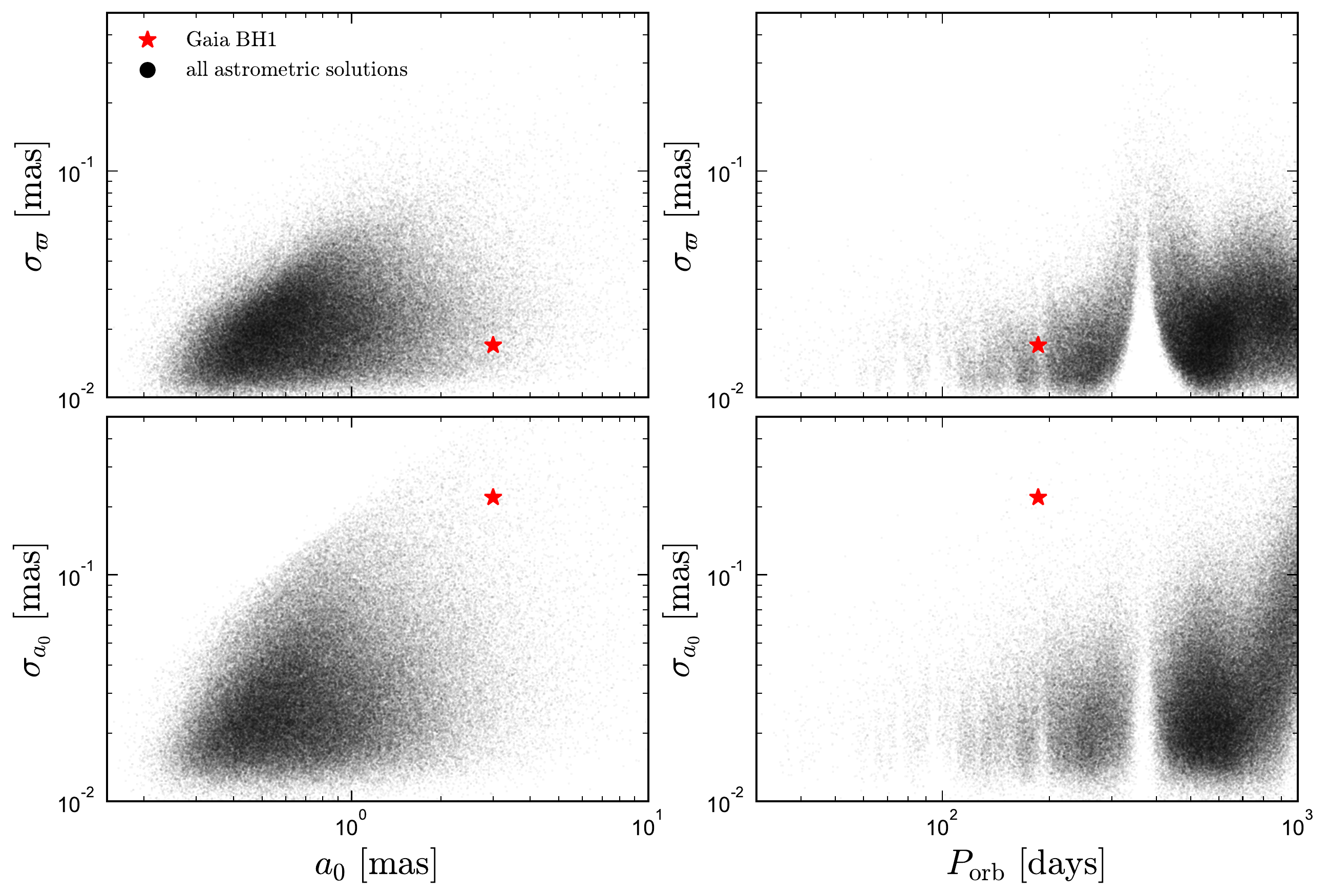}
    \caption{Uncertainties in parallax (top) and photocenter semi-major axis ($a_0$; bottom) for all sources with astrometric solutions published in DR3. Left column shows uncertainties as a function of $a_0$; right column shows them as a function of $P_{\rm orb}$. Red star shows Gaia BH1. A correlation is apparent between the astrometric uncertainties and $a_0$. This is in part because sources with small $a_0$ and large uncertainties were removed from the catalog through period-dependent quality cuts (Equations~\ref{eq:cut}-\ref{eq:ecut}). More structure is apparent as a function of $P_{\rm orb}$: binaries with periods close to 0.5 years and 1 year, as well as harmonics of the 63 day precession period, have larger uncertainties. Uncertainties in $a_0$ also increase at $P_{\rm orb} > 700$ days, likely due to incomplete phase coverage. } 
    \label{fig:a0_err_parallax}
\end{figure*}

It is not immediately obvious whether this trend reflects an actual dependence of the astrometric uncertainties on $a_0$ among all solutions that were initially fit, or is mainly a result of the astrometric SNR cuts imposed on the published solutions (Equations~\ref{eq:cut}-\ref{eq:ecut}). Figure~\ref{fig:a0_err_parallax} shows that removal of solutions with larger uncertainties and short periods likely has a significant effect on trends with $a_0$: there is a ``wedge'' of missing solutions with small $a_0$ and larger uncertainties. 

To determine whether an inherent dependence of the astrometric uncertainties on $a_0$ (or another parameter that scales with it) is required to explain the trends in Figure~\ref{fig:a0_err}, we preformed the following experiment. Using the simulation setup described in Appendix~\ref{sec:other_cands}, we begin with all sources in the \texttt{gaia\_source} catalog with \texttt{parallax\_over\_error} > 10 and create a simulated binary population that we then subject to the DR3 SNR cuts. To predict the uncertainties in $a_0$ and $\varpi$ for each simulated binary, we first tabulate the mean and standard deviation of $\sigma_{a_0}$ and $\sigma_{\varpi}$ as a function of apparent $G-$band magnitude, {\it multiply this mean by an $a_0$-dependent scaling factor}, and then draw uncertainties for each source from a Gaussian with the appropriate mean and standard deviation. Finally, we compare the simulated distribution of $\sigma_{a_0}$ and $\sigma_{\varpi}$ for binaries surviving the quality cuts to the observed distribution shown in Figure~\ref{fig:a0_err}.

We parameterize the scaling factor as $f = a\log(a_0/{\rm mas})+b$. We find that when $a=0$ and $b=1$ (i.e., no dependence of uncertainties on $a_0$), the predicted correlation between $a_0$ and the astrometric uncertainties is weaker than observed. We can best reproduce the observed distribution for $\sigma_{a_0}$ when $a=1.3$ and $b=1.4$; for $\sigma_{\varpi}$, we find $a =0.64$ and $b=1.2$. We adopt these values when predicting the expected astrometric uncertainties of a simulated binary population in Section~\ref{sec:howmany}. 

\bsp	
\label{lastpage}
\end{document}